\documentclass[a4paper,11pt]{article}
\usepackage{jheppub}
\usepackage{amsmath,amssymb,slashed,braket}
\usepackage{dsfont}
\usepackage{amsfonts}
\usepackage{bm,bbm}
\usepackage{graphicx}
\usepackage[dvipsnames,table]{xcolor}
\usepackage[normalem]{ulem}
\usepackage{hyperref}
\hypersetup{colorlinks,citecolor= blue,linkcolor= blue, urlcolor=blue}
\usepackage{array,boldline,makecell,booktabs}
\usepackage{verbatim}
\usepackage{epsfig}
\usepackage{multirow}
\usepackage{hhline}
\usepackage{bbold}
\usepackage{float}
\usepackage[version=4]{mhchem}
\usepackage{cleveref}
\usepackage{orcidlink}
\usepackage{subcaption}
\usepackage{caption}
\usepackage{placeins} 
\usepackage{arydshln}
\usepackage{pifont}
\allowdisplaybreaks[4]
\newcommand{\xmark}{\ding{55}}
\newcommand{\tr}{\mbox{Tr}}
\newcommand{\calO}{\mathcal{O}}

\newcommand{\C}{ {\tt C} }

\newcommand{\tL}{ {\tt L} }
\newcommand{\tR}{ {\tt R} }

\title{Investigation of nucleon decays into one lepton plus two mesons}

\author[a,b,c]{Wei-Qi Fan\,\orcidlink{0009-0001-5778-2571},}
\author[a,b]{Yi Liao\,\orcidlink{0000-0002-1009-5483}}
\emailAdd{liaoy@m.scnu.edu.cn}
\author[a,b]{and Xiao-Dong Ma\,\orcidlink{0000-0001-7207-7793}}
\emailAdd{maxid@scnu.edu.cn}
\affiliation[a]{State Key Laboratory of Nuclear Physics and
Technology, Institute of Quantum Matter, \\
South China Normal
University, \\
Guangzhou 510006, China}
\affiliation[b]{Guangdong Basic Research Center of Excellence for
Structure and Fundamental Interactions of Matter, 
Guangdong Provincial Key Laboratory of Nuclear Science, \\
Guangzhou 510006, China}
\affiliation[c]{School of Physics, Nankai University, \\
Tianjin 300071, China}

\abstract{ 
We systematically investigate baryon-number-violating (BNV) nucleon decays into one lepton ($e,\mu,\nu/\bar\nu$) and two pseudoscalar mesons ($\pi\pi,\pi\eta,\pi K$) within the low-energy effective field theory (LEFT) framework. 
By employing chiral perturbation theory, we obtain general expressions for the decay widths of these three-body nucleon decay modes induced by dimension-6 BNV LEFT operators and express them in terms of the associated Wilson coefficients. 
Since the same set of LEFT operators contribute to the experimentally well-constrained two-body nucleon decays, we then utilize the experimental bounds on them to constrain the relevant Wilson coefficients.
From the obtained constraints, we derive improved limits on the occurrence of 22 three-body modes involving a charged lepton and 9 modes containing a neutrino or an antineutrino,
with the new partial lifetime bounds being orders of magnitude stronger than the existing experimental limits. 
Our results serve as useful benchmarks for future experimental searches for these nucleon decays.
}


\makeatletter
\gdef\@fpheader{}
\makeatother
\begin{document} 

\maketitle
\setcounter{page}{2}

\section{Introduction}

Baryon number violation (BNV) is one of the three key conditions necessary to explain the matter-antimatter asymmetry observed in the Universe~\cite{Sakharov:1967dj}.
It may also play a role in generating tiny neutrino masses and addressing the dark matter conundrum, as proposed in various scenarios beyond the standard model (SM), see, e.g., Refs.~\cite{Barbieri:1979ag, 
Babu:1992ia,Fornal:2018eol,Elor:2018twp,Ge:2024lzy,Broussard:2025opd}. 
Experimentally, searches for nucleon decays provide the most feasible 
avenue to test this fundamental conservation law and unveil the nature of underlying BNV interactions.   

Over the past several decades, stringent experimental bounds on conventional two-body nucleon decays involving a lepton and a pseudoscalar meson have been established~\cite{ParticleDataGroup:2024cfk}, with partial lifetime bounds of order ${\calO}(10^{32\text{---}34})\,{\rm yr}$.
These limits will be further improved by ongoing and upcoming large-fiducial-mass neutrino experiments, such as Super-Kamiokande (Super-K), JUNO~\cite{JUNO:2015zny}, Hyper-Kamiokande~\cite{Hyper-Kamiokande:2018ofw}, DUNE~\cite{DUNE:2020ypp}, and two proposed detectors Theia~\cite{Theia:2019non} and ESSnuSB~\cite{ESSnuSB:2023ogw}. 

Theoretically, the same underlying BNV interactions responsible for two-body nucleon decays ${\tt N}\to lM$ will inevitably also give rise to three-body decay modes with an additional pseudoscalar meson ${\tt N}\to lM_1 M_2$,
where the nucleon ${\tt N}=p,\,n$, the lepton $l=e^\pm, \mu^\pm, \nu/\bar\nu$, and the pseudoscalar mesons $M,\,M_{1,2}=\pi^\pm,\,\pi^0,\,\eta,K^\pm, \bar K^0, K^0$.
Moreover, given the nature of the strong interactions involved in producing a second meson, one would not expect the decay rates for these three-body modes to be significantly suppressed by phase space alone. 
Despite their inevitability, these three-body processes remain largely unexplored theoretically, and only an estimate of ${\tt N} \to l\pi\pi$ was given in~\cite{Wise:1980ch}. 
Experimentally, they have also attracted far less attention than their two-body counterparts.
The old IMB-3 experiment conducted a search for 16 three-body modes 27 years ago~\cite{McGrew:1999nd}, with the resulting partial lifetime bounds that are generally one to two orders of magnitude weaker than those on two-body modes. 
Very recently, the Super-K experiment~\cite{Kamiokande:2026zwu} has taken a crucial step forward, improving the IMB-3 limits on the two modes 
$p\to (e^+,\mu^+)\pi^0\pi^0$ by 
more than one order of magnitude. 
In \cref{tab:data}, we summarize the experimental status of all kinematically allowed three-body nucleon decay modes into a single lepton and two pseudoscalar mesons. 
Note that the experimental bounds marked with a dagger are derived from inclusive searches reported in~\cite{Learned:1979gp,Cherry:1981uq}.

\begin{table}[t]
\centering
\resizebox{\linewidth}{!}{
\renewcommand{\arraystretch}{1.1}
\begin{tabular}{|c|l|c|l|c|l|c|l|c|}
\hline
\multirow{2}*{Class}
& \multicolumn{4}{c|}{Proton decay}  
& \multicolumn{4}{c|}{Neutron decay} 
\\
\cline{2-9}
& \multicolumn{1}{c|}{Charged lepton} 
& \multicolumn{1}{c|}{$\Gamma_{\tt exp}^{-1}\,[\rm yr]$}  
& \multicolumn{1}{c|}{(Anti)neutrino} 
& \multicolumn{1}{c|}{$\Gamma_{\tt exp}^{-1}\,[\rm yr]$} 
& \multicolumn{1}{c|}{Charged lepton} 
& \multicolumn{1}{c|}{$\Gamma_{\tt exp}^{-1}\,[\rm yr]$} 
& \multicolumn{1}{c|}{(Anti)neutrino} 
& \multicolumn{1}{c|}{$\Gamma_{\tt exp}^{-1}\,[\rm yr]$}  
\\
\hline
\multirow{14}*{\rotatebox[origin=c]{90}{$\Delta(B-L)=0$}}
& $p\to e^+ \pi^+ \pi^-$ & $8.2\times 10^{31}$~\cite{McGrew:1999nd}
& $p\to \bar \nu \pi^+ \pi^0$ & ---
& $n\to e^+ \pi^- \pi^0$ & $5.2\times 10^{31}$~\cite{McGrew:1999nd}
& $n\to \bar \nu \pi^+ \pi^-$ & ---
\\ 
& $p\to e^+ \pi^0 \pi^0$ & $7.2\times 10^{33}$~\cite{Kamiokande:2026zwu}
& $p\to \bar \nu \pi^+ \eta$ & ---
& $n\to e^+ \pi^- \eta $ 
& $0.6\times 10^{30}\,(\dagger)$~\cite{Learned:1979gp}
& $n\to \bar \nu \pi^0\pi^0 $  & ---
\\ 
& $p\to e^+ \pi^0 \eta$ 
& $0.6\times 10^{30}\,(\dagger)$~\cite{Learned:1979gp}
& $p\to \bar \nu \pi^+ K^0 $ & --- 
& $n\to e^+ \pi^- K^0 $  & $1.8\times 10^{31}$~\cite{Frejus:1990myz}
& $n\to \bar \nu \pi^0 \eta$ & ---
\\ 
& $p\to e^+ \pi^0 K^0$ 
&  $0.6\times 10^{30}\,(\dagger)$~\cite{Learned:1979gp}
& $p\to \bar \nu \pi^+ \bar K^0 $\,(\xmark) & ---
& $n\to e^+ \pi^- \bar K^0$\,(\xmark) 
& $0.6\times 10^{30}\,(\dagger)$~\cite{Learned:1979gp}
& $n\to \bar \nu \pi^+ K^-$\,(\xmark) & ---
\\ 
& $p\to e^+ \pi^0 \bar K^0$\,(\xmark) 
& $0.6\times 10^{30}\,(\dagger)$~\cite{Learned:1979gp}
& $p\to \bar \nu \pi^0 K^+ $ & ---
& $n\to e^+ \pi^0 K^- $\,(\xmark) 
& $0.6\times 10^{30}\,(\dagger)$~\cite{Learned:1979gp}
& $n\to \bar \nu \pi^- K^+$ & ---
\\ 
\cline{6-7}
& $p\to e^+ \pi^+ K^-$\,(\xmark) & $ 7.5\times 10^{31}$~\cite{McGrew:1999nd}
&  &
& $n\to \mu^+ \pi^- \pi^0$ & $7.4\times 10^{31}$~\cite{McGrew:1999nd}
& $n\to \bar \nu \pi^0 K^0$ & ---
\\
& $p\to e^+ \pi^- K^+$ & $ 7.5\times 10^{31}$~\cite{McGrew:1999nd}
& & 
& $n\to \mu^+ \pi^- \eta$ 
& $12\times 10^{30}\,(\dagger)$~\cite{Cherry:1981uq}
& $n\to \bar \nu \pi^0 \bar K^0$\,(\xmark) & ---
\\
\cline{2-3}
& $p\to \mu^+ \pi^+ \pi^-$ & $1.33\times 10^{32}$~\cite{McGrew:1999nd}
& & 
& $n\to \mu^+ \pi^- K^0$ 
& $12\times 10^{30}\,(\dagger)$~\cite{Cherry:1981uq}
& & 
\\ 
& $p\to \mu^+ \pi^0 \pi^0$ & $4.5\times 10^{33}$~\cite{Kamiokande:2026zwu}
& & 
& $n\to \mu^+ \pi^- \bar K^0$\,(\xmark) 
& $12\times 10^{30}\,(\dagger)$~\cite{Cherry:1981uq}
& &  
\\ 
& $p\to \mu^+ \pi^0 \eta$
& $12\times 10^{30}\,(\dagger)$~\cite{Cherry:1981uq}
&  & 
& $n\to \mu^+ \pi^0 K^-$\,(\xmark) 
& $12\times 10^{30}\,(\dagger)$~\cite{Cherry:1981uq}
&  & 
\\ 
& $p\to \mu^+ \pi^0 K^0$ 
& $12\times 10^{30}\,(\dagger)$~\cite{Cherry:1981uq}
&  & 
&  & 
&  & 
\\ 
& $p\to \mu^+ \pi^0 \bar K^0$\,(\xmark)
& $12\times 10^{30}\,(\dagger)$~\cite{Cherry:1981uq}
&  & 
&  & 
&  & 
\\ 
& $p\to \mu^+ \pi^+ K^-$\,(\xmark) & $ 2.45\times 10^{32}$~\cite{McGrew:1999nd}
& & 
& & 
& & 
\\
& $p\to \mu^+ \pi^- K^+$ & $ 2.45\times 10^{32}$~\cite{McGrew:1999nd}
& & 
& & 
& & 
\\
\hline
\multirow{10}*{\rotatebox[origin=c]{90}{$\Delta(B+L)=0$}}
& $p\to e^- \pi^+ \pi^+$\,(\xmark)  & $8.2 \times10^{31}$~\cite{McGrew:1999nd}
& $p\to \nu \pi^+\pi^0$ & ---
& $n\to e^- \pi^+ \pi^0 $\,(\xmark) & $5.2 \times10^{31}$~\cite{McGrew:1999nd}
& $n\to \nu \pi^+\pi^-$& --- 
\\ 
& $p \to e^- \pi^+ K^+$ & $7.5\times10^{31}$~\cite{McGrew:1999nd}
& $p\to \nu \pi^+ \eta $ & ---
& $n\to e^- \pi^+\eta $\,(\xmark) & ---
& $n\to \nu \pi^0\pi^0 $ & --- 
\\\cline{2-3}
& $p\to \mu^- \pi^+ \pi^+ $\,(\xmark) & $1.33\times10^{32}$~\cite{McGrew:1999nd}  
& $p\to \nu \pi^+ K^0 $& ---
& $n\to e^- \pi^+ K^0 $ & ---
& $n\to \nu \pi^0 \eta $ & --- 
\\ 
& $p \to \mu^- \pi^+ K^+$ & $2.45\times10^{32}$~\cite{McGrew:1999nd} 
& $p\to \nu \pi^+ \bar K^0 $\,(\xmark) & --- 
& $n\to e^- \pi^+ \bar K^0 $\,(\xmark) & ---  
& $n\to \nu \pi^+ K^- $\,(\xmark) & ---  
\\ 
& &
& $p\to \nu \pi^0 K^+ $ & --- 
& $n\to e^- \pi^0 K^+$ & ---  
& $n\to \nu \pi^- K^+ $ & --- 
\\\cline{6-7}
& &
& & 
& $n\to \mu^- \pi^+ \pi^0 $\,(\xmark) & $7.4 \times10^{31}$~\cite{McGrew:1999nd}
& $n\to \nu \pi^0 K^0 $ & ---
\\ 
& &
& & 
& $n\to \mu^- \pi^+\eta $\,(\xmark) & ---
& $n\to \nu \pi^0 \bar K^0 $\,(\xmark) & ---
\\ 
& &
& & 
& $n\to \mu^- \pi^+ K^0 $ & ---
& & 
\\ 
& &
& &
& $n\to \mu^- \pi^+ \bar K^0 $\,(\xmark) & --- 
& & 
\\ 
& &
& & 
& $n\to \mu^- \pi^0 K^+ $ & ---  
& & 
\\
\hline
\end{tabular}}
\caption{
Summary of all possible kinematically allowed three-body nucleon decay modes involving a lepton and two pseudoscalar mesons, together with their existing experimental 90\,\% lower bounds on the partial lifetimes. Experimental bounds from inclusive searches are indicated with a dagger. 
The decay modes marked with a `\xmark' cannot be induced by dim-6 operators at leading order, and therefore are neglected in this work.
For the four decay modes ${\tt N}\to(e^-,\mu^-)\pi\pi$ marked with `\xmark' in the $\Delta(B+L)=0$ sector, we adopt the experimental bounds on ${\tt N}\to(e^+,\mu^+)\pi\pi$ reported in~\cite{McGrew:1999nd}, as the two sets of modes give rise to similar Cherenkov signatures in experimental searches.
}
\label{tab:data}
\end{table}

In hindsight, once a positive signal is observed in two-body decays, 
the corresponding three-body modes would become equally important.
Because both classes of processes depend on the same set of parameters,  
three-body decays are essential to fully determine the interaction structure and uncover the nature of the underlying BNV scenario. Motivated by future experimental sensitivity and the close relationship between two- and three-body modes in characterizing underlying BNV interactions, we aim to fill this gap by undertaking a systematic study of these three-body modes. 

This work has two primary goals.
First, we establish general model-independent expressions for the decay widths of three-body modes within the framework of low-energy effective field theory (LEFT)~\cite{Jenkins:2017jig,Liao:2020zyx}, thereby facilitating future experimental searches and theoretical studies. 
Second, we correlate three- and two-body modes within the same LEFT setup to derive improved bounds on the former from the best existing experimental bounds on the latter, providing a benchmark for further investigations. 

We take the leading-order dimension-6 (dim-6) BNV LEFT operators as our starting point and employ the chiral perturbation theory (ChPT) framework developed for BNV interactions in~\cite{Claudson:1981gh,Liao:2025vlj,Liao:2025sqt} to handle nonperturbative QCD effects. This allows the hadronic matrix elements to be calculated systematically in a perturbative manner, with their effects captured by a few low energy constants (LECs).
Within the ChPT framework, we derive the relevant BNV lepton-baryon-meson interaction vertices up to the quadratic order in the pseudoscalar meson fields. 
Combining these with the standard baryon ChPT vertices, we derive general expressions for the matrix element squared on the basis of the leading-order Feynman diagrams. After performing the phase space integration numerically, we finally obtain the decay width expressions for 31 three-body nucleon decay modes, all parametrized in terms of the Wilson coefficients (WCs) of the BNV LEFT operators. 

Since the same set of LEFT operators contributes to experimentally well-constrained two-body modes, it is possible to derive improved bounds on three-body modes by leveraging  their correlations. 
In this paper, we employ two methods for this analysis. 
The first is the usual single-operator-dominance assumption widely used in the literature. In this approach, 
for each given LEFT operator or WC, the decay widths of the associated two- and three-body decay modes are proportional to each other. 
Therefore, by selecting the two-body mode that imposes the most severe constraint on the WC under consideration, one can derive a very stringent, yet operator-dependent, bound on the partial lifetime of each three-body mode. 
For three-body modes involving a charged lepton $e^+$ or $\mu^+$, this method yields partial lifetime bounds that are three to eight orders of magnitude stronger than the previous experimental limits, with the exact improvement depending on the specific mode and the LEFT operator involved.  

The second method avoids this ad hoc assumption and retains all relevant WCs in each decay mode through a global analysis~\cite{Fan:2026fqo}.
For each subset of LEFT operators or WCs that share the same field content but differ in chiral structures, they contribute to the same set of two- and three-body processes.
The decay widths for these processes 
take a quadratic polynomial form in the WCs, differing only in their numerical prefactors. 
Owing to these distinct dependencies,  
we can employ a subset of experimentally best-constrained two-body modes to derive closed-region bounds on the multidimensional WC parameter space. 
From these we obtain conservative assumption-free bounds on the partial lifetimes of three-body modes by varying the WCs in the allowed region.
As anticipated, the bounds obtained in this way are relatively weaker than those derived using the first method.
Nevertheless, they remain far more stringent than the existing IMB-3 bounds by more than three orders of magnitude.  

Although the global approach described above is more robust, 
in practice it cannot be applied to cases where there is insufficient two-body data to constrain multidimensional WCs to a closed region. The three-body nucleon decay modes involving a kaon are a concrete example of this limitation.
In such cases, the single-operator-dominance analysis takes over. Our newly derived partial lifetime bounds on these three-body kaon decays
are on the order of $\calO(10^{33\text{---}37}\,\rm yr)$. 

The remainder of this paper is organized as follows. In \cref{sec:EFT}, we collect the relevant dim-6 BNV LEFT operators and derive their corresponding hadronic counterparts within the ChPT framework, with the detailed chiral Lagrangian terms relevant to our analysis given in appendix \ref{app:chira_lLag}. 
The derivation of the general decay width expressions arising from leading-order Feynman diagrams is presented in \cref{sec:decaywidth}, and the resulting numerical decay widths expressed in terms of the LEFT WCs are summarized in appendix \ref{app:DeW_three_body}. 
In \cref{sec:newbounds}, we provide the numerical analysis of both methods and present our final improved bounds on these three-body nucleon decay modes. 
Our conclusions are given in 
\cref{sec:summary}.
In appendix \ref{app:vectorM}, we estimate the neglected vector-meson contributions, while in appendix \ref{app:DeW_two_body} we collect the general decay width results for the related two-body nucleon decay modes.  

\section{Effective field theory description}
\label{sec:EFT}

\subsection{Low-energy effective field theory}

As nucleon decays occur at an energy scale of $\calO(1\,{\rm GeV})$, LEFT offers the most powerful and convenient framework for systematically describing various decay modes in terms of a small number of WCs.  
In this work, we focus on the lowest dimension-6 (dim-6) BNV operators involving light $u,d,s$ quarks and containing at most one strange quark, as they provide the dominant contributions to nucleon decays into a lepton and one or two pseudoscalar mesons (decay modes without `\xmark' listed in \cref{tab:data}).
Note that dim-7 and higher-dimensional LEFT operators could, in principle, become relevant if the contributions from dim-6 LEFT operators are suppressed. 
However, such suppression would generally require cancellations or fine-tuning in the underlying theory, as no fundamental symmetry forbids the generation of dim-6 LEFT contributions. We therefore do not consider this possibility in the present work.

For convenience, we denote the triple-quark part of a general dim-6 BNV operator as ${\cal N}_{yzw}^{\tt XY} \equiv q_{{\tt X} y}^\alpha (\overline{ q_{{\tt Y} z}^{\beta \tt C} } q_{{\tt Y} w}^\gamma)\epsilon_{\alpha \beta \gamma}$, where $\tt X/Y=\tL, \tR$ indicate the chiralities of quark fields and $y/z/w=u,d,s$ label the three light quark flavors. The Greek letters $\alpha,\beta,\gamma$ are contracted color indices. 
With this notation, together with the charged lepton $\ell_{\tL(\tR)}$ and neutrino $\nu_\tL$ fields, all relevant dim-6 BNV operators can be classified into five classes in terms of quark field configurations:  
\begin{subequations}
\label{eq:dim6ope}
\begin{align}
\ell uud(4):\quad 
&\calO_{\ell uud}^{\tL\tR(\tL\tL)} = \overline{\ell_{\tL}^\C}{\cal N}_{uud}^{\tL\tR(\tL\tL)},~
\calO_{\ell uud}^{\tR\tL(\tR\tR)} = \overline{\ell_{\tR}^\C}{\cal N}_{uud}^{\tR\tL(\tR\tR)};
\\
\ell uus(4):\quad 
&\calO_{\ell usu}^{\tL\tR(\tL\tL)} = \overline{\ell_{\tL}^\C}{\cal N}_{usu}^{\tL\tR(\tL\tL)},~ 
\calO_{\ell usu}^{\tR\tL(\tR\tR)} = \overline{\ell_{\tR}^\C}{\cal N}_{usu}^{\tR\tL(\tR\tR)};
\\
\bar\ell dds(4):\quad
&\calO_{\bar\ell dds}^{\tL\tR(\tL\tL)} = \overline{\ell_{\tR}}{\cal N}_{dds}^{\tL\tR(\tL\tL)},~ 
\calO_{\bar\ell dds}^{\tR\tL(\tR\tR)} = \overline{\ell_{\tL}}{\cal N}_{dds}^{\tR\tL(\tR\tR)};
\\
\hat\nu udd(2+2):\quad 
&\calO_{\nu dud}^{\tL\tR(\tL\tL)} = \overline{\nu_{\tL}^\C}{\cal N}_{dud}^{\tL\tR(\tL\tL)};~
\calO_{\bar\nu dud}^{\tR\tL(\tR\tR)} = \overline{\nu_{\tL}}{\cal N}_{dud}^{\tR\tL(\tR\tR)}; 
\\
\hat\nu uds(5+5):\quad 
&\calO_{\nu uds}^{\tL\tR} = \overline{\nu_{\tL}^\C}{\cal N}_{uds}^{\tL\tR}, \, 
\calO_{\nu dsu}^{\tL\tR(\tL\tL)} = \overline{\nu_{\tL}^\C}{\cal N}_{dsu}^{\tL\tR(\tL\tL)}, \,
\calO_{\nu sud}^{\tL\tR(\tL\tL)} = \overline{\nu_{\tL}^\C}{\cal N}_{sud}^{\tL\tR(\tL\tL)}; 
\nonumber\\
&\calO_{\bar\nu uds}^{\tR\tL} = \overline{\nu_{\tL}}{\cal N}_{uds}^{\tR\tL},~ 
\calO_{\bar\nu dsu}^{\tR\tL(\tR\tR)} = \overline{\nu_{\tL}}{\cal N}_{dsu}^{\tR\tL(\tR\tR)},~
\calO_{\bar\nu sud}^{\tR\tL(\tR\tR)} = \overline{\nu_{\tL}}{\cal N}_{sud}^{\tR\tL(\tR\tR)}.
\end{align}
\end{subequations}
Here, $\hat\nu$ denotes a neutrino or an antineutrino, and the number of independent operators in each case is given in parentheses.
Experimentally, neither neutrinos nor antineutrinos are directly detectable, so decay modes involving them are treated on the same footing. 
Note that operators in $\ell uud$, $\ell uus$, $\nu udd$, and $\nu uds$ conserve the baryon-minus-lepton quantum number $\Delta (B-L)=0$, while those in $\bar \ell dds$, 
$\bar\nu udd$, and $\bar\nu uds$ conserve the baryon-plus-lepton quantum number $(B+L)=0$.

Within the standard model effective field theory (SMEFT) framework, 
these LEFT operators with $\Delta(B-L)=0$ and 
$\Delta(B+L)=0$ arise at leading order from the dim-6~\cite{Jenkins:2017jig} and dim-7~\cite{Liao:2020zyx} BNV SMEFT operators, respectively. 
Analyses of conventional two-body $\Delta B=1$ nucleon decays and $\Delta B=2$ processes induced by higher-dimensional BNV SMEFT operators, together with their matching onto the LEFT and baryon ChPT frameworks, can be found in~\cite{Beneito:2023xbk,He:2021mrt,BasiBeneito:2026feb}.
The SMEFT framework, in turn, offers a direct pathway to interpreting low-energy results in terms of specific ultraviolet (UV) scenarios. 

All of the operators in \cref{eq:dim6ope} belong to isospin $I=0,\,1/2,\,\text{or}\,1$.
Consequently, they cannot mediate processes such as $p\to \ell^-\pi^+\pi^+$ and $n\to \ell^- \pi^+(\pi^0,\eta)$ at the leading order, which require an isospin change of $3/2$ units. 
Moreover, decay modes involving a $\bar K^0(\bar ds)$ or $K^-(\bar us)$ cannot be generated by these operators either, as they carry the wrong strangeness quantum number. 
These processes can arise only from dim-7 or higher operators, or through additional SM weak vertices, and are therefore neglected  in this work.  
For completeness, we list them in \cref{tab:data} but mark them with a `\xmark'. 

In \cref{tab:opeandprocess}, we summarize, for each set of operators in \cref{eq:dim6ope},  the associated two- and three-body processes they induce at the leading order. 
From the table, we see that for each charged lepton flavor, the four operators $\calO_{\ell uud}^{\tL\tR/\tL\tL/\tR\tL/\tR\tR}$ induce three two-body and five three-body modes, whereas each of the two sets of four operators $\calO_{\ell usu}^{\tL\tR/\tL\tL/\tR\tL/\tR\tR}$ and  $\calO_{\bar \ell dds}^{\tL\tR/\tL\tL/\tR\tL/\tR\tR}$ induces one two-body process and three three-body processes. 
For the (anti)neutrino case,  
the two operators $\calO_{\nu dud}^{\tL\tR/\tL\tL}$ ($\calO_{\bar\nu dud}^{\tR\tL/\tR\tR}$) induce three two-body  and five three-body modes containing an antineutrino (a neutrino),
whereas the five operators  $\calO_{\nu uds}^{\tL\tR},\,\calO_{\nu dsu}^{\tL\tR/\tL\tL},\,\calO_{\nu sud}^{\tL\tR/\tL\tL}$ ($\calO_{\bar\nu uds}^{\tR\tL},\,\calO_{\bar\nu dsu}^{\tR\tL/\tR\tR},\,\calO_{\bar\nu sud}^{\tR\tL/\tR\tR}$) induce two two-body and four three-body processes. 
Such correlations between two- and three-body modes make it possible to derive improved bounds on the latter from the experimentally well-constrained limits on the former. Before doing so, we must calculate the decay widths and express them in terms of the WCs of the corresponding LEFT operators. 
This can be carried out systematically within the ChPT framework, as discussed below. 

\begin{table}[t]
\center
\resizebox{\linewidth}{!}{
\renewcommand{\arraystretch}{1.1}
\begin{tabular}{|l|c c|c c|c c|c c|c c c c c| }
\hline
\multirow{3}*{\rotatebox[origin=c]{90}{Operators}}
& \multicolumn{6}{c|}{Charged lepton case}
& \multicolumn{7}{c|}{Neutrino or antineutrino case}
\\\cline{2-14}
&$~\calO_{\ell uud}^{\tL\tR}~$ 
&$~\calO_{\ell uud}^{\tL\tL}~$ 
&$~\calO_{\ell usu}^{\tL\tR}~$ 
&$~\calO_{\ell usu}^{\tL\tL}~$ 
&$~\calO_{\bar\ell dds}^{\tL\tR}~$
&$~\calO_{\bar\ell dds}^{\tL\tL}~$
&$~\calO_{\nu dud}^{\tL\tR}~$ 
&$~\calO_{\nu dud}^{\tL\tL}~$
&$~\calO_{\nu uds}^{\tL\tR}~$ 
&$~\calO_{\nu dsu}^{\tL\tR}~$ 
&$~\calO_{\nu sud}^{\tL\tR}~$
&$~\calO_{\nu dsu}^{\tL\tL}~$ 
&$~\calO_{\nu sud}^{\tL\tL}~$
\\
&$~\calO_{\ell uud}^{\tR\tL}~$ 
&$~\calO_{\ell uud}^{\tR\tR}~$ 
&$~\calO_{\ell usu}^{\tR\tL}~$ 
&$~\calO_{\ell usu}^{\tR\tR}~$ 
&$~\calO_{\bar\ell dds}^{\tR\tL}~$
&$~\calO_{\bar\ell dds}^{\tR\tR}~$
&$\calO_{\bar\nu dud}^{\tR\tL}~$ 
&$\calO_{\bar\nu dud}^{\tR\tR}~$
&$~\calO_{\bar\nu uds}^{\tR\tL}~$ 
&$~\calO_{\bar\nu dsu}^{\tR\tL}~$ 
&$~\calO_{\bar\nu sud}^{\tR\tL}~$
&$~\calO_{\bar\nu dsu}^{\tR\tR}~$ 
&$~\calO_{\bar\nu sud}^{\tR\tR}~$
\\\hline
\multirow{3}*{\rotatebox[origin=c]{90}{${\tt N}\to l\, M$}}
&\multicolumn{2}{c|}{$p\to \ell^+\pi^0 $} 
&\multicolumn{2}{c|}{$p\to \ell^+ K^{0}$}
&\multicolumn{2}{c|}{$n\to \ell^- K^+$}
&\multicolumn{2}{c|}{$p\to \hat\nu \pi^{+}$}
&\multicolumn{5}{c|}{$p\to \hat\nu K^{+}$}
\\%
&\multicolumn{2}{c|}{$p\to \ell^+\eta $} 
&\multicolumn{2}{c|}{}
&\multicolumn{2}{c|}{}
&\multicolumn{2}{c|}{$n\to \hat\nu \pi^0$}
&\multicolumn{5}{c|}{$n\to \hat\nu K^0$} 
\\%
& \multicolumn{2}{c|}{$n\to \ell^+\pi^- $}
& & & &
& \multicolumn{2}{c|}{$n\to \hat\nu \eta $}
& & & & &
\\\hline
\multirow{5}*{\rotatebox[origin=c]{90}{${\tt N}\to l\, M_1\, M_2$}} 
& \multicolumn{2}{c|}{$p\to \ell^+\pi^+\pi^- $}
& \multicolumn{2}{c|}{$p\to \ell^+\pi^- K^+ $}
& \multicolumn{2}{c|}{$p\to \ell^-\pi^+ K^+ $}
& \multicolumn{2}{c|}{$p\to \hat\nu \pi^+\pi^0 $} 
& \multicolumn{5}{c|}{$p\to \hat\nu \pi^+ K^0$}
\\%
& \multicolumn{2}{c|}{$p\to \ell^+\pi^0\pi^0 $}
& \multicolumn{2}{c|}{$p\to \ell^+\pi^0 K^0 $}
& \multicolumn{2}{c|}{$n\to \ell^-\pi^+ K^0 $}
& \multicolumn{2}{c|}{$p\to \hat\nu \pi^+\eta $} 
& \multicolumn{5}{c|}{$p\to \hat\nu \pi^0 K^+$}
\\%
& \multicolumn{2}{c|}{$p\to \ell^+\pi^0\eta $}
& \multicolumn{2}{c|}{$n\to \ell^+\pi^- K^0 $}
& \multicolumn{2}{c|}{$n\to \ell^-\pi^0 K^+ $}
& \multicolumn{2}{c|}{$n\to \hat\nu \pi^+\pi^- $}
& \multicolumn{5}{c|}{$n\to \hat\nu \pi^- K^+$}
\\%
& \multicolumn{2}{c|}{$n\to \ell^+\pi^-\pi^0 $}
& \multicolumn{2}{c|}{}
& \multicolumn{2}{c|}{}
& \multicolumn{2}{c|}{$n\to \hat\nu \pi^0\pi^0 $}
& \multicolumn{5}{c|}{$n\to \hat\nu \pi^0 K^0$}
\\
& \multicolumn{2}{c|}{$n\to \ell^+\pi^-\eta $}
& \multicolumn{2}{c|}{} 
& \multicolumn{2}{c|}{}
& \multicolumn{2}{c|}{$n\to \hat\nu \pi^0\eta $}
& \multicolumn{5}{c|}{}
\\
\hline
\end{tabular} }
\caption{Summary of each set of BNV LEFT operators and their induced two- and three-body BNV nucleon decay modes.} 
\label{tab:opeandprocess}
\end{table}

\subsection{Chiral perturbation theory}
\label{subsec:ChPT}

To compute the hadronic matrix elements, we use chiral perturbation theory to match the quark-level operators in LEFT onto their hadronic counterparts.  
In ChPT, the relevant hadronic degrees of freedom are the octet baryon 
and pseudoscalar meson fields. 
We denote them as $B(x)$ and $\Sigma(x) = \xi^2(x) = \exp[i\sqrt{2}\Pi(x)/F_0]$, 
whose explicit forms are given by
\begin{align}
\Pi =   
\begin{pmatrix}
\frac{\pi^0}{\sqrt{2}}+\frac{\eta}{\sqrt{6}} & \pi^+ & K^+
\\
\pi^- & -\frac{\pi^0}{\sqrt{2}}+\frac{\eta}{\sqrt{6}} & K^0
\\
K^- & \bar{K}^0 & -\sqrt{\frac{2}{3}}\eta
\end{pmatrix},~
B =
\begin{pmatrix}
{\Sigma^{0}\over \sqrt{2}}+{\Lambda^0 \over \sqrt{6}}  & \Sigma^+ & p \\
\Sigma^- & -{\Sigma^{0} \over \sqrt{2}}+{\Lambda^0 \over \sqrt{6}} &  n \\ 
\Xi^- & \Xi^0 & - \sqrt{2\over 3}\Lambda^0
\end{pmatrix},
\end{align}
where $F_0=f_{\pi}/\sqrt{2}$ denotes the pion decay constant in the chiral limit, with a numerical value $f_{\pi}=130.41(20)~\rm MeV$~\cite{ParticleDataGroup:2024cfk}.

\textbf{Baryon-number-violating part}: At leading chiral order, the matching results for the dim-6 BNV LEFT operators take the following form~\cite{Claudson:1981gh,Fan:2024gzc,Liao:2025vlj}:
\begin{align}
{\cal L}_{\tt \slashed{B}}^{\tt ChPT}
& =
c_1 \tr\big[ 
{\cal P}_{\bar{\pmb{3}}_\tL \otimes \pmb{3}_\tR}
\xi B_{\tL}\xi
-{\cal P}_{\pmb{3}_\tL \otimes \bar{\pmb{3}}_\tR}
\xi^{\dagger} B_{\tR}\xi^{\dagger} \big]
\nonumber
\\
&\quad +c_2 \tr \big[
{\cal P}_{\pmb{8}_{\tL} \otimes \pmb{1}_{\tR}}
\xi B_{\tL}\xi^{\dagger} -{\cal P}_{\pmb{1}_\tL \otimes \pmb{8}_\tR}
\xi^{\dagger}B_{\tR}\xi\big]+\text{h.c.},
\label{eq:BNV_ChPT}
\end{align}
where $B_{\tL(\tR)}\equiv P_{\tL(\tR)} B$ represent the chiral baryon fields with $P_{\tL}=(1-\gamma_5)/2$ and $P_{\tR}=(1+\gamma_5)/2$. $c_1$ and $c_2$ are hadronic LECs, and the lattice QCD calculations give $c_1=-0.01257(111)\,\text{GeV}^3$ and $c_2=0.01269(107)\,\text{GeV}^3$~\cite{Yoo:2021gql}. 
${\cal P}_{i\otimes j}$ is a spurion field matrix constructed from the product of the lepton field and the WC associated with each BNV LEFT operator. It transforms as an irreducible representation $i\otimes j$ under the QCD chiral group ${\rm SU(3)}_\tL\otimes {\rm SU(3)}_\tR$. 
For the LEFT operators in \cref{eq:dim6ope}, the spurion field
matrices take the form~\cite{Fan:2024gzc}:
{\fontsize{10.3pt}{13.6pt}\selectfont
\begin{subequations}
\label{eq:spu}
\begin{align}
{\cal P}_{\pmb{8}_\tL \otimes \pmb{1}_\tR} 
& =
\begin{pmatrix}
0  &  {\cellcolor{gray!15} C^{\tL\tL,x}_{\bar{\ell} dds} \overline{\ell_{\tR x}} }
& -
\\[4pt]
C^{\tL\tL,x}_{\ell usu} \overline{\ell_{\tL x}^{\C}} 
& C^{\tL\tL,x}_{\nu dsu} \overline{\nu_{\tL x}^{\C}} 
& - 
\\[4pt]
C^{\tL\tL,x}_{\ell uud} \overline{\ell_{\tL x}^{\C}} 
& C^{\tL\tL,x}_{\nu dud} \overline{\nu_{\tL x}^{\C}} 
& C^{\tL\tL,x}_{\nu sud} \overline{\nu_{\tL x}^{\C}} 
\end{pmatrix},\,
& {\cal P}_{\pmb{1}_\tL \otimes \pmb{8}_\tR} 
& =
\begin{pmatrix}
0  &  {\cellcolor{gray!15} C^{\tR\tR,x}_{\bar{\ell} dds} \overline{\ell_{\tL x}} }
& - \\[4pt]
C^{\tR\tR,x}_{\ell usu} \overline{\ell_{\tR x}^{\C}} 
& {\cellcolor{gray!15} C^{\tR\tR,x}_{\bar{\nu} dsu}  \overline{\nu_{\tL x}} }
& - \\[4pt]
C^{\tR\tR,x}_{\ell uud} \overline{\ell_{\tR x}^{\C}} 
& {\cellcolor{gray!15} C^{\tR\tR,x}_{\bar{\nu} dud}  \overline{\nu_{\tL x}} }
& {\cellcolor{gray!15} C^{\tR\tR,x}_{\bar{\nu} sud} \overline{\nu_{\tL x}} }
\end{pmatrix},
\\
{\cal P}_{\pmb{3}_\tL \otimes \bar{\pmb{3}}_\tR} 
& = 
\begin{pmatrix}	
{\cellcolor{gray!15} C^{\tR\tL,x}_{\bar{\nu}uds}\overline{\nu_{\tL x}}  } 
& {\cellcolor{gray!15} C^{\tR\tL,x}_{\bar{\ell} dds} \overline{\ell_{\tL x}} }
& - \\[4pt]
C^{\tR\tL,x}_{\ell usu} \overline{\ell_{\tR x}^{\C}}
& {\cellcolor{gray!15} C^{\tR\tL,x}_{\bar{\nu} dsu}  \overline{\nu_{\tL x}} }
& -\\[4pt]
C^{\tR\tL,x}_{\ell uud} \overline{\ell_{\tR x}^{\C}} 
&{\cellcolor{gray!15} C^{\tR\tL,x}_{\bar{\nu} dud}  \overline{\nu_{\tL x}} }
&{\cellcolor{gray!15} C^{\tR\tL,x}_{\bar{\nu} sud} \overline{\nu_{\tL x}} }
\end{pmatrix},\,
& {\cal P}_{\bar{\pmb{3}}_\tL \otimes \pmb{3}_\tR} 
& =
\begin{pmatrix}
C^{\tL\tR,x}_{\nu uds}\overline{\nu_{\tL x}^{\C}} &
{\cellcolor{gray!15} C^{\tL\tR,x}_{\bar{\ell} dds} \overline{\ell_{\tR x}} }
& - \\[4pt]
C^{\tL\tR,x}_{\ell usu} \overline{\ell_{\tL x}^{\C}}
& C^{\tL\tR,x}_{\nu dsu}  \overline{\nu_{\tL x}^{\C}} 
& - \\[4pt]
C^{\tL\tR,x}_{\ell uud} \overline{\ell_{\tL x}^{\C}} 
& C^{\tL\tR,x}_{\nu dud}  \overline{\nu_{\tL x}^{\C}} 
&C^{\tL\tR,x}_{\nu sud} \overline{\nu_{\tL x}^{\C}} 
\end{pmatrix},
\end{align}
\end{subequations}}%
where we have omitted entries involving two strange quarks, denoted by `--'. 
The index $x$ labels the lepton flavor and is summed over $e$ and $\mu$ for the charged lepton case and over $e,\mu,\tau$ for the (anti)neutrino case.  
The white and gray elements correspond to $\Delta(B - L) = 0$ and $\Delta(B + L) = 0$ interactions, respectively.

Taking \cref{eq:spu} into \cref{eq:BNV_ChPT} and performing a Taylor expansion of the pseudoscalar meson matrix, 
we can obtain various interaction vertices involving a lepton, a baryon, and an arbitrary number of mesons. 
For the two- and three-body nucleon decay modes,
only the baryon-lepton mixing term ${\cal L}_{Bl}$ and the three- and four-point terms ${\cal L}_{BlM}$ and ${\cal L}_{BlMM}$ involving one and two mesons are relevant to the leading-order tree-level Feynman diagrams.  
These terms are summarized in appendix \ref{app:BNV_terms}.

\textbf{Baryon-number-conserving part}: In addition to the BNV vertices given above, the baryon-number-conserving (BNC) chiral interactions between octet baryons and mesons that originate from QCD strong interactions are also needed. The leading-order Lagrangian is~\cite{Jenkins:1990jv,Bijnens:1985kj} 
\begin{align}
\label{eq:LChPT}
{\cal L}_{\tt ChPT}^B & = 
{\rm Tr}[\bar B (i \slashed{D} -M) B] 
+ \frac{D}{2} {\rm Tr}(\bar B \gamma^\mu \gamma_5\{u_\mu,B\}) 
+ \frac{F}{2} {\rm Tr}(\bar B \gamma^\mu \gamma_5 [u_\mu,B]),
\end{align}
where $u_{\mu}=i\left[\xi(\partial_{\mu}-ir_{\mu})\xi^{\dagger}
-\xi^{\dagger}(\partial_{\mu}-il_{\mu})\xi\right]= - i \xi^\dagger (D_\mu\Sigma )\xi^\dagger$.
The covariant derivatives are defined as
 $D_\mu \Sigma = \partial_\mu \Sigma - i l_\mu \Sigma +i\Sigma r_\mu$ 
and $D_\mu B = \partial_\mu B + [\Gamma_\mu ,B]-i v_\mu^{(s)}B$, respectively, 
where $\Gamma_\mu$ is the chiral connection defined as
$\Gamma_{\mu}= \frac{1}{2}
\big[ \xi(\partial_{\mu}-ir_{\mu})\xi^{\dagger}+\xi^{\dagger}(\partial_{\mu}-il_{\mu})\xi \big]$.
Here, $v_\mu^{(s)}$, $l_{\mu}$, and $r_{\mu}$ are external source fields composed of non-QCD objects. They are irrelevant in our discussion and can be set to zero. 
$D$ and $F$ are LECs, and the recent lattice calculation yields $ D=0.730(11)$ and $F=0.447^{6}_{7}$~\cite{Bali:2022qja}. 

For our purpose, the required interactions include the three-point vertices $\bar BBM$ involving one meson and the four-point vertices $\bar B {\tt N} MM$ involving two mesons, where ${\tt N}$, $B$, and $M$ represent nucleons, generic octet baryons, and pseudoscalar mesons, respectively.
These terms are obtained by expanding \cref{eq:LChPT} to the second order in the pseudoscalar meson fields, and the relevant terms are collected in appendix \ref{app:BNC_terms}.

\section{Calculation of decay width}
\label{sec:decaywidth}

\begin{figure}[t]
\centering
\includegraphics[width=0.95\textwidth]{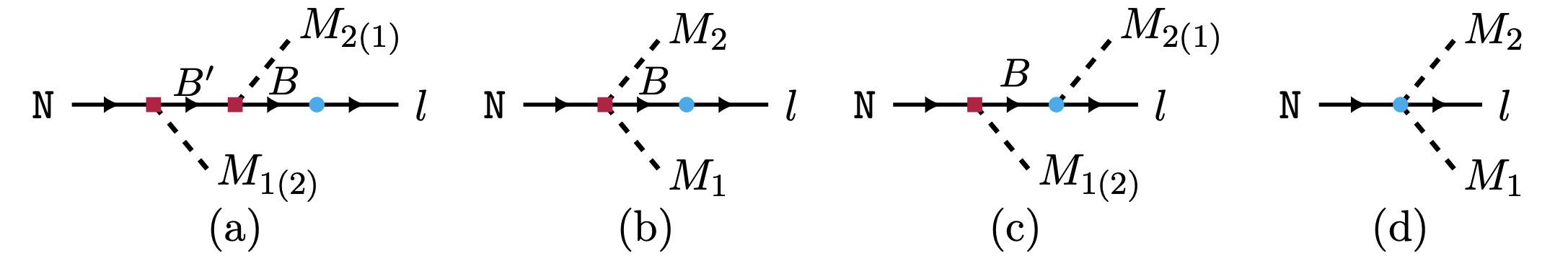}
\caption{Leading-order Feynman diagrams contributing to three-body nucleon decays into a lepton and two pseudoscalar mesons. The cyan blobs and purple squares represent the dim-6 BNV vertices and the SM strong interaction vertices, respectively.}
\label{fig:Feyndiagram}
\end{figure}

In this work, we focus on leading-order contributions to three-body nucleon decays mediated by the BNV dim-6 operators plus SM strong interactions. Contributions arising from additional QED or weak vertices are neglected, as they are suppressed relative to the former.
Taking into account the BNV and BNC hadronic interactions in \cref{subsec:ChPT}, 
we can form four tree-level Feynman diagrams that contribute to a generic three-body process ${\tt N}\to l M_1M_2$ at leading order, as shown in \cref{fig:Feyndiagram}.
The general Lagrangian terms entering into \cref{fig:Feyndiagram} can be parameterized as
\begin{align}
{\cal L}_{{\tt N} \to l M_1M_2} = &
\mathbb A_{B_1B_2 M}(\overline{B_2}\gamma^\mu\gamma_5 B_1)
\partial_\mu \bar M
+\mathbb A_{{\tt N} B M_1 M_2} (\overline{B}\gamma^\mu {\tt N}) \bar M_1 i \overleftrightarrow{\partial_\mu} \bar M_2 
\nonumber\\
&+\overline{l}(\mathbb B_{Bl}^\tL P_\tL + \mathbb B_{Bl}^\tR P_\tR) B
+ \overline{l}(i\,\mathbb B_{BlM}^\tL P_\tL + i\,\mathbb B_{BlM}^\tR P_\tR) B \bar M
\nonumber\\
& + \frac{1}{1+\delta_{M_1,M_2}}\overline{l}(\mathbb B_{{\tt N}lM_1 M_2}^\tL P_\tL
+ \mathbb B_{{\tt N}lM_1 M_2}^\tR P_\tR) {\tt N} \bar M_1 \bar M_2,
\label{eq:LN2lMM}
\end{align}
where $\bar M$ and $\bar M_{1,2}$ denote the conjugate meson fields.
The terms in the first line represent the BNC interactions, and the three- and four-point couplings $\mathbb A_{B_1B_2 M}$ and $\mathbb A_{{\tt N} B M_1 M_2} $ can be extracted from \cref{eq:LBBM,eq:LBNMM}, respectively, once the field configurations are specified.
The terms in the second and third lines denote two-, three-, and four-point BNV interactions, and the corresponding couplings $\mathbb B_{Bl}^{\tL(\tR)}$, $\mathbb B_{BlM}^{\tL(\tR)}$, and $\mathbb B_{{\tt N}lM_1 M_2}^{\tL(\tR)}$ for nonvanishing vertices are easily read off from \cref{eq:VertexBl,eq:VertexBlM,eq:VertexNBlM}, respectively. 
Note that $\overline{l} =\overline{\ell^\C}{\rm~or~}\overline{\nu_\tL^\C}$ if the interactions belong to the $\Delta(B-L)=0$ sector, and $\overline{l} =\overline{\ell}{\rm~or~}\overline{\nu_\tL}$ for the $\Delta(B+L)=0$ case.

In terms of the above general parametrization,
the amplitude for the decay process ${\tt N}(p)\to  l(k) M_1(p_1) M_2(p_2)$ due to the four diagrams in \cref{fig:Feyndiagram} takes the form
\begin{align}
{\cal M} 
=\,& \Big[ \overline{u_l}(\mathbb B_{B l}^\tL P_\tL + \mathbb B_{B l}^\tR P_\tR)
\frac{\slashed{k}+m_{B}}{m_{B}^2-k^2}
\mathbb A_{B' B M_2} \slashed{p}_2\gamma_5 
\frac{\slashed{k}+\slashed{p}_2+m_{B'}}{(k+p_2)^2- m_{B'}^2}
\mathbb A_{{\tt N} B' M_1} \slashed{p}_1\gamma_5
u_{\tt N}
\notag\\
&
+(M_1,p_1)\leftrightarrow (M_2,p_2) \Big]
\notag\\
&+\overline{u_l} (\mathbb B_{Bl}^\tL P_\tL + \mathbb B_{Bl}^\tR P_\tR)
\frac{\slashed{k}+m_{B}}{k^2- m_{B}^2}
\mathbb A_{{\tt N} B M_1 M_2}(\slashed{p}_2 - \slashed{p}_1) u_{\tt N}
\notag\\
& + \Big[\overline{u_l} (\mathbb B_{BlM_2}^\tL P_\tL + \mathbb B_{BlM_2}^\tR P_\tR)
\frac{\slashed{k}+\slashed{p}_2+m_{B}}{(k+p_2)^2- m_{B}^2}
\mathbb A_{{\tt N} B M_1} \slashed{p}_1\gamma_5
u_{\tt N} + (M_1,p_1)\leftrightarrow (M_2,p_2) \Big]
\notag\\
& + \overline{u_l} (\mathbb B_{{\tt N}lM_1 M_2}^\tL P_\tL + \mathbb B_{{\tt N}lM_1 M_2}^\tR P_\tR) u_{\tt N}, 
\label{eq:ampfull}
\end{align}
where the four terms correspond, in order, to the four diagrams in \cref{fig:Feyndiagram}. Implicit summations are taken over all possible intermediate baryon states $B$ and $B'$. 
Note that for both \cref{fig:Feyndiagram}\,(a) and \cref{fig:Feyndiagram}\,(c), two distinct diagrams arise from exchanging the two mesons (excluding the modes ${\tt N}\to l \pi^+\pi^-$).
In the case where $M_1\neq M_2$ (excluding the modes ${\tt N}\to l\pi^0\eta$), the intermediate baryon states appearing in these crossed contributions generally differ from those in the original diagrams and therefore require careful treatment.

\begin{table}[t]
\centering
\resizebox{\linewidth}{!}{
\renewcommand{\arraystretch}{1.4}
\begin{tabular}{|l|l|l|c|l|c|c|}
\hline
\multirow{2}*{\quad~Mode}
& \multicolumn{3}{c|}{Diagrams (a) and (b)} 
& \multicolumn{2}{c|}{Diagram (c)}
& \multicolumn{1}{c|}{Diagram (d)}
\\\cline{2-7}
& \multicolumn{1}{c|}{$\mathbb A_{{\tt N}B'M_{1(2)}}\mathbb A_{B'BM_{2(1)}}$} 
& \multicolumn{1}{c|}{$\mathbb A_{{\tt N}BM_1 M_2}$}  
& \multicolumn{1}{c|}{$\mathbb B_{B\ell}$} 
& \multicolumn{1}{c|}{$\mathbb A_{{\tt N}BM_{1(2)}}$} 
& \multicolumn{1}{c|}{$\mathbb B_{B\ell M_{2(1)}}$} 
& \multicolumn{1}{c|}{$\mathbb B_{{\tt N}\ell M_1 M_2}$} 
\\\hline\hline
~$p\to \ell^+ \pi^+ \pi^-$ 
& $\mathbb A_{pn\pi^+}\mathbb A_{np\pi^-} = \frac{(D+F)^2}{2 F_0^2}$
&~$\mathbb A_{pp\pi^+\pi^-} =\frac{-1}{4F_0^2}$
& $\mathbb B_{p\ell^+}$
&~$\mathbb A_{pn\pi^+} = \frac{D+F}{\sqrt{2} F_0} $
& $\mathbb B_{n\ell^+ \pi^-} $
& $\mathbb B_{p\ell^+\pi^+\pi^-}$
\\\hline
~$p\to \ell^+ \pi^0 \pi^0$ 
& $\mathbb A_{pp\pi^0}\mathbb A_{pp\pi^0} = \frac{(D+F)^2}{4 F_0^2}$
& \multicolumn{1}{c|}{---}
& $\mathbb B_{p\ell^+}$
&~$\mathbb A_{pp\pi^0}= \frac{D+F}{2 F_0}$
& $\mathbb B_{p\ell^+ \pi^0}$
& $\mathbb B_{p\ell^+\pi^0\pi^0}$
\\\hline
~$p\to \ell^+ \pi^0 \eta$
& \makecell[l]{$\mathbb A_{pp\pi^0}\mathbb A_{pp\eta} = \frac{(D+F)(3F-D)}{4\sqrt{3} F_0^2}$ \\
$\mathbb A_{pp\eta}\mathbb A_{pp\pi^0} = \frac{(D+F)(3F-D)}{4\sqrt{3} F_0^2}$ }
& \multicolumn{1}{c|}{---}
& $\mathbb B_{p\ell^+}$
& \makecell[l]{~$\mathbb A_{pp\pi^0} = \frac{D+F}{2 F_0}$ \\
~$\mathbb A_{pp\eta}= \frac{3F-D}{2\sqrt{3} F_0}$ }
& \makecell[l]{ $\mathbb B_{p \ell^+ \eta}$ \\ 
$\mathbb B_{p \ell^+ \pi^0}$ }
& $\mathbb B_{p\ell^+\pi^0\eta}$
\\\hline
~$n\to \ell^+ \pi^- \pi^0$
&\makecell[l]{ 
$\mathbb A_{np\pi^-}\mathbb A_{pp\pi^0} = \frac{(D+F)^2}{2\sqrt{2}F_0^2}$ \\
$\mathbb A_{nn\pi^0}\mathbb A_{np\pi^-} = \frac{-(D+F)^2}{2\sqrt{2}F_0^2}$  }
&~$\mathbb A_{np\pi^-\pi^0} = \frac{-1}{2\sqrt{2} F^2_0} $
& $\mathbb B_{p \ell^+}$
& \makecell[l]{~$\mathbb A_{np\pi^-} = 
\frac{D+F}{\sqrt{2} F_0}$ \\ 
~$\mathbb A_{nn\pi^0} = 
- \frac{D+F}{2 F_0}$}
& \makecell[l]{
$\mathbb B_{p\ell^+\pi^0}$\\
${\mathbb B}_{n\ell^+\pi^-}$}
& ---
\\\hline
~$n\to \ell^+ \pi^- \eta $
& \makecell[l]{ 
$\mathbb A_{np\pi^-}\mathbb A_{pp \eta} = \frac{(D+F)(3F-D)}{2\sqrt{6}F_0^2}$ \\
$\mathbb A_{nn\eta}\mathbb A_{np\pi^-} = \frac{(D+F)(3F-D)}{2\sqrt{6}F_0^2}$  }
& \multicolumn{1}{c|}{---}
& $\mathbb B_{p \ell^+}$
& \makecell[l]{~$\mathbb A_{np\pi^-} = 
\frac{D+F}{\sqrt{2} F_0}$ \\ 
~$\mathbb A_{nn\eta} = \frac{3F-D}{2\sqrt{3} F_0}$}
& \makecell[l]{
$\mathbb B_{p\ell^+\eta}$\\
${\mathbb B}_{n\ell^+\pi^-}$}
& $\mathbb B_{n\ell^+\pi^- \eta}$
\\\hline\hline
~$p\to \ell^+ \pi^0 K^0$
& \makecell[l]{$\mathbb A_{pp\pi^0}\mathbb A_{p\Sigma^+ K^0} = \frac{D^2-F^2}{2\sqrt{2} F_0^2}$ \\
$\mathbb A_{p\Sigma^+ K^0}\mathbb A_{\Sigma^+ \Sigma^+ \pi^0} = \frac{(D-F)F}{\sqrt{2} F_0^2}$ }
&~$\mathbb A_{p\Sigma^+\pi^0 K^0} = \frac{-1}{4\sqrt{2}F_0^2}$
& $\mathbb B_{\Sigma^+ \ell^+}$
& \makecell[l]{~$\mathbb A_{pp\pi^0} = \frac{D+F}{2 F_0}$ \\
~$\mathbb A_{p\Sigma^+ K^0}= \frac{D-F}{\sqrt{2} F_0}$ }
& \makecell[l]{$\mathbb B_{p \ell^+ K^0}$ \\ 
$\mathbb B_{\Sigma^+ \ell^+ \pi^0}$ }
& $\mathbb B_{p\ell^+\pi^0 K^0}$
\\\hline
~$p\to \ell^+ \pi^- K^+$
& \makecell[l]{$\mathbb A_{p\Lambda^0 K^+}\mathbb A_{\Lambda^0 \Sigma^+ \pi^-} = \frac{-D(D+3F)}{6 F_0^2}$ \\
$\mathbb A_{p\Sigma^0 K^+}\mathbb A_{\Sigma^0 \Sigma^+ \pi^-} = \frac{(F-D)F}{2 F_0^2}$ }
&~$\mathbb A_{p\Sigma^+\pi^- K^+} = \frac{1}{4 F_0^2}$
& $\mathbb B_{\Sigma^+ \ell^+}$
& \makecell[l]{~$\mathbb A_{p\Lambda^0 K^+} = -\frac{D+3F}{2\sqrt{3} F_0}$ \\
~$\mathbb A_{p\Sigma^0 K^+} = \frac{D-F}{2 F_0}$ }
& \makecell[l]{ $\mathbb B_{\Lambda^0 \ell^+ \pi^-}$ \\ 
$\mathbb B_{\Sigma^0 \ell^+ \pi^-}$ }
& $\mathbb B_{p\ell^+\pi^- K^+}$
\\\hline
~$n\to \ell^+ \pi^- K^0 $
& \makecell[l]{ 
$\mathbb A_{n p \pi^-}\mathbb A_{p \Sigma^+ K^0} = \frac{D^2-F^2}{2 F_0^2}$ \\
$\mathbb A_{n \Lambda^0 K^0}\mathbb A_{\Lambda^0 \Sigma^+ \pi^-} = \frac{-(D+3F)D}{6 F_0^2}$ \\
$\mathbb A_{n \Sigma^0 K^0}\mathbb A_{\Sigma^0 \Sigma^+ \pi^-} = \frac{(D-F)F}{2 F_0^2}$ }
& \multicolumn{1}{c|}{---}
& $\mathbb B_{\Sigma^+ \ell^+}$
& \makecell[l]{~$\mathbb A_{np\pi^-} = \frac{D+F}{\sqrt{2} F_0}$ \\ 
~$\mathbb A_{n\Lambda^0 K^0} = -\frac{D+3F}{2\sqrt{3} F_0}$ \\ 
~$\mathbb A_{n\Sigma^0 K^0} = \frac{F-D}{2 F_0}$}
& \makecell[l]{
$\mathbb B_{p\ell^+ K^0}$\\
${\mathbb B}_{\Lambda^0 \ell^+\pi^-}$\\
${\mathbb B}_{\Sigma^0 \ell^+\pi^-}$}
& $\mathbb B_{n\ell^+\pi^- K^0}$
\\\hline\hline
~$p \to \ell^- \pi^+ K^+$
& \makecell[l]{ 
$\mathbb A_{p n \pi^+}\mathbb A_{n \Sigma^- K^+} = \frac{D^2-F^2}{2 F_0^2}$ \\
$\mathbb A_{p \Lambda^0 K^+}\mathbb A_{\Lambda^0 \Sigma^- \pi^+} = \frac{-(D+3F)D}{6 F_0^2}$ \\
$\mathbb A_{p \Sigma^0 K^+}\mathbb A_{\Sigma^0 \Sigma^- \pi^+} = \frac{(D-F)F}{2 F_0^2}$ }
& \multicolumn{1}{c|}{---}
& $\mathbb B_{\Sigma^- \ell^-}$
& \makecell[l]{~$\mathbb A_{pn\pi^+} = \frac{D+F}{\sqrt{2} F_0}$ \\ 
~$\mathbb A_{p\Lambda^0 K^+} = -\frac{D+3F}{2\sqrt{3} F_0}$ \\ 
~$\mathbb A_{p\Sigma^0 K^+} = \frac{D-F}{2 F_0}$}
& \makecell[l]{
$\mathbb B_{n\ell^- K^+}$\\
${\mathbb B}_{\Lambda^0 \ell^-\pi^+}$\\
${\mathbb B}_{\Sigma^0 \ell^-\pi^+}$}
& $\mathbb B_{p\ell^-\pi^+ K^+}$
\\\hline
~$n\to \ell^- \pi^+ K^0 $
& \makecell[l]{ 
$\mathbb A_{n \Lambda^0 K^0}\mathbb A_{\Lambda^0 \Sigma^- \pi^+} = \frac{-(D+3F)D}{6 F_0^2}$ \\
$\mathbb A_{n\Sigma^0 K^0}\mathbb A_{\Sigma^0 \Sigma^- \pi^+} = \frac{(F-D)F}{2 F_0^2}$ }
&~$\mathbb A_{n\Sigma^-\pi^+ K^0} = \frac{1}{4 F_0^2}$
& $\mathbb B_{\Sigma^- \ell^-}$
& \makecell[l]{~$\mathbb A_{n\Lambda^0 K^0} = -\frac{D+3F}{2\sqrt{3} F_0}$ \\ 
~$\mathbb A_{n\Sigma^0 K^0} = \frac{F-D}{2 F_0}$}
& \makecell[l]{
${\mathbb B}_{\Lambda^0 \ell^-\pi^+}$\\
${\mathbb B}_{\Sigma^0 \ell^-\pi^+}$}
& $\mathbb B_{n\ell^-\pi^+ K^0}$
\\\hline
~$n\to \ell^- \pi^0 K^+$
& \makecell[l]{ 
$\mathbb A_{n n \pi^0}\mathbb A_{n \Sigma^- K^+} = \frac{F^2-D^2}{2\sqrt{2} F_0^2}$ \\
$\mathbb A_{n\Sigma^- K^+}\mathbb A_{\Sigma^- \Sigma^- \pi^0} = \frac{(F-D)F}{\sqrt{2} F_0^2}$ }
&~$\mathbb A_{n\Sigma^-\pi^0 K^+} = \frac{1}{4\sqrt{2} F_0^2}$
& $\mathbb B_{\Sigma^- \ell^-}$
& \makecell[l]{~$\mathbb A_{nn \pi^0} = -\frac{D+F}{2 F_0}$ \\ 
~$\mathbb A_{n\Sigma^- K^+} = \frac{D-F}{\sqrt{2} F_0}$}
& \makecell[l]{
${\mathbb B}_{n \ell^- K^+}$\\
${\mathbb B}_{\Sigma^- \ell^-\pi^0}$}
& $\mathbb B_{n\ell^-\pi^0 K^+}$
\\
\hline
\end{tabular}}
\caption{
The relevant couplings $\mathbb A_i$ and $\mathbb B_i$ in each specific decay mode involving a charged lepton $\ell=e,\mu$. Note that we omit the superscripts $\tL$ and $\tR$ in $\mathbb B_i$ for simplicity.}
\label{tab:ell_mode}
\end{table}

\begin{table}[t]
\centering
\resizebox{\linewidth}{!}{
\renewcommand{\arraystretch}{1.4}
\begin{tabular}{|l|l|l|c|l|c|c|}
\hline
\multirow{2}*{\quad~Mode}
& \multicolumn{3}{c|}{Diagrams (a) and (b)} 
& \multicolumn{2}{c|}{Diagram (c)}
& \multicolumn{1}{c|}{Diagram (d)}
\\\cline{2-7}
& \multicolumn{1}{c|}{$\mathbb A_{{\tt N}B'M_{1(2)}}\mathbb A_{B'BM_{2(1)}}$} 
& \multicolumn{1}{c|}{$\mathbb A_{{\tt N}BM_1 M_2}$}  
& \multicolumn{1}{c|}{$\mathbb B_{B\ell}$} 
& \multicolumn{1}{c|}{$\mathbb A_{{\tt N}BM_{1(2)}}$} 
& \multicolumn{1}{c|}{$\mathbb B_{B\ell M_{2(1)}}$} 
& \multicolumn{1}{c|}{$\mathbb B_{{\tt N}\ell M_1 M_2}$} 
\\
\hline\hline
~$p\to \hat \nu \pi^+ \pi^0$
&\makecell[l]{ 
$\mathbb A_{pn\pi^+}\mathbb A_{nn\pi^0} = \frac{-(D+F)^2}{2\sqrt{2}F_0^2}$ \\
$\mathbb A_{pp\pi^0}\mathbb A_{pn\pi^+} = \frac{(D+F)^2}{2\sqrt{2}F_0^2}$  }
&~$\mathbb A_{pn\pi^+\pi^0} = \frac{1}{2\sqrt{2} F^2_0} $
& $\mathbb B_{n \hat \nu}$
& \makecell[l]{~$\mathbb A_{pn\pi^+} = 
\frac{D+F}{\sqrt{2} F_0}$ \\ 
~$\mathbb A_{pp\pi^0} = 
\frac{D+F}{2 F_0}$}
& \makecell[l]{
$\mathbb B_{n \hat\nu \pi^0}$\\
${\mathbb B}_{p \hat \nu\pi^+}$}
& ---
\\\hline
~$p\to \hat \nu \pi^+ \eta$
& \makecell[l]{ 
$\mathbb A_{pn\pi^+}\mathbb A_{nn\eta} = \frac{(D+F)(3F-D)}{2\sqrt{6}F_0^2}$ \\
$\mathbb A_{pp\eta}\mathbb A_{pn\pi^+} = \frac{(D+F)(3F-D)}{2\sqrt{6}F_0^2}$  }
& \multicolumn{1}{c|}{---}
& $\mathbb B_{n \hat \nu}$
& \makecell[l]{~$\mathbb A_{pn\pi^+} = \frac{D+F}{\sqrt{2} F_0}$ \\ 
~$\mathbb A_{pp\eta} = \frac{3F-D}{2\sqrt{3} F_0}$}
& \makecell[l]{
$\mathbb B_{n \hat\nu \eta}$\\
${\mathbb B}_{p \hat \nu\pi^+}$}
& $\mathbb B_{p \hat \nu \pi^+ \eta}$
\\\hline
~$n\to \hat \nu \pi^+ \pi^-$
& $\mathbb A_{np\pi^-}\mathbb A_{pn \pi^+} = \frac{(D+F)^2}{2 F_0^2}$  
&~$\mathbb A_{nn\pi^+ \pi^-} = \frac{1}{4F^2_0} $
& $\mathbb B_{n \hat \nu}$ 
&~$\mathbb A_{np\pi^-} = \frac{D+F}{\sqrt{2} F_0}$
& ${\mathbb B}_{p \hat \nu\pi^+}$
& $\mathbb B_{n \hat \nu \pi^+ \pi^-}$
\\\hline
~$n\to \hat \nu \pi^0\pi^0 $
& $\mathbb A_{nn\pi^0}\mathbb A_{nn \pi^0} = \frac{(D+F)^2}{4 F_0^2}$ 
& \multicolumn{1}{c|}{---}
& $\mathbb B_{n \hat \nu}$ 
&~$\mathbb A_{nn\pi^0} = -\frac{D+F}{2 F_0}$
& ${\mathbb B}_{n \hat \nu \pi^0}$
& $\mathbb B_{n \hat \nu \pi^0 \pi^0}$
\\\hline
~$n\to \hat \nu \pi^0 \eta$
&\makecell[l]{$\mathbb A_{nn\pi^0}\mathbb A_{nn \eta} = \frac{(D+F)(D-3F)}{4\sqrt{3} F_0^2}$\\
$\mathbb A_{nn\eta}\mathbb A_{nn \pi^0} = \frac{(D+F)(D-3F)}{4\sqrt{3} F_0^2}$ }
& \multicolumn{1}{c|}{---}
& $\mathbb B_{n \hat \nu}$ 
& \makecell[l]{~$\mathbb A_{nn\pi^0} = -\frac{D+F}{2 F_0}$ \\
~$\mathbb A_{nn\eta} = \frac{3F-D}{2\sqrt{3} F_0}$ }
& \makecell[l]{ ${\mathbb B}_{n \hat \nu \eta}$ \\
${\mathbb B}_{n \hat \nu \pi^0}$ }
& $\mathbb B_{n \hat \nu \pi^0 \eta}$
\\\hline\hline
\multirow{3}*{~$p\to \hat \nu \pi^+ K^0 $ }
& \makecell[l]{ $\mathbb A_{pn\pi^+}\mathbb A_{n\Lambda^0 K^0} = \frac{-(D+F)(D+3F)}{2\sqrt{6}F_0^2}$ \\
$\mathbb A_{p\Sigma^+ K^0}\mathbb A_{\Sigma^+ \Lambda^0 \pi^+} = \frac{(D-F)D}{\sqrt{6}F_0^2}$  }
&~$\mathbb A_{p\Lambda^0 \pi^+ K^0} =\frac{\sqrt{3}}{4\sqrt{2} F^2_0} $
& $\mathbb B_{\Lambda^0 \hat \nu}$ 
&~$\mathbb A_{pn\pi^+} = \frac{D+F}{\sqrt{2} F_0}$
& $\mathbb B_{n \hat\nu K^0}$
& \multirow{3}*{ $\mathbb B_{p \hat \nu \pi^+ K^0}$ }
\\\cdashline{2-4}
& \makecell[l]{
$\mathbb A_{pn\pi^+}\mathbb A_{n\Sigma^0 K^0} = \frac{F^2-D^2}{2\sqrt{2}F_0^2}$ \\
$\mathbb A_{p\Sigma^+ K^0}\mathbb A_{\Sigma^+ \Sigma^0 \pi^+} = \frac{(F-D)F}{\sqrt{2}F_0^2}$  }
&~$\mathbb A_{p\Sigma^0 \pi^+ K^0} = \frac{1}{4\sqrt{2} F^2_0} $
& $\mathbb B_{\Sigma^0 \hat \nu}$  
&~$\mathbb A_{p\Sigma^+ K^0} = \frac{D-F}{\sqrt{2} F_0}$
& ${\mathbb B}_{\Sigma^+ \hat \nu\pi^+}$
&
\\\hline
\multirow{3}*{~$p\to \hat \nu \pi^0 K^+ $}
& \makecell[l]{ $\mathbb A_{pp\pi^0}\mathbb A_{p\Lambda^0 K^+} = \frac{-(D+F)(D+3F)}{4\sqrt{3}F_0^2}$ \\
$\mathbb A_{p\Sigma^0 K^+}\mathbb A_{\Sigma^0 \Lambda^0 \pi^0} = \frac{(D-F)D}{2\sqrt{3}F_0^2}$   }
&~$\mathbb A_{p\Lambda^0 \pi^0 K^+} = \frac{\sqrt{3}}{8F^2_0} $ 
& $\mathbb B_{\Lambda^0 \hat \nu}$ 
& \multirow{3}*{~\makecell[l]{$\mathbb A_{pp\pi^0} = \frac{D+F}{2 F_0}$ 
\vspace{0.5em}\\ 
$\mathbb A_{p\Lambda^0 K^+} = \frac{-(D+3F)}{2\sqrt{3} F_0}$ 
\vspace{0.5em}\\
$\mathbb A_{p\Sigma^0 K^+} = \frac{D-F}{2 F_0}$} }
& \multirow{3}*{ \makecell[l]{
$\mathbb B_{p \hat\nu K^+}$ 
\vspace{0.5em}\\
${\mathbb B}_{\Lambda^0 \hat \nu\pi^0}$ 
\vspace{0.5em}\\
${\mathbb B}_{\Sigma^0 \hat \nu\pi^0}$} }
& \multirow{3}*{$\mathbb B_{p \hat \nu \pi^0 K^+}$ }
\\\cdashline{2-4}
& \makecell[l]{
$\mathbb A_{pp\pi^0}\mathbb A_{p\Sigma^0 K^+} = \frac{D^2-F^2}{4 F_0^2}$ \\
$\mathbb A_{p\Lambda^0 K^+}\mathbb A_{\Lambda^0 \Sigma^0 \pi^0} = \frac{-(D+3F)D}{6F_0^2}$ }
&~$\mathbb A_{p\Sigma^0 \pi^0 K^+} = \frac{1}{8 F^2_0} $
& $\mathbb B_{\Sigma^0 \hat \nu}$ 
& 
&
&
\\\hline
\multirow{3}*{~$n\to \hat \nu \pi^- K^+$}
& \makecell[l]{ $\mathbb A_{np\pi^-}\mathbb A_{p\Lambda^0 K^+} = \frac{-(D+F)(D+3F)}{2\sqrt{6}F_0^2}$ \\
$\mathbb A_{n\Sigma^- K^+}\mathbb A_{\Sigma^- \Lambda^0 \pi^-} = \frac{(D-F)D}{\sqrt{6}F_0^2}$   }
&~$\mathbb A_{n\Lambda^0 \pi^- K^+} = \frac{\sqrt{3}}{4\sqrt{2}F^2_0}$ 
& $\mathbb B_{\Lambda^0 \hat \nu}$ 
&~$\mathbb A_{np\pi^-} = \frac{D+F}{\sqrt{2} F_0}$ 
& $\mathbb B_{p \hat\nu K^+}$ 
& \multirow{3}*{ $\mathbb B_{n \hat \nu \pi^- K^+}$ }
\\\cdashline{2-4}
& \makecell[l]{
$\mathbb A_{np\pi^-}\mathbb A_{p\Sigma^0 K^+} = \frac{D^2-F^2}{2\sqrt{2} F_0^2}$ \\
$\mathbb A_{n\Sigma^- K^+}\mathbb A_{\Sigma^- \Sigma^0 \pi^-} = \frac{(D-F)F}{\sqrt{2}F_0^2}$ }
&~$\mathbb A_{n\Sigma^0 \pi^- K^+} = \frac{-1}{4\sqrt{2} F^2_0} $ 
& $\mathbb B_{\Sigma^0 \hat \nu}$ 
&~$\mathbb A_{n\Sigma^- K^+} = \frac{D-F}{\sqrt{2} F_0}$
& ${\mathbb B}_{\Sigma^- \hat \nu \pi^-}$
&
\\\hline
\multirow{3}*{~$n\to \hat \nu \pi^0 K^0$ }
& \makecell[l]{ $\mathbb A_{nn\pi^0}\mathbb A_{n\Lambda^0 K^0} = \frac{(D+F)(D+3F)}{4\sqrt{3}F_0^2}$ \\
$\mathbb A_{n\Sigma^0 K^0}\mathbb A_{\Sigma^0 \Lambda^0 \pi^0} = \frac{(F-D)D}{2\sqrt{3}F_0^2}$   }
&~$\mathbb A_{n\Lambda^0 \pi^0 K^0} = -\frac{\sqrt{3}}{8F^2_0} $
& $\mathbb B_{\Lambda^0 \hat \nu}$ 
& \multirow{3}*{~\makecell[l]{
$\mathbb A_{nn\pi^0} = -\frac{D+F}{2 F_0}$ 
\vspace{0.5em}\\ 
$\mathbb A_{n\Lambda^0 K^0} = -\frac{D+3F}{2\sqrt{3} F_0}$ \vspace{0.5em}\\ 
$\mathbb A_{n\Sigma^0 K^0} = \frac{F-D}{2 F_0}$ } }
& \multirow{3}*{\makecell[l]{
$\mathbb B_{n \hat\nu K^0}$ 
\vspace{0.5em}\\
${\mathbb B}_{\Lambda^0 \hat \nu \pi^0}$ 
\vspace{0.5em}\\
${\mathbb B}_{\Sigma^0 \hat \nu \pi^0}$ } }
& \multirow{3}*{ $\mathbb B_{n \hat \nu \pi^0 K^0}$}
\\\cdashline{2-4}
& \makecell[l]{
$\mathbb A_{nn\pi^0}\mathbb A_{n\Sigma^0 K^0} = \frac{D^2-F^2}{4 F_0^2}$ \\
$\mathbb A_{n\Lambda^0 K^0}\mathbb A_{\Lambda^0 \Sigma^0 \pi^0} = \frac{-(D+3F)D}{6F_0^2}$ }
&~$\mathbb A_{n\Sigma^0 \pi^0 K^0} = \frac{1}{8 F^2_0} $
& $\mathbb B_{\Sigma^0 \hat \nu}$
& 
&
&
\\
\hline
\end{tabular}}
\caption{
Similar to \cref{tab:ell_mode}, but for decay modes involving a neutrino or an antineutrino.}
\label{tab:hatnu_mode}
\end{table}

For each three-body nucleon decay mode listed in \cref{tab:opeandprocess}, 
the explicit expressions for the coefficients $\mathbb A_i$ and $\mathbb B_i$, corresponding to each diagram in \cref{fig:Feyndiagram}, are summarized in \cref{tab:ell_mode,tab:hatnu_mode}.
Since \cref{fig:Feyndiagram}\,(a) contains two BNC three-point vertices, we present the product of the two $\mathbb A_i$ coefficients in \cref{tab:ell_mode,tab:hatnu_mode}, with $\mathbb A_{{\tt N}B'M_{1,2}}$, which involves the initial nucleon state, always listed first.  
From the tables, the intermediate baryon states associated with each diagram can be identified by the subscripts of the corresponding 
$\mathbb A_i$s and $\mathbb B_i$s.
In the first column of the table, the order of the two nonidentical mesons is important: the first meson in the middle position and the second meson in the last position correspond to $M_1$ and $M_2$, respectively, in the generic process ${\tt N}\to l M_1 M_2$. 
Following this convention, one can easily identify the two distinct contributions in 
\cref{fig:Feyndiagram}\,(a) and (c) from \cref{tab:ell_mode,tab:hatnu_mode}.

Using the Dirac equations and properties of gamma matrices, the amplitude in \cref{eq:ampfull} can be reduced into the following form:
\begin{align}
{\cal M}
= \overline{u_l}\left[
\mathbb D_{\texttt{N};l M_1M_2}^{{\tt S}\tL} P_\tL 
+\mathbb  D_{\texttt{N};l M_1M_2}^{{\tt S}\tR} P_\tR 
+\mathbb  D_{\texttt{N};l M_1M_2}^{{\tt V}\tL} \slashed{q} P_\tL 
+\mathbb  D_{\texttt{N};l M_1M_2}^{{\tt V}\tR} \slashed{q} P_\tR 
\right]u_\texttt{N},
\label{eq:amp3body}
\end{align}
where $q=p_2-p_1$.  
Defining the invariant masses for pairs of final state particles as $s\equiv(p_1+p_2)^2$, $t\equiv (p_2+k)^2$, and $u\equiv (p_1+k)^2$, 
the $\mathbb D$-parameters in \cref{eq:amp3body} are given by
\begin{subequations}
\label{eq:Dcoeffs}
\begin{align}
\mathbb D_{\texttt{N};l M_1M_2}^{{\tt S}\tL} =\,&
\frac{\mathbb A_{{\tt N} B' M_1} \mathbb A_{B' B M_2}}{2(m_l^2 - m_{B}^2) (t - m_{B'}^2)}
\big\{m_l (m_l \mathbb B_{B l}^\tL + m_{B}\mathbb B_{B l}^\tR) (m_\texttt{N}^2 - t)
\nonumber\\
&
+(m_{B}\mathbb B_{B l}^\tL+ m_l \mathbb B_{B l}^\tR)
[m_{B'}(m_\texttt{N}^2 - t)- (m_\texttt{N}+m_{B'})(t-m_l^2)] \big\}
\nonumber\\
&+ 
\frac{\mathbb A_{{\tt N} B' M_2} \mathbb A_{B' B M_1}}{2(m_l^2 - m_{B}^2) (u - m_{B'}^2)}
\big\{ m_l (m_l \mathbb B_{B l}^\tL + m_{B}\mathbb B_{B l}^\tR) (m_\texttt{N}^2 - u)
\nonumber\\
&
+ (m_{B}\mathbb B_{B l}^\tL+ m_l \mathbb B_{B l}^\tR)
[m_{B'}(m_\texttt{N}^2 - u)- (m_\texttt{N}+m_{B'})(u-m_l^2)]
\big\}
\nonumber\\
&
-\frac{\mathbb A_{{\tt N} B M_1}}{2(t - m_B^2)}
\big[(m_\texttt{N}^2 - m_\texttt{N}m_B - 2 t) \mathbb B_{BlM_2}^\tL
-m_l (m_\texttt{N} + m_B) \mathbb B_{BlM_2}^\tR \big]
\nonumber\\
& - \frac{\mathbb A_{{\tt N} B M_2}}{2(u - m_B^2)}
\big[(m_\texttt{N}^2 - m_\texttt{N}m_B - 2 u) \mathbb B_{BlM_1}^\tL
-m_l (m_\texttt{N} + m_B) \mathbb B_{BlM_1}^\tR \big]
\nonumber\\
&+ \mathbb B_{{\tt N}lM_1 M_2}^\tL, 
\\
\mathbb  D_{\texttt{N};l M_1M_2}^{{\tt S}\tR} =\,&  
\frac{\mathbb A_{{\tt N} B' M_1} \mathbb A_{B' B M_2}}{2(m_l^2 - m_{B}^2) (t - m_{B'}^2)}
\big\{ m_l (m_l \mathbb B_{B l}^\tR + m_{B}\mathbb B_{B l}^\tL) (m_\texttt{N}^2 - t)
\nonumber\\
&
+ (m_{B}\mathbb B_{B l}^\tR+ m_l \mathbb B_{B l}^\tL)
[m_{B'}(m_\texttt{N}^2 - t)- (m_\texttt{N}+m_{B'})(t-m_l^2)]
\big\}
\nonumber\\
&+ 
\frac{\mathbb A_{{\tt N} B' M_2} \mathbb A_{B' B M_1}}{2(m_l^2 - m_{B}^2) (u - m_{B'}^2)}
\big\{m_l (m_l \mathbb B_{B l}^\tR + m_{B}\mathbb B_{B l}^\tL) (m_\texttt{N}^2 - u)
\nonumber\\
&
+(m_{B}\mathbb B_{B l}^\tR+ m_l \mathbb B_{B l}^\tL)
[m_{B'}(m_\texttt{N}^2 - u)- (m_\texttt{N}+m_{B'})(u-m_l^2)]
\big\}
\nonumber\\
&%
+\frac{\mathbb A_{{\tt N} B M_1}}{2(t - m_B^2)}
\big[(m_\texttt{N}^2 - m_\texttt{N}m_B - 2 t) \mathbb B_{BlM_2}^\tR
-m_l (m_\texttt{N} + m_B) \mathbb B_{BlM_2}^\tL \big]
\nonumber\\
& + \frac{\mathbb A_{{\tt N} B M_2}}{2(u - m_B^2)}
\big[(m_\texttt{N}^2 - m_\texttt{N}m_B - 2 u) \mathbb B_{BlM_1}^\tR
-m_l (m_\texttt{N} + m_B) \mathbb B_{BlM_1}^\tL \big]
\nonumber\\
&+ \mathbb B_{{\tt N}lM_1 M_2}^\tR, 
\\
\mathbb  D_{\texttt{N};l M_1M_2}^{{\tt V}\tL} =\,&  
\frac{\mathbb A_{{\tt N} B' M_1} \mathbb A_{B' B M_2}}{2(m_l^2 - m_{B}^2) (t - m_{B'}^2)}
\big\{ (m_l \mathbb B_{B l}^\tL + m_{B}\mathbb B_{B l}^\tR)(m_\texttt{N} m_{B'}+t)
\nonumber\\
&+m_l(m_{B}\mathbb B_{B l}^\tL+m_l \mathbb B_{B l}^\tR) (m_\texttt{N} +m_{B'})
\big\}
\nonumber\\
&-
\frac{\mathbb A_{{\tt N} B' M_2} \mathbb A_{B' B M_1}}{2(m_l^2 - m_{B}^2) (u - m_{B'}^2)}
\big\{ (m_l \mathbb B_{B l}^\tL + m_{B}\mathbb B_{B l}^\tR)(m_\texttt{N} m_{B'}+u)
\nonumber\\
&+m_l(m_{B}\mathbb B_{B l}^\tL+m_l \mathbb B_{B l}^\tR) (m_\texttt{N} +m_{B'})
\big\}
\nonumber\\
&%
+(m_\texttt{N}+m_B) \left[ \frac{\mathbb A_{{\tt N} B M_1} \mathbb B_{BlM_2}^\tR}{2(t - m_B^2)}
-\frac{\mathbb A_{{\tt N} B M_2} \mathbb B_{BlM_1}^\tR}{2(u - m_B^2)}\right]
\nonumber\\
&+ \mathbb A_{{\tt N} B M_1 M_2} \frac{m_l \mathbb B_{Bl}^\tL + m_B \mathbb B_{Bl}^\tR }{m_l^2- m_{B}^2},
\\
\mathbb  D_{\texttt{N};l M_1M_2}^{{\tt V}\tR}=\,& 
\frac{\mathbb A_{{\tt N} B' M_1} \mathbb A_{B' B M_2}}{2(m_l^2 - m_{B}^2) (t - m_{B'}^2)}
\big\{ (m_l \mathbb B_{B l}^\tR + m_{B}\mathbb B_{B l}^\tL)(m_\texttt{N} m_{B'}+t)
\nonumber\\
&+m_l(m_{B}\mathbb B_{B l}^\tR+m_l \mathbb B_{B l}^\tL) (m_\texttt{N} +m_{B'})
\big\}
\nonumber\\
&-
\frac{\mathbb A_{{\tt N} B' M_2} \mathbb A_{B' B M_1}}{2(m_l^2 - m_{B}^2) (u - m_{B'}^2)}
\big\{ (m_l \mathbb B_{B l}^\tR + m_{B}\mathbb B_{B l}^\tL) (m_\texttt{N} m_{B'}+u)
\nonumber\\
&+m_l(m_{B}\mathbb B_{B l}^\tR+m_l \mathbb B_{B l}^\tL) (m_\texttt{N} +m_{B'})
\big\}
\nonumber\\
&-(m_\texttt{N}+m_B) \left[ \frac{\mathbb A_{{\tt N} B M_1} \mathbb B_{BlM_2}^\tL}{2(t - m_B^2)}
-\frac{\mathbb A_{{\tt N} B M_2} \mathbb B_{BlM_1}^\tL}{2(u - m_B^2)}\right]
\nonumber\\
&
+ \mathbb A_{{\tt N} B M_1 M_2} \frac{m_l \mathbb B_{Bl}^\tR + m_B \mathbb B_{Bl}^\tL }{m_l^2- m_{B}^2},
\end{align}
\end{subequations}
where, again, summations over all possible intermediate baryon states $B$ and $B'$ are implied. These states have been explicitly incorporated into the coefficients in \cref{tab:ell_mode,tab:hatnu_mode}, from which one can recognize the specific baryon states and the corresponding expressions relevant to each term above.

From \cref{eq:amp3body}, the spin-averaged and -summed matrix element squared can be expressed compactly as 
\begin{align}
\overline{|{\cal M}|^2} 
& =  {1\over 2}(m_\texttt{N}^2 + m_l^2 - s)
\left(|\mathbb D_{\texttt{N};lM_1M_2}^{{\tt S}\tL}|^2 + |\mathbb D_{\texttt{N};lM_1M_2}^{{\tt S}\tR}|^2\right)
\nonumber
\\
& + {1\over 2} \left[(m_\texttt{N}^2 +m_l^2 - s)(s-2m_1^2 - 2 m_2^2) 
+(t-u)^2 - (m_1^2-m_2^2)^2\right] 
\nonumber
\\
&\times \left(|\mathbb D_{\texttt{N};lM_1M_2}^{{\tt V}\tL}|^2 
+ |\mathbb D_{\texttt{N};lM_1M_2}^{{\tt V}\tR}|^2\right)
\nonumber
\\
& + 2 m_l m_\texttt{N} \left[ \Re\left(
\mathbb D_{\texttt{N};lM_1M_2}^{{\tt S}\tL}
\mathbb D_{\texttt{N};lM_1M_2}^{{\tt S}\tR\,*}  \right)
- (s-2m_1^2 - 2 m_2^2) \, 
\Re\left(\mathbb D_{\texttt{N};lM_1M_2}^{{\tt V}\tL}
\mathbb D_{\texttt{N};lM_1M_2}^{{\tt V}\tR\,*} \right) \right]
\nonumber
\\
& + m_l(t-u-m_1^2 +m_2^2)\,\Re\left(
\mathbb D_{\texttt{N};lM_1M_2}^{{\tt S}\tL}
\mathbb D_{\texttt{N};lM_1M_2}^{{\tt V}\tL\,*} 
+\mathbb D_{\texttt{N};lM_1M_2}^{{\tt S}\tR}
\mathbb D_{\texttt{N};lM_1M_2}^{{\tt V}\tR\,*}
\right) 
\nonumber
\\
& +m_\texttt{N} (t-u+m_1^2-m_2^2)\,\Re\left(
\mathbb D_{\texttt{N};lM_1M_2}^{{\tt S}\tL}
\mathbb D_{\texttt{N};lM_1M_2}^{{\tt V}\tR\,*} 
+\mathbb D_{\texttt{N};lM_1M_2}^{{\tt S}\tR}
\mathbb D_{\texttt{N};lM_1M_2}^{{\tt V}\tL\,*} \right). 
\label{eq:Msquared}
\end{align}
Finally, the decay width is
\begin{align}
\Gamma_{\texttt{N}\to lM_1 M_2}=\frac{1}{1+\delta_{M_1 M_2}}\frac{1}{256\pi^3 m_{\texttt N}^3}  \int_{s_-}^{s_+} d s \int_{t_-}^{t_+} d t \quad\overline{|{\cal M}_{\texttt{N}\to lM_1 M_2}|^2},
\end{align}
where the integration limits are
\begin{align}
s_- & = (m_1 +m_2)^2,\quad 
s_+ = (m_{\texttt N} - m_l)^2,
\nonumber\\
t_\pm & = (E_2^* + E_3^*)^2 - \Big(\sqrt{E_2^{*2} - m_2^2} \mp \sqrt{E_3^{*2} - m_l^2} \Big)^2, \quad 
\nonumber\\
E_2^* & = 
\frac{s - m_1^2 + m_2^2}{2\sqrt{s} },\quad
E_3^* = 
\frac{ m_{\texttt N}^2 - s - m_l^2}{2\sqrt{s}}.
\end{align}

Based on the above formalism, we compute 
the decay width for each three-body nucleon decay as a function of the LEFT WCs.
Numerically, we use the central values of particle masses and pion decay constant from the Particle Data Group~\cite{ParticleDataGroup:2024cfk}, and
the LECs $c_{1,2}$ and $D,F$ from the lattice QCD determinations in~\cite{Yoo:2021gql,Bali:2022qja}. 
The final decay width expressions are summarized in appendix \ref{app:DeW_three_body}.

From the analytical results derived in this section, one can also analyze the momentum and invariant-mass distributions of the final-state charged leptons and mesons. 
These distributions are essential for more realistic experimental searches of these three-body decay modes, especially when compared to the assumption of a flat phase space distribution used in the previous experimental searches. 
Moreover, they provide a bridge for computations in specific new physics scenarios. Given a model, the LEFT WCs can be expressed in terms of its parameters by integrating out its heavy degrees of freedom.
Substituting them into the formulas in appendix \ref{app:DeW_three_body} then allows more definite predictions of these processes. 

In the above calculation, we neglected contributions mediated by octet vector mesons due to their large uncertainties. 
Fortunately, these contributions can be estimated within the framework of resonance ChPT.
A detailed evaluation of this additional contribution is presented in appendix \ref{app:vectorM}. From the final numerical results, 
we find that vector-meson contributions account for less than 30\,\% for decay modes involving two nonidentical pions, whereas for several modes involving a kaon and an $e^\pm$ or a (anti)neutrino, they can be several times larger than the contributions given in appendix \ref{app:DeW_three_body}.
For completeness, the total contributions, including the vector-meson component, are provided in appendix \ref{app:vectorM} for reference. 

Finally, we emphasize that our calculation is based on the leading-order baryon ChPT Lagrangian. 
The convergence of $\rm SU(3)$ baryon ChPT may be slower for decay modes involving a kaon or an $\eta$ because of their relatively large masses and the momenta over the three-body phase space.
In addition, cancellations or interference among contact, baryon-pole, and vector-meson contributions may enhance the sensitivity of certain decay modes to higher-order effects in specific regions of phase space.
To investigate this sensitivity, 
we have examined the corrections associated with the baryon-mass terms in the next-to-leading-order chiral Lagrangian, 
$\mathcal{L}\supset  
b_D \overline{B}\{M_+, B\} 
+ b_F \overline{B}[M_+,B]$~\cite{Bijnens:1985kj}, where $M_+ = \xi^\dagger M_q \xi^\dagger + \xi M_q^\dagger \xi$, with $M_q = {\rm diag}(m_u,m_d,m_s)$
the light-quark-mass matrix, and the LECs $b_D=0.226$ and $b_F=-0.811$~\cite{Buchalla:1995vs,He:2025wxn}.
We find that these terms induce changes in the decay widths of up to $50\,\%$ for certain operators in kaon-related modes, while the changes are below $5\,\%$ for pion-pair or $\eta$-related modes.
This exercise provides a partial indication of the sensitivity of our predictions to higher-order effects, rather than a complete estimate of the theoretical uncertainties.
A systematic assessment would require the full set of relevant higher-order contributions, including additional chiral operators and loop corrections, together with improved control of the associated LECs.
We therefore regard our results as leading-order estimates and note that the theoretical uncertainties may be more significant for certain channels, particularly those involving a kaon or an $\eta$ meson.

\section{Improved bounds on three-body nucleon decays ${\tt N}\to l M_1 M_2$}
\label{sec:newbounds}

Having established the general decay width expressions for all relevant three-body nucleon decay modes,
we now proceed to derive improved bounds on them by employing the experimentally well-constrained two-body modes that arise from the same set of dim-6 BNV operators. 
To this end, we calculate the corresponding decay widths for the two-body modes.
Following the general formalism of Ref.~\cite{Liao:2025sqt}, we obtain  their explicit expressions, which are summarized in appendix \ref{app:DeW_two_body}.
Using the experimental bounds on the two-body processes as inputs, we can derive constraints on the associated WCs of the dim-6 operators, from which new improved lifetime bounds on the three-body modes can be set. 

In this work, we pursue two approaches to achieve this goal. The first relies on the single-operator-dominance assumption, where we activate one operator at a time.
Consequently, the decay width of a three-body mode becomes proportional to that of the associated two-body modes. This allows us to translate experimental bounds on the two-body modes into bounds on the three-body mode, and the most stringent among them is chosen as the final limit. The merit of this approach lies in its ability to yield a very tight bound on each corresponding three-body mode for every relevant operator. 
Its main limitation, however, is that the resulting bounds are operator-dependent.  

The second method adopts a global analysis 
that treats all operators on the same footing. Unlike the approach that constrains a single WC from one two-body mode, we simultaneously use several two-body modes to confine the multidimensional WC parameter space into a closed, bounded region. We then evaluate the partial lifetimes for every three-body mode described by the same set of operators by varying the WCs within this allowed region and select the smallest value as the final conservative limit.   
Clearly, the partial lifetime bound obtained in this way is generally weaker, by up to orders of magnitude,  than that from the single-operator-dominance analysis.
Nevertheless, this conservative bound is assumption-free and therefore more robust. In the following, we discuss the results of the two approaches one by one.  

\subsection{Single operator analysis} 

We begin with the first approach, considering one operator at a time.
By requiring that the theoretical decay width of each two-body mode in appendix \ref{app:DeW_two_body} be less than the inverse of its experimental lower partial lifetime limit, 
we derive constraints on the WCs. 
For each LEFT operator $\calO_i^j$, we define a dimensionless WC through $\tilde C_i^j\equiv
(10^{15}\,{\rm GeV})^2\,C_i^j$, where $C_i^j$ is its dimensionful WC. 
Similarly to Table S4 in Ref.~\cite{Liao:2025sqt}, \cref{tab:BoundonWC} presents the one-operator-at-a-time bounds on the magnitudes of the dimensionless WCs, obtained from the current best experimental limits on two-body nucleon decays listed in the first and second columns. 
In the table, the most stringent bound on each WC is highlighted in gray. Operators related by a chirality flip $\tL\leftrightarrow \tR$ yield identical bounds and are therefore grouped together. For operators involving a charged lepton, $\ell$, the superscript $x$ denotes either $e$ or $\mu$, depending on the specific process; while for operators involving a neutrino or an antineutrino, $x$ can represent any of the three flavors.

\begin{table}[t]
\center
\resizebox{\linewidth}{!}{
\renewcommand{\arraystretch}{1.1}
\begin{tabular}{|l|c|c|c|c|c|c|c|c|c|c|c|c|c|c|}
\hline
\multirow{3}*{\quad~Mode}  
& \multirow{3}*{ $\makecell{ \Gamma^{-1}_{\rm exp.} 
\\{[}10^{33}\,\rm yr]}$}
&\multicolumn{13}{c|}{Constraint on the dimensionless WC $|\tilde C_i^j|\equiv
(10^{15}\,{\rm GeV})^2|C_i^j|$ }
\\\cline{3-15}
&
&$~\calO_{\ell uud}^{\tL\tR,x}~$ 
&$~\calO_{\ell uud}^{\tL\tL,x}~$ 
&$~\calO_{\ell usu}^{\tL\tR,x}~$ 
&$~\calO_{\ell usu}^{\tL\tL,x}~$
&$~\calO_{\bar\ell dds}^{\tL\tR,x}~$
&$~\calO_{\bar\ell dds}^{\tL\tL,x}~$
&$~\calO_{\nu dud}^{\tL\tR,x}~$ 
&$~\calO_{\nu dud}^{\tL\tL,x}~$
&$~\calO_{\nu uds}^{\tL\tR,x}~$ 
&$~\calO_{\nu dsu}^{\tL\tR,x}~$ 
&$~\calO_{\nu dsu}^{\tL\tL,x}~$ 
&$~\calO_{\nu sud}^{\tL\tR,x}~$
&$~\calO_{\nu sud}^{\tL\tL,x}~$
\\\cline{3-15}
& 
&$~\calO_{\ell uud}^{\tR\tL,x}~$ 
&$~\calO_{\ell uud}^{\tR\tR,x}~$ 
&$~\calO_{\ell usu}^{\tR\tL,x}~$ 
&$~\calO_{\ell usu}^{\tR\tR,x}~$
&$~\calO_{\bar\ell dds}^{\tR\tL,x}~$
&$~\calO_{\bar\ell dds}^{\tR\tR,x}~$
&$\calO_{\bar\nu dud}^{\tR\tL,x}~$ 
&$\calO_{\bar\nu dud}^{\tR\tR,x}~$
&$~\calO_{\bar\nu uds}^{\tR\tL,x}~$ 
&$~\calO_{\bar\nu dsu}^{\tR\tL,x}~$ 
&$~\calO_{\bar\nu dsu}^{\tR\tR,x}~$ 
&$~\calO_{\bar\nu sud}^{\tR\tL,x}~$
&$~\calO_{\bar\nu sud}^{\tR\tR,x}~$
\\\hline
$~p\to e^+ \pi^0 $ & $24$~\cite{Super-Kamiokande:2020wjk} 
&\cellcolor{gray!50}$0.066$ &\cellcolor{gray!50}$0.065$ & $-$ & $-$ & $-$ & $-$ & $-$ & $-$ & $-$ & $-$ & $-$ & $-$ & $-$
\\
$~p\to \mu^+ \pi^0 $ & $16$~\cite{Super-Kamiokande:2020wjk}
&\cellcolor{gray!50}$0.082$ &\cellcolor{gray!50}$0.081$ & $-$ & $-$ & $-$ & $-$ & $-$ & $-$ & $-$ & $-$ & $-$ & $-$ & $-$
\\
$~p\to e^+ \eta $ & $14$\,\cite{Super-Kamiokande:2024qbv}
& $1.25$ & $0.13$ & $-$ & $-$ & $-$ & $-$ & $-$ & $-$ & $-$ & $-$ & $-$ & $-$ & $-$
\\
$~p\to \mu^+ \eta $ & $7.3$\,\cite{Super-Kamiokande:2024qbv}
& $1.55$ & $0.19$ & $-$ & $-$ & $-$ & $-$ & $-$ & $-$ & $-$ & $-$ & $-$ & $-$ & $-$
\\
$~n\to e^+ \pi^-$ & $5.3$~\cite{Super-Kamiokande:2017gev}
& $0.1$ & $0.099$ & $-$ & $-$ & $-$ & $-$ & $-$ & $-$ & $-$ & $-$ & $-$ & $-$ & $-$
\\
$~n\to \mu^+ \pi^-$ & $3.5$~\cite{Super-Kamiokande:2017gev}
& $0.12$ & $0.12$ & $-$ & $-$ & $-$ & $-$ & $-$ & $-$ & $-$ & $-$ & $-$ & $-$ & $-$
\\\hline
$~p\to e^+ K^{0} $ & $1.0$~\cite{Super-Kamiokande:2005lev}
& $-$ & $-$ 
&\cellcolor{gray!50}$0.56$ &\cellcolor{gray!50}$0.87$
& $-$ &  $-$  & $-$ & $-$ & $-$ & $-$ & $-$ & $-$ & $-$
\\
$~p\to \mu^+ K^{0} $ & $4.5$~\cite{ParticleDataGroup:2024cfk}
& $-$ & $-$ 
&\cellcolor{gray!50}$0.27$ &\cellcolor{gray!50}$0.41$
& $-$ & $-$ & $-$ & $-$ & $-$ & $-$ & $-$ & $-$ & $-$
\\\hline
$~n\to e^- K^+$ & $0.032$~\cite{Frejus:1991ben}
& $-$ & $-$ & $-$ & $-$
&\cellcolor{gray!50}$3.09$  &\cellcolor{gray!50}$4.82$ 
& $-$ & $-$ & $-$ & $-$ & $-$ & $-$ & $-$ 
\\
$~n\to \mu^- K^+$ & $0.057$~\cite{Frejus:1991ben}
& $-$ & $-$ & $-$ & $-$
&\cellcolor{gray!50}$2.35$ &\cellcolor{gray!50}$3.59$
& $-$ & $-$ & $-$ & $-$ & $-$ & $-$ & $-$ 
\\\hline\hline
$~p\to \hat\nu_x \pi^{+} $ & $0.39$~\cite{Super-Kamiokande:2013rwg}
& $-$ & $-$ & $-$ & $-$ & $-$ & $-$ 
& $0.37$ & $0.36$ 
& $-$ & $-$ & $-$ & $-$ & $-$
\\
$~n\to \hat\nu_x \pi^0 $ & $1.4$\,\cite{Super-Kamiokande:2025lxa}
& $-$ & $-$ & $-$ & $-$ & $-$ & $-$
&\cellcolor{gray!50}$0.27$ &\cellcolor{gray!50}$0.27$ 
& $-$ & $-$ & $-$ & $-$ & $-$
\\
$~n\to  \hat\nu_x\eta $ & $0.158$~\cite{McGrew:1999nd}
& $-$ & $-$ & $-$ & $-$  & $-$ & $-$
& $11.76$ & $1.25$ 
& $-$ & $-$ & $-$ & $-$ & $-$
\\\hline
$~p\to \hat\nu_x K^{+}~$ & $6.61$\,\cite{Mine:2016mxy}
& $-$ & $-$ & $-$ & $-$ & $-$ & $-$ & $-$ & $-$
&\cellcolor{gray!50}$0.32$ &\cellcolor{gray!50}$0.66$ &\cellcolor{gray!50}$0.65$ &\cellcolor{gray!50}$0.17$ &\cellcolor{gray!50}$0.16$
\\
$~n\to \hat\nu_x K^0 $ & $0.78$\,\cite{Super-Kamiokande:2025ibz}
& $-$ & $-$ & $-$ & $-$ & $-$ & $-$ & $-$ & $-$ 
& $1.92$ & $0.94$ & $0.65$ & $0.49$ & $0.48$ 
\\\hline
\end{tabular}}
\caption{
Bounds on the WC associated with each dim-6 BNV operator in the LEFT, derived from the two-body nucleon decays involving a pseudoscalar meson. }
\label{tab:BoundonWC}
\end{table}

\begin{figure}[t]
\centering
\includegraphics[width=0.49\textwidth]{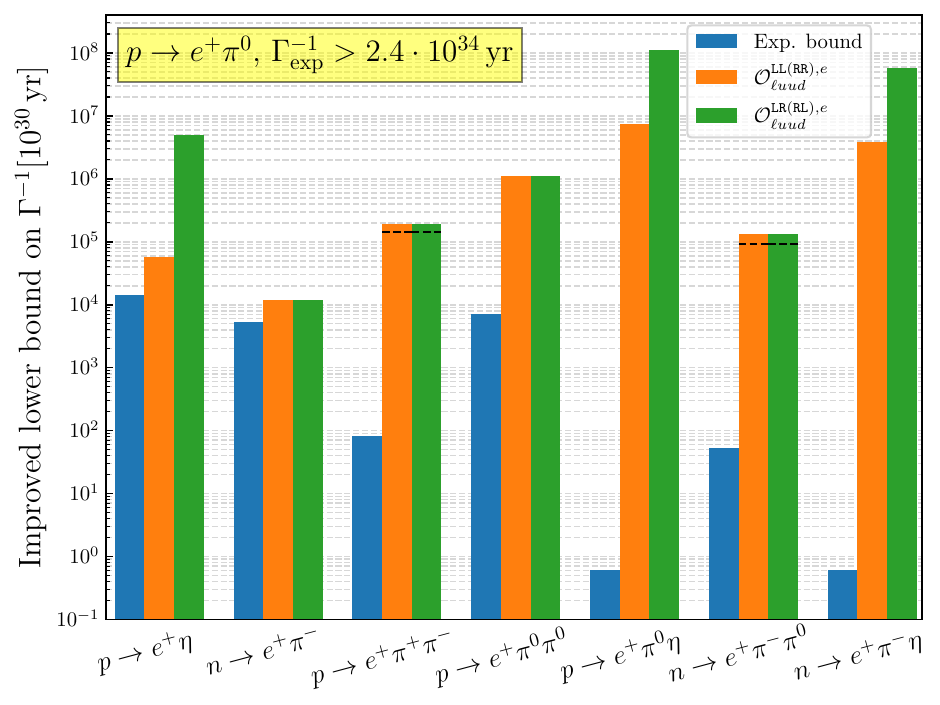}
\includegraphics[width=0.49\textwidth]{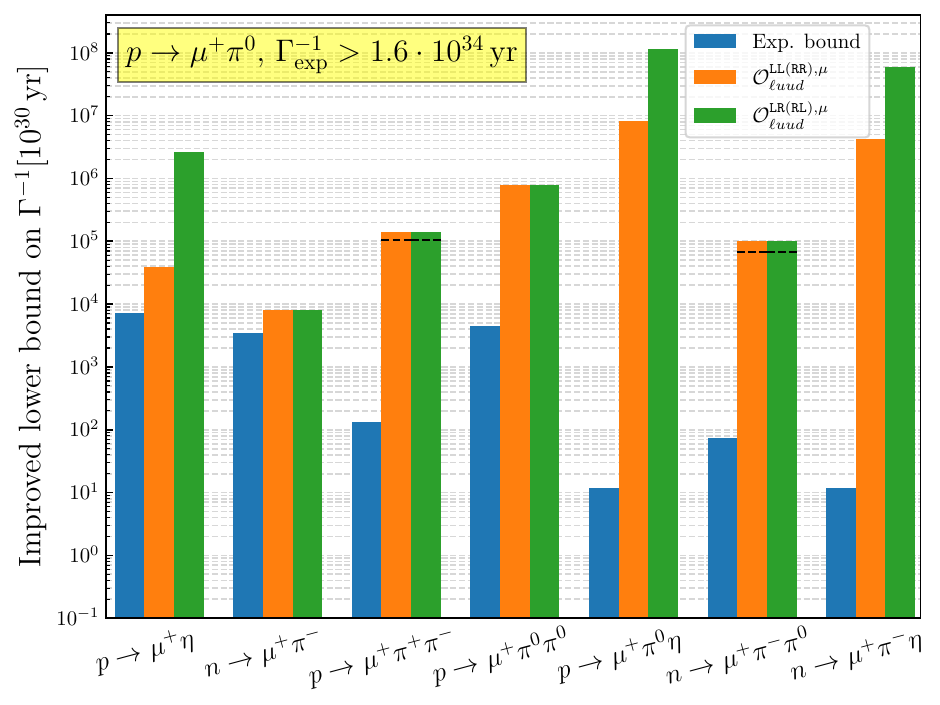}
\\
\includegraphics[width=0.49\textwidth]{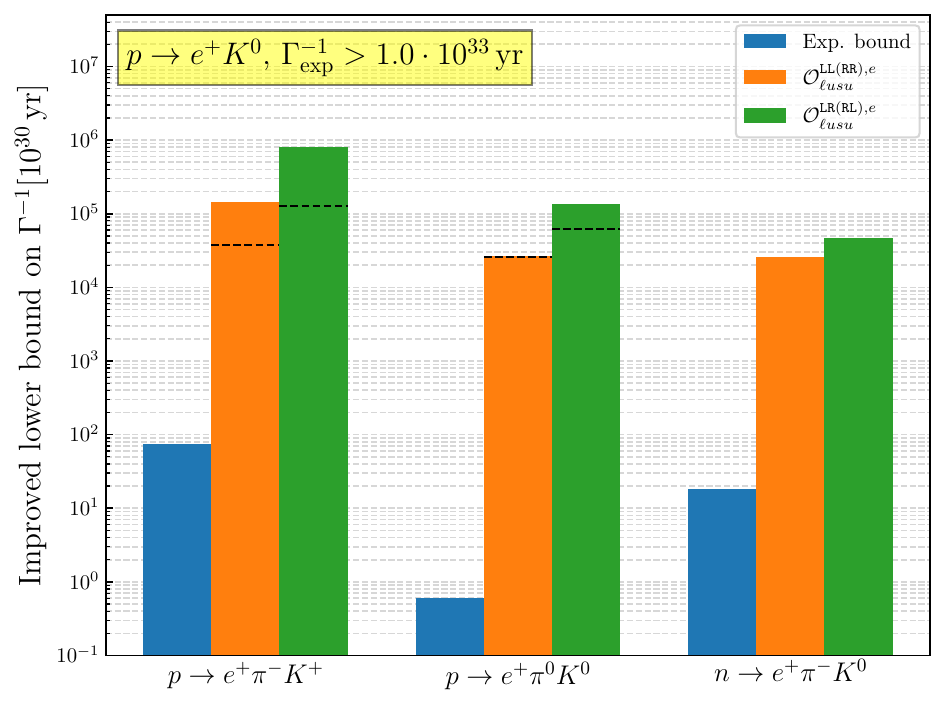}
\includegraphics[width=0.49\textwidth]{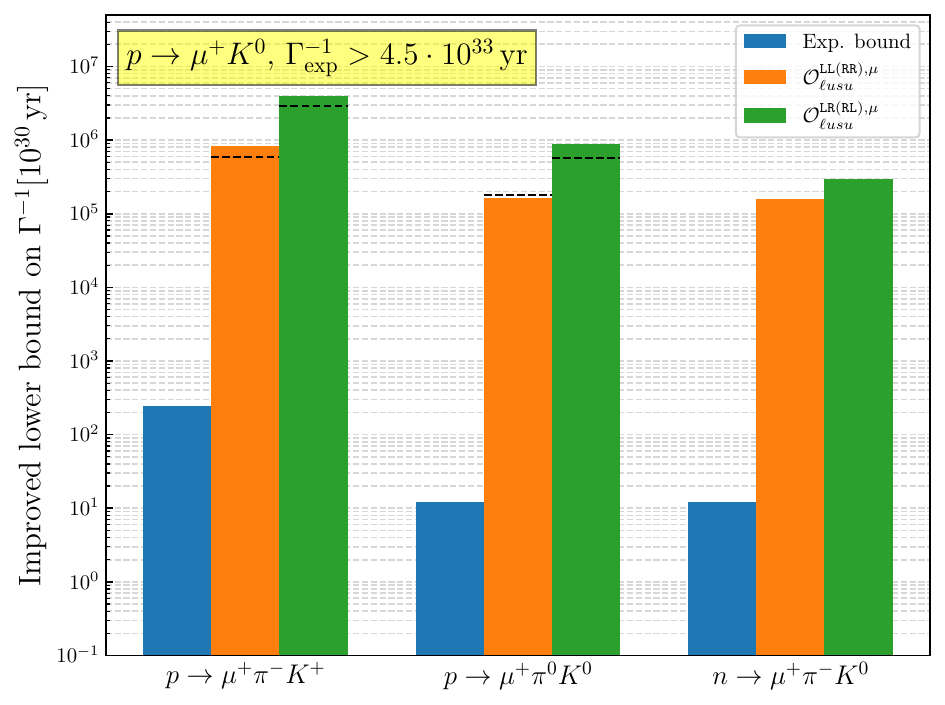}
\\
\includegraphics[width=0.49\textwidth]{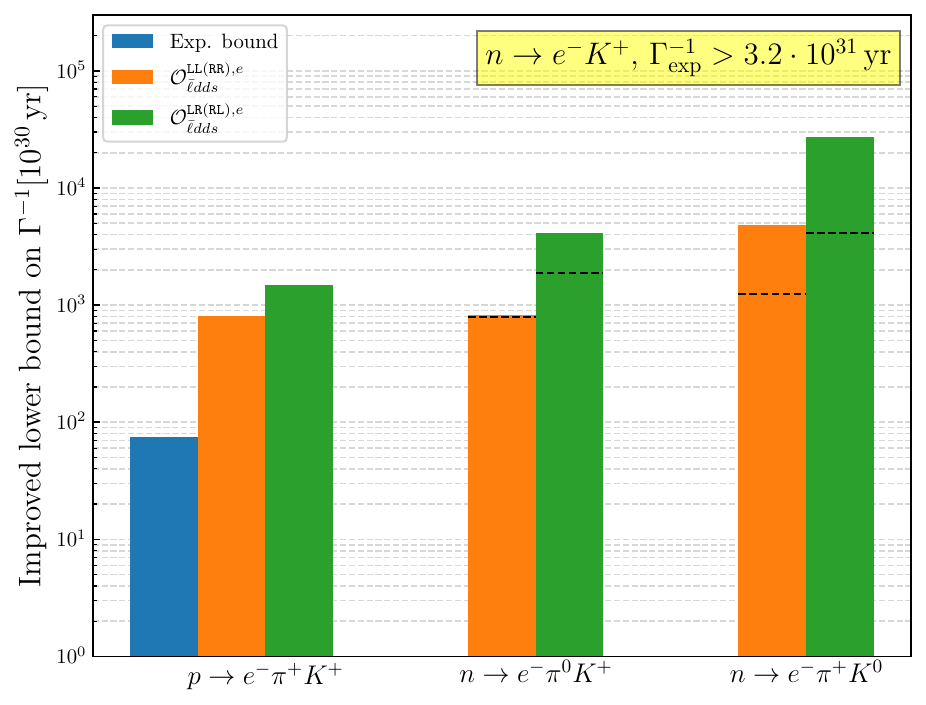}
\includegraphics[width=0.49\textwidth]{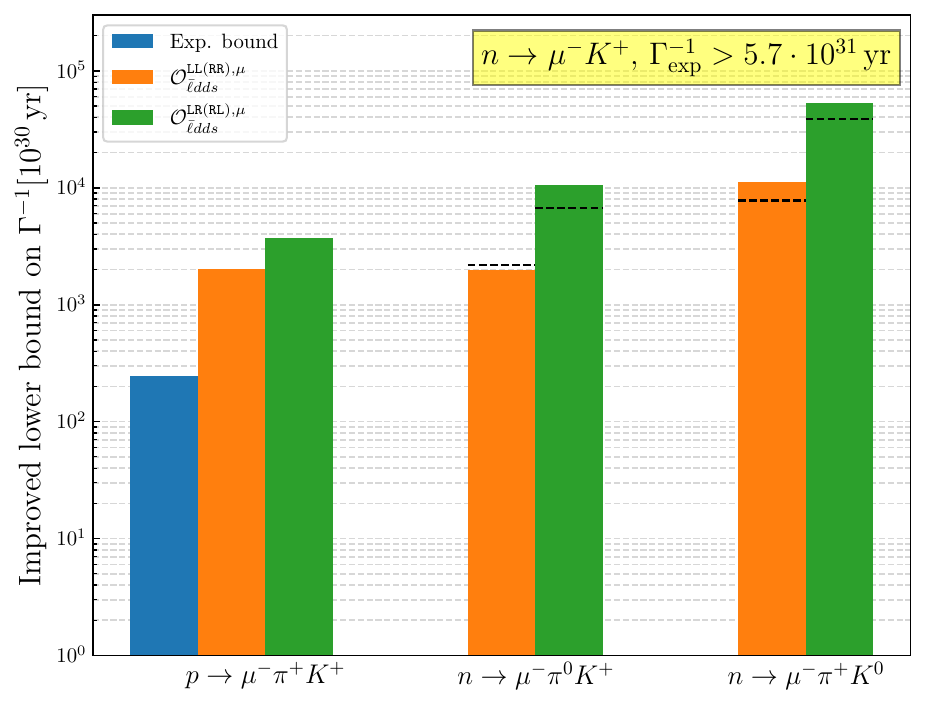}
\caption{The derived improved single-operator-dominance bounds on the three-body nucleon decay modes involving a charged lepton $e$ or $\mu$. The dashed black lines indicate the bounds obtained after including vector-meson-mediated contributions.}
\label{fig:newbound1}    
\end{figure}

\begin{figure}[t]
\centering
\includegraphics[width=0.49\textwidth]{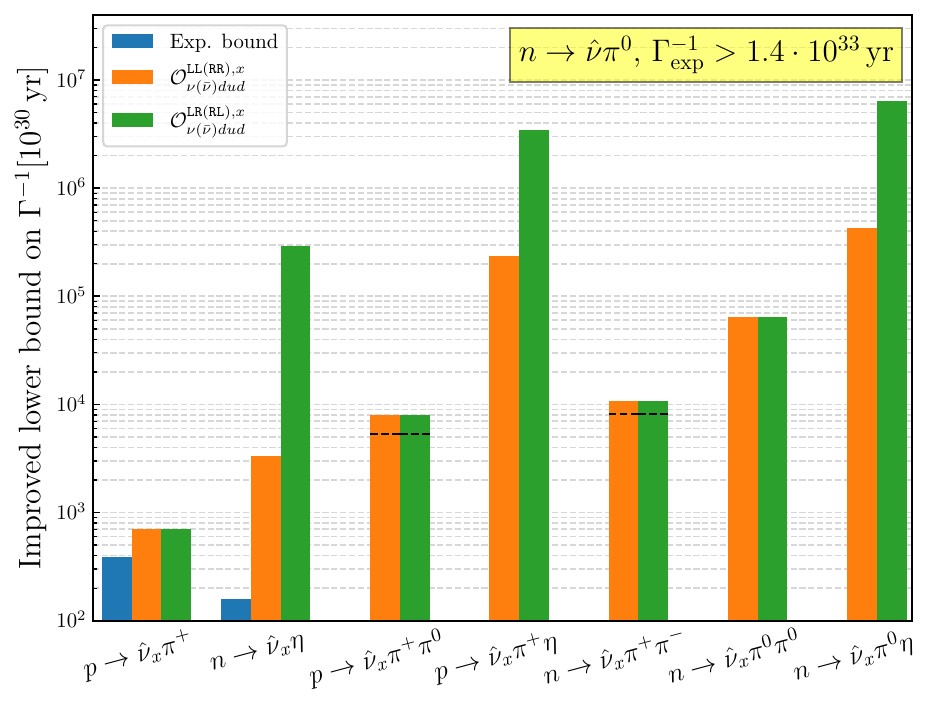}
\includegraphics[width=0.49\textwidth]{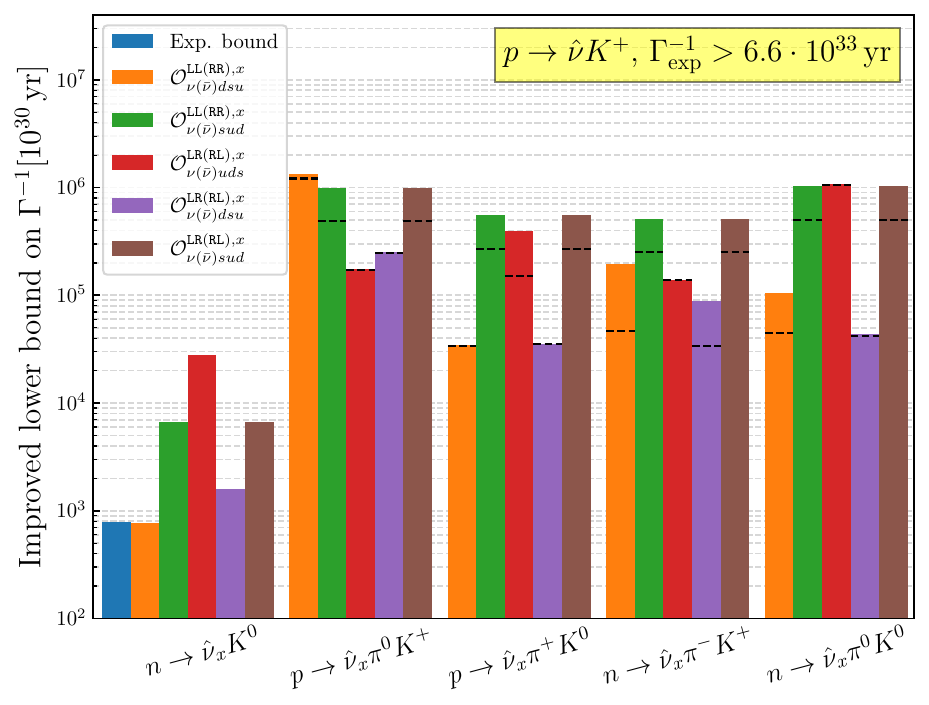}
\caption{Same as \cref{fig:newbound1}, but for the decay modes involving a neutrino or an antineutrino.}
\label{fig:newbound2}    
\end{figure}

By incorporating the most stringent bound on each WC (shown in gray in \cref{tab:BoundonWC}) into the decay width expressions for the three-body modes given in appendix \ref{app:DeW_three_body}, we establish improved lower limits on their partial lifetimes. 
The resulting limits are summarized in \cref{fig:newbound1} for modes involving a charged lepton and \cref{fig:newbound2} for those involving a neutrino or an antineutrino, respectively.
In these figures, the dashed black lines denote bounds evaluated using the decay widths from appendix \ref{app:vectorM}, which include the vector-meson-mediated contributions.   
In each panel, we show in inset the two-body mode along with its experimental bound used to derive these improved limits, while the existing experimental bounds on the three-body modes are shown as blue bars for comparison.
Additionally, this single-operator analysis also leads to improved bounds on seven two-body modes $p\to (e^+,\mu^+)\eta$, $n\to(e^+,\mu^+) \pi^-$, $p\to \hat\nu\pi^+$, and $n\to \hat\nu(\eta,K^0)$, which we include in the plots for completeness.  

The top two panels in \cref{fig:newbound1} show the improved bounds on the ten three-body modes involving $\pi\pi$ or $\pi\eta$ and the four two-body modes $p\to(e^+,\mu^+)\eta$ and $n\to (e^+,\mu^+)\pi^-$ that are induced by the $\ell uud$-configuration operators. As observed, the new limits on $p\to (e^+,\mu^+)\pi^+\pi^-$ and $n\to (e^+,\mu^+)\pi^-\pi^0$ are approximately three orders of magnitude stronger than the current experimental bounds, while those on $p\to (e^+,\mu^+)\pi^0\eta$ and $n\to (e^+,\mu^+)\pi^-\eta$ improve by more than six orders of magnitude relative to the inclusive limits.  
For the ${\tt N}\to \ell^+ \pi K$ mediated by the $\ell uus$-type operators (middle panels), 
the improvement over the current experimental bounds also exceeds three orders of magnitude, 
reaching $\calO(10^{34\text{---}36}\,\rm yr)$.
For the $\Delta(B+L)=0$ processes 
${\tt N}\to \ell^- \pi K$, mediated by operators of the $\bar\ell dds$ type (bottom row), 
we establish lower bounds on the four modes $n \to (e^-,\mu^-) \pi^0 K^+$ and $n \to (e^-,\mu^-)  \pi^+ K^0$ for the first time, with $\Gamma^{-1} \gtrsim 10^{33\text{---}34}$ yr.

Figure\,\ref{fig:newbound2} presents the bounds on nucleon decay modes involving a neutrino or an antineutrino. 
The left panel shows the improved bounds on five three-body modes involving $\pi\pi$ or $\pi\eta$ and two two-body modes ($p\to\hat\nu \pi^+$ and $n\to\hat\nu \eta$) induced by the $\nu(\bar \nu) dud$-type operators. 
The lower limits on $\Gamma^{-1}(p\to\hat\nu\pi^+\pi^0)$ and $\Gamma^{-1}(n\to\hat\nu\pi^+\pi^-)$ are around $\calO(10^{34}\,\rm yr)$, while 
for both $p\to\hat\nu \pi^+\eta$ and $n\to\hat\nu \pi^0\eta$ the bounds increase to more than $\calO(10^{35}\,\rm yr)$ and $\calO(10^{36}\,\rm yr)$ for the two different groups of operators, respectively. 
Notably, the bound on the two-body mode $n\to \hat\nu \eta$ improves by more than an order of magnitude compared to its current experimental limit.
For processes mediated by the $\nu(\bar \nu) uds$-type operators (right panel), the bounds on the three-body modes involving $\pi K$ are generally around $\calO(10^{34\text{---}36}\,\rm yr)$. It should be emphasized that our results provide the first bounds on these three-body modes involving a neutrino or an antineutrino. 
Finally, as indicated by the black dashed lines in \cref{fig:newbound1,fig:newbound2}, vector-meson contributions can weaken the bounds on modes with $(e^+,\hat\nu)\pi K$ final states by up to a factor of 5, while having a minor impact on the remaining modes.
This comparison suggests that the bounds on kaon-related modes may be subject to an order-one theoretical uncertainties from vector-meson contributions.

\subsection{Global analysis}

We now turn to the global analysis approach by simultaneously considering all WCs that share the same field configuration but differ only in chiralities. To confine the multidimensional WC parameter space into a bounded region and derive improved bounds on the three-body modes, several two-body modes must be used. 
First of all, this is not possible for the $\ell uus$- and $\bar\ell dds$-type operators, since only one two-body nucleon decay mode is available for each type. 
Furthermore, for the $\hat\nu uds$-type operators, the two two-body modes are insufficient either to constrain the multidimensional WC space into a bounded region, owing to more operators involved. 
This implies that, in the multi-operator analysis, the corresponding three-body decay modes involving $\pi K$ mesons are not bounded by the two-body data at all. For example, for the $\ell uus$-type operators, 
the decay width of $p\to e^+(\mu^+) K^0$ vanishes along the directions $C^{\tL\tR(\tR\tL)}_{e(\mu) usu}/C^{\tL\tL(\tR\tR)}_{e(\mu) usu} \approx -0.643 (-0.654)$, whereas the corresponding three-body decay widths remain nonzero.
Similar flat directions exist in the multidimensional WC spaces of the $\bar\ell dds$- and $\hat\nu uds$-type operators, allowing the three-body modes to be enhanced while suppressing the corresponding two-body modes.
Consequently, we are left with the $\ell uud$- and $\nu(\bar\nu)udd$-type operators for a global analysis. Each of these types corresponds to three two-body and five three-body modes, as shown in \cref{tab:opeandprocess}.

For each of the three cases ($euud$, $\mu uud$, and $\hat\nu udd$), we find that two two-body decays, one involving a $\pi$ and the other an $\eta$, are sufficient to constrain the multidimensional parameter space to a closed, bounded region. 
This is possible because the $\pi$- and $\eta$-related modes have significantly different numerical prefactors for each quadratic WC term. 
In fact, the approximate isospin symmetry implies $\Gamma(n\to \ell^+\pi^-)\approx2\Gamma(p\to \ell^+\pi^0)$ 
and $\Gamma(p\to \hat\nu\pi^+)\approx 2\Gamma(n\to \hat\nu\pi^0)$.
For this reason, we choose as input channels the six experimentally best-constrained $\pi$- and $\eta$-related modes: $p \to e^+ (\pi^0,\eta)$, $p \to \mu^+ (\pi^0,\eta)$, and $n \to \hat\nu (\pi^0,\eta)$ (see \cref{tab:data}). 

To establish the constraints,  
we parameterize each of the four $\ell uud$-type WCs by a dimensionless magnitude and a phase,
\begin{align}
C_{\ell uud}^{\tL\tL,x} 
&\equiv | \tilde{C}_{\ell uud}^{\tL\tL,x} | e^{i\alpha_0} /(10^{15}\,{\rm GeV})^2,  
&
C_{\ell uud}^{\tR\tR,x} 
&\equiv | \tilde{C}_{\ell uud}^{\tR\tR,x} | e^{i\alpha_1} /(10^{15}\,{\rm GeV})^2, 
\nonumber\\
C_{\ell uud}^{\tL\tR,x} 
&\equiv | \tilde{C}_{\ell uud}^{\tL\tR,x} | e^{i\alpha_2}/(10^{15}\,{\rm GeV})^2,
&
C_{\ell uud}^{\tR\tL,x} 
&\equiv | \tilde{C}_{\ell uud}^{\tR\tL,x} | e^{i\alpha_3}/(10^{15}\,{\rm GeV})^2.
\end{align}
Given that the decay widths depend only on phase differences, there are three independent phase parameters, chosen as follows: 
$\theta_1\equiv\alpha_1-\alpha_0,\,  \theta_2\equiv\alpha_2-\alpha_0,\, 
\theta_3\equiv\alpha_3-\alpha_0$.
Together with the four positive dimensionless magnitudes $\{| \tilde{C}_{\ell uud}^{\tL\tL,x} |,\, | \tilde{C}_{\ell uud}^{\tR\tR,x} |,\, | \tilde{C}_{\ell uud}^{\tL\tR,x} |,\, | \tilde{C}_{\ell uud}^{\tR\tL,x} |\}$, there are 
seven independent real parameters to describe eight two- and three-body processes involving a charged lepton $\ell$.
Similarly, for the two WCs involving a neutrino field, we parameterize them as
\begin{align}
C^{\tL\tL,x}_{ \nu dud}\equiv
|\tilde C^{\tL\tL,x}_{ \nu dud}| e^{i\beta_0}/(10^{15}\,{\rm GeV})^2,
\quad
C^{\tL\tR,x}_{ \nu dud}\equiv|\tilde C^{\tL\tR,x}_{ \nu dud}| e^{i\beta_1}/(10^{15}\,{\rm GeV})^2, 
\end{align}
with the phase difference $\theta\equiv\beta_1-\beta_0$. In this case, only three real parameters ( $|\tilde C^{\tL\tL,x}_{ \nu dud}|$, $|\tilde C^{\tL\tR,x}_{ \nu dud}|$, $\theta$) are required to describe eight antineutrino-related processes. Due to the invisible nature of neutrinos and antineutrinos, the situation for neutrino-related modes is exactly analogous to that for antineutrinos. Therefore, we present only the results for the antineutrino case.

\begin{figure}[t]
\centering
\includegraphics[width=1.0\textwidth]{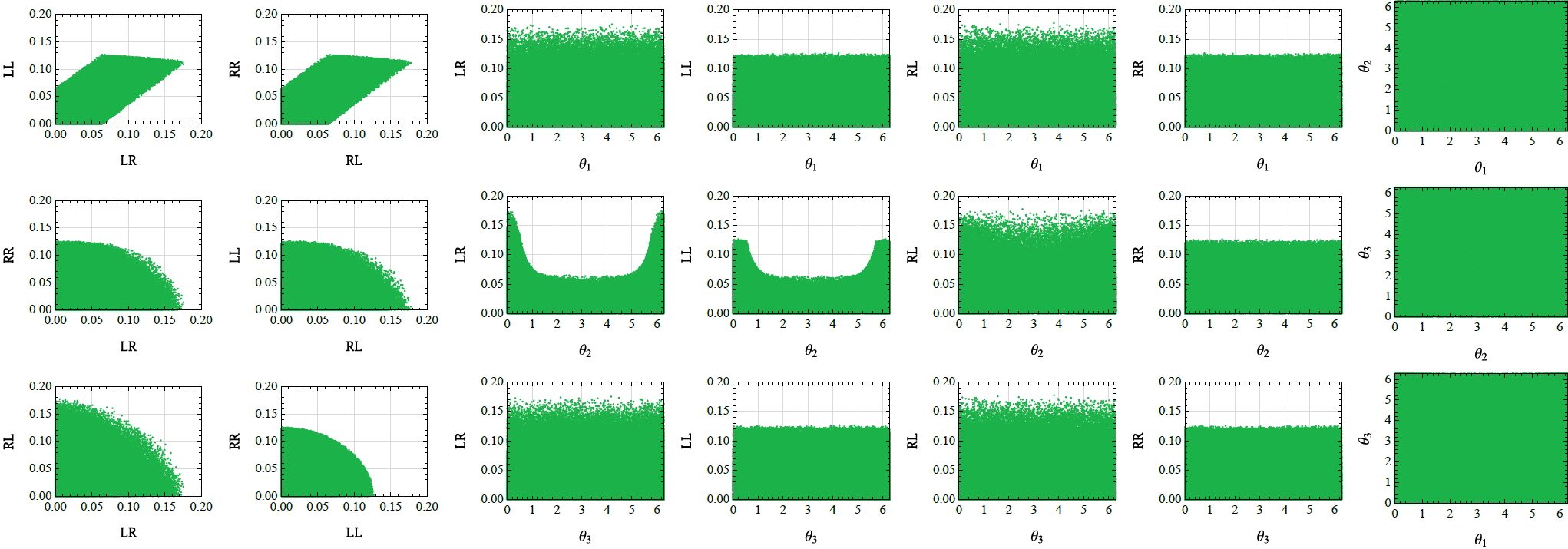}
\caption{
Various 2D plots are shown for the surviving points that satisfy constraints from the $p\to e^+ \pi^0$ and $p\to  e^+\eta$ channels, with each plot spanned by a pair of independent parameters. For simplicity, the magnitudes $| \tilde{C}_{\ell uud}^{\tL\tL,e}|,\, |\tilde{C}_{\ell uud}^{\tR\tR,e}|,\, |\tilde{C}_{\ell uud}^{\tL\tR,e}|,\, |\tilde{C}_{\ell uud}^{\tR\tL,e}|$ are represented by $\tL\tL, \tR\tR, \tL\tR, \tR\tL$, respectively. }
\label{fig:2Deuud}
\end{figure}

\begin{figure}[t]
\includegraphics[width=\textwidth]{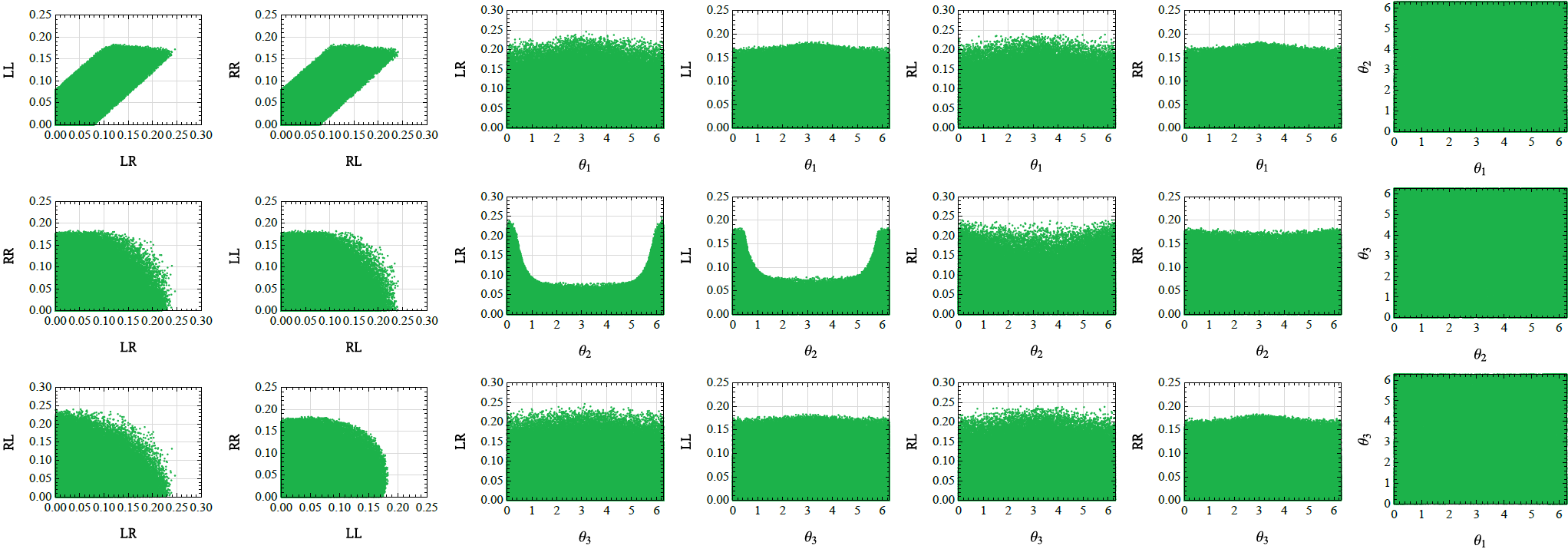}
\caption{Same as \cref{fig:2Deuud} but for WCs associated with the $\mu uud$-type operators.}
\label{fig:2Dmuuud}
\end{figure}

\begin{figure}[t]
\centering
\includegraphics[width=\textwidth]{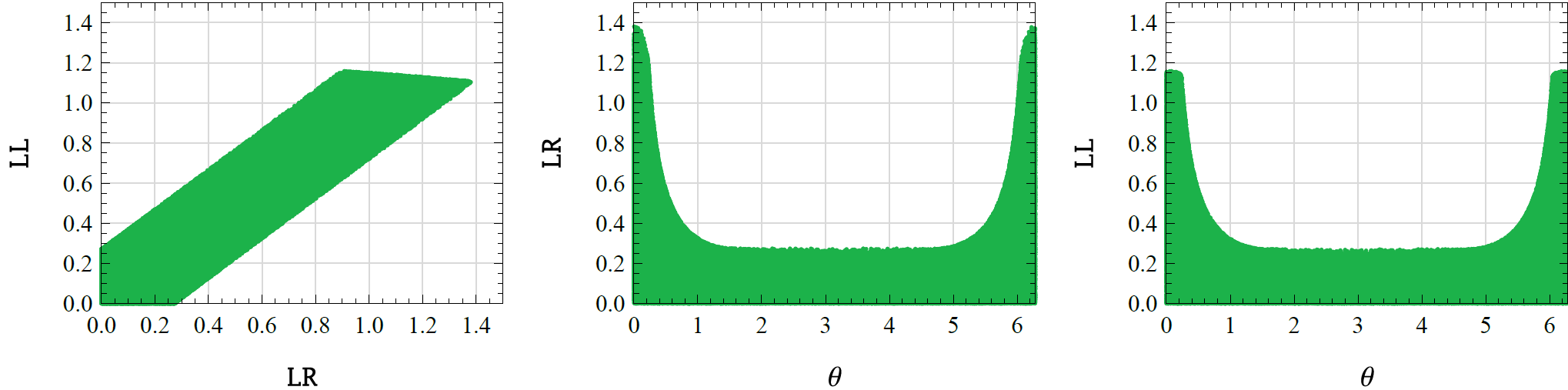}
\caption{Same as \cref{fig:2Deuud} but for the two WCs associated with operators $\calO_{\nu dud}^{\tL\tR(\tL\tL),x}$.}
\label{fig:2Dvudd}
\end{figure}

For each of the three groups of parameters (seven parameters for the $e^+$ modes, seven for the $\mu^+$ modes, and three for the $\bar\nu$ modes.), 
we randomly generate $\calO(10^7)$ points by 
requiring the phases $\theta,\,\theta_i \in [0,\,2\pi]$ and the WC magnitudes $|\tilde C_i^j|\in[0,\,2]$.
We then retain only those points that simultaneously satisfy the partial lifetime limits on the two two-body decays. 
The surviving points in various 2D planes spanned by pairs of independent parameters are shown in \cref{fig:2Deuud,fig:2Dmuuud,fig:2Dvudd} for the WCs associated with the $euud$-, $\mu uud$-, and $\nu udd$-type operators, respectively. 
In all three cases, the WC magnitudes are clearly constrained to lie below the initially chosen upper values, confirming that the two two-body modes ${\tt N}\to l (\pi,\,\eta)$ related to each field configuration case indeed confine the corresponding multidimensional WC space to a closed, bounded region. 
To test the adequacy of the adopted scan range, we extended the scan of the WC magnitudes to $|\tilde C_i^j|\in[0,\,10]$ and found no surviving points with $|\tilde C_i^j|>2$ satisfying the two-body constraints.   
We also varied the number of sampled points from $10^6$ to $10^8$ and repeated the analysis with different random seeds to ensure the numerical stability of our results.

Note that we impose the experimentally published 
90\,\% CL limits on two-body nucleon decays as hard constraints to determine the allowed WC parameter space.
A rigorous likelihood-based analysis would require the underlying experimental likelihoods, including the relevant statistical and systematic uncertainties, correlations, and nuisance parameters, which cannot generally be uniquely reconstructed from the published lower limits alone.
Such a statistical treatment is beyond the scope of the present work.
Throughout the paper, the term ``global analysis" refers to the simultaneous consideration of multiple WCs and experimental constraints, without assuming single-operator dominance, rather than to a statistically defined global fit.

\begin{figure}[t]
\centering
\includegraphics[width=0.9\textwidth]{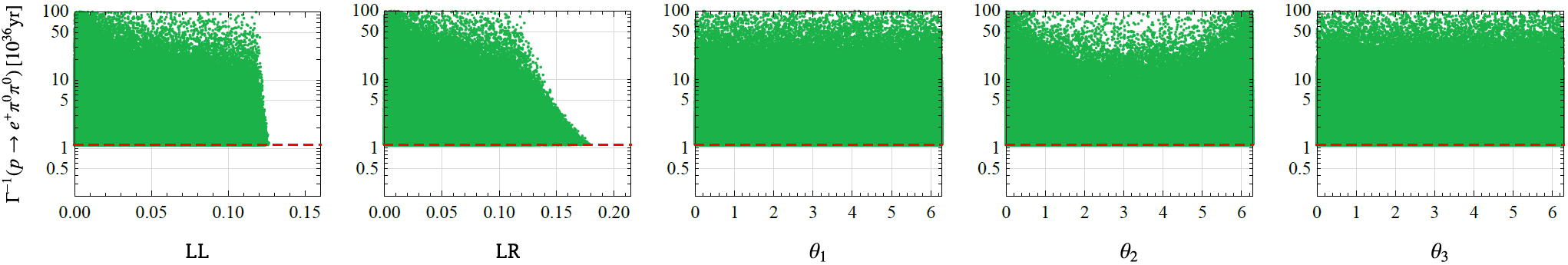}
\includegraphics[width=0.9\textwidth]{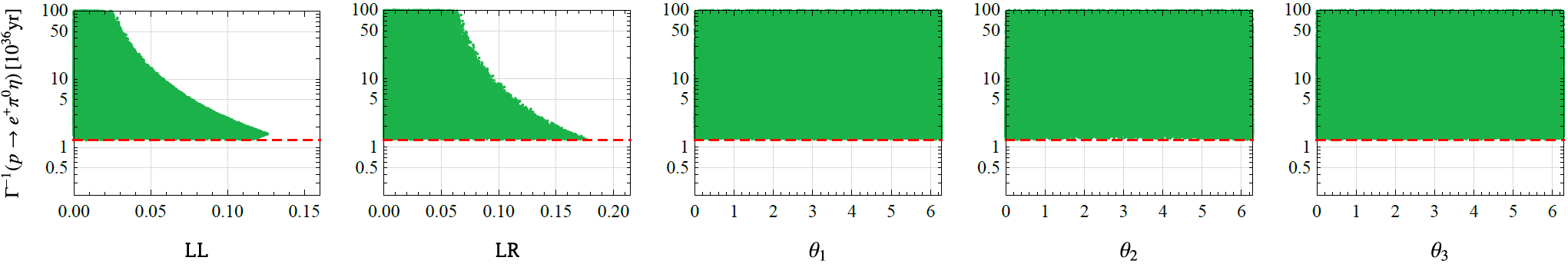}
\caption{The derived partial lifetimes as functions of WC magnitudes and phases for the processes $p \to e^+ \pi^0 \pi^0$ (upper) and $p \to e^+ \pi^0 \eta$ (lower) in the allowed multidimensional WC region. The final conservative bound is marked by the red dashed line.}
\label{fig:p2epi0pi0oreta}
\end{figure}

\begin{figure}[h]
\centering
\includegraphics[width=0.9\textwidth]{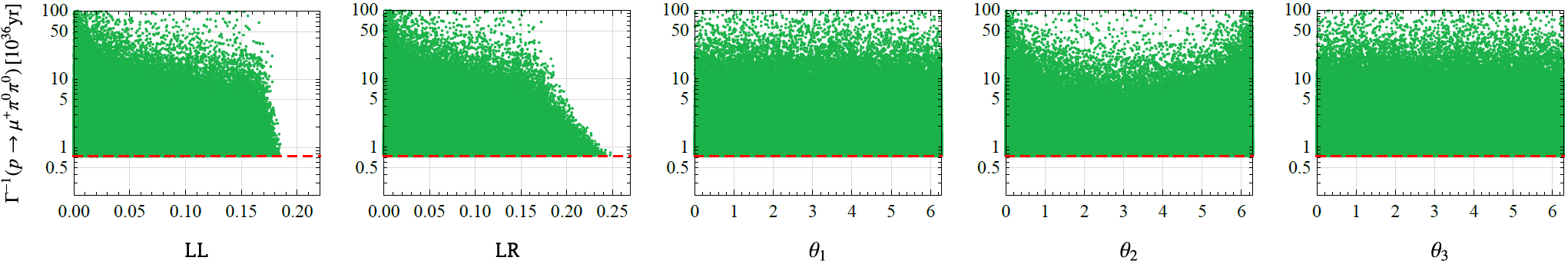}
\includegraphics[width=0.9\textwidth]{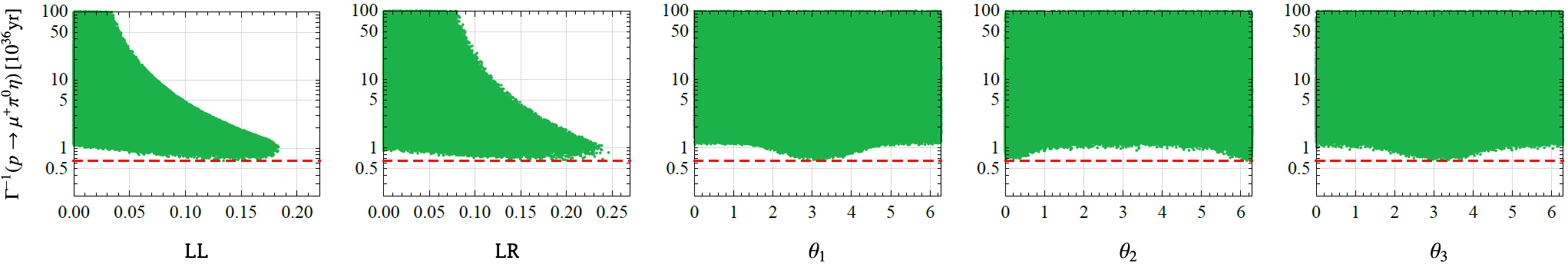}
\caption{Similar to \cref{fig:p2epi0pi0oreta}, but for $p \to \mu^+ \pi^0 \pi^0$ (upper) and $p \to \mu^+ \pi^0 \eta$ (lower).}
\label{fig:p2mupi0pi0oreta}
\end{figure}

\begin{figure}[h]
\centering
\includegraphics[width=0.9\textwidth]{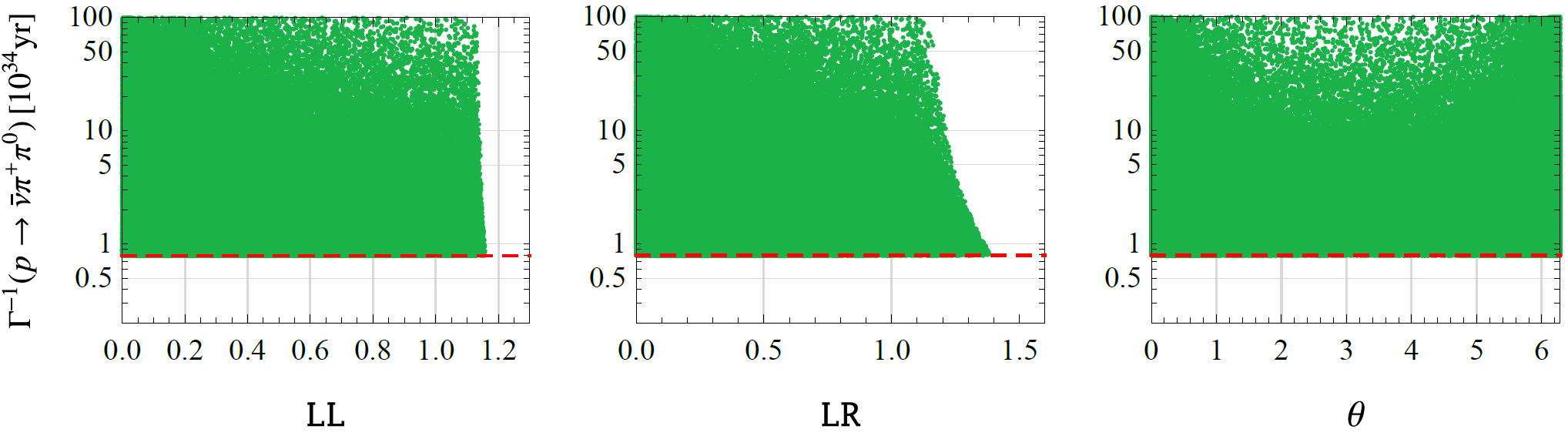}
\includegraphics[width=0.9\textwidth]{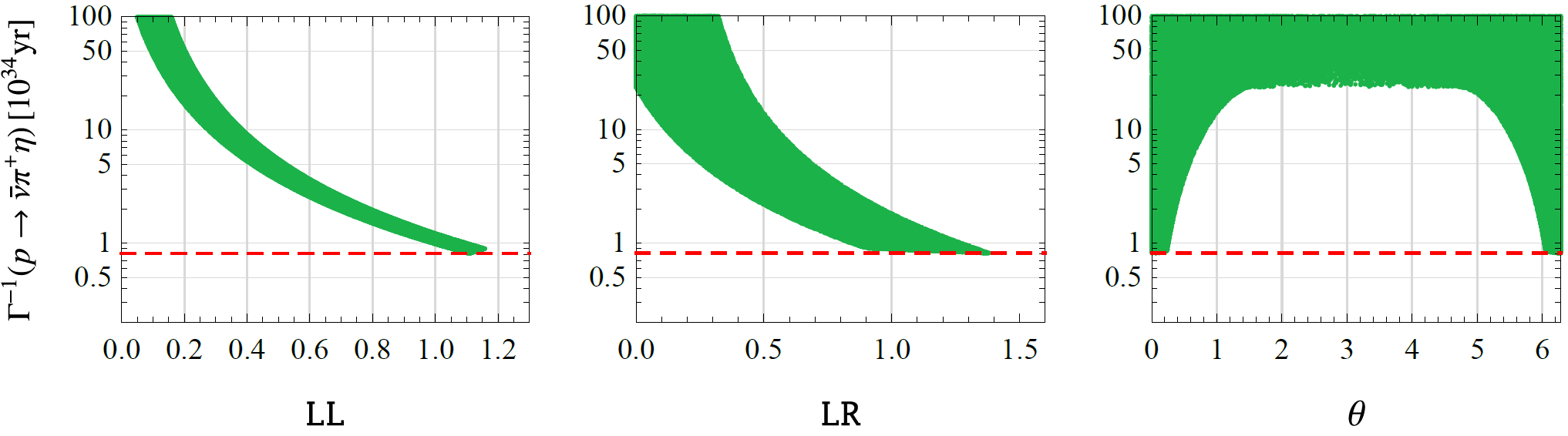}
\caption{Similar to \cref{fig:p2epi0pi0oreta}, but for $p \to \bar\nu \pi^+ \pi^0$ (upper) and $p \to \bar\nu \pi^+ \eta$ (lower).}
\label{fig:p2nupipi0oreta}
\end{figure}

Using the decay width expressions for the three-body nucleon decays into $\pi\pi$ and $\pi\eta$ provided in appendix \ref{app:DeW_three_body}, we calculate the partial lifetimes for these three-body modes by varying the WCs in the allowed regions.
For each of the three lepton cases ($l=e^+,\mu^+, \bar\nu$), we select two  representative three-body modes, $p\to l\pi(\pi^0,\,\eta)$, to illustrate the calculated partial lifetimes as functions of the parameters in the allowed WC region. 
The results are shown in \cref{fig:p2epi0pi0oreta,fig:p2mupi0pi0oreta,fig:p2nupipi0oreta} for $p\to e^+\pi^0(\pi^0,\,\eta)$, $p\to \mu^+\pi^0(\pi^0,\,\eta)$, and $p\to \bar\nu\pi^+(\pi^0,\,\eta)$, respectively.
For the $e^+$ and $\mu^+$ cases, the partial lifetimes against $C_{\ell uud}^{\tR\tL(\tR\tR)}$ are similar to those against $C_{\ell uud}^{\tL\tR(\tL\tL)}$ and are therefore omitted, as the decay widths are invariant under interchanging $\tL\leftrightarrow\tR$ in the WCs. 

Figures\,\ref{fig:p2epi0pi0oreta} to \ref{fig:p2nupipi0oreta} clearly show that each decay mode has a minimal partial lifetime value, indicated by the red dashed line. 
As a conservative estimate, we adopt this minimum as the final limit. 
We checked the stability of the extracted minimum lifetimes against variations in the sampled points and random seeds, finding that the minimum values remain unchanged within the numerical precision of our analysis.
The complete lifetime limits on all these 15 three-body nucleon decay modes involving $\pi\pi$ or $\pi\eta$ from this analysis are reported in \cite{Fan:2026fqo}. 
These results show that the bounds obtained from this global analysis are around $\calO(10^{34\text{---}36})$ yr, generally weaker than those from the single-operator-dominance analysis presented in the previous subsection. 
For the 9 modes into $\pi\pi$, ${\tt N} \to l\pi\pi$, the two approaches yield nearly identical bounds. This arises from the fact that every BNV vertex involving only nucleons and pions depends on the single WC combination $c_1 C_x^{\tL\tR} + c_2 C_x^{\tL\tL}$ ($x=\ell uud, \nu dud$) or its chirality partner, see \cref{eq:VertexBl,eq:VertexBlM,eq:VertexNBlM}. Neglecting the lepton-mass-suppressed interference terms between the two chirality combinations in each mode, the decay width expressions in appendices \ref{app:DeW_three_body} and \ref{app:DeW_two_body} lead to the following WC-independent ratios: 
\begin{subequations}
\begin{align}
\frac{\Gamma(p\to e^+\pi^+\pi^-)}{\Gamma(p\to e^+\pi^0)} &\approx 0.13,&
\frac{\Gamma(p\to e^+\pi^0\pi^0)}{\Gamma(p\to e^+\pi^0)} &\approx 0.021,& 
\frac{\Gamma(n\to e^+\pi^-\pi^0)}{\Gamma(p\to e^+\pi^0)} & \approx 0.18,
\\
\frac{\Gamma(p\to \mu^+\pi^+\pi^-)}{\Gamma(p\to \mu^+\pi^0)} &\approx 0.12,&
\frac{\Gamma(p\to \mu^+\pi^0\pi^0)}{\Gamma(p\to \mu^+\pi^0)} &\approx 0.020,& 
\frac{\Gamma(n\to \mu^+\pi^-\pi^0)}{\Gamma(p\to \mu^+\pi^0)} & \approx 0.16,
\\ 
\frac{\Gamma(n\to \hat\nu\pi^+\pi^-)}{\Gamma(n\to \hat\nu\pi^0)} &\approx 0.13,& 
\frac{\Gamma(n\to \hat\nu\pi^0\pi^0)}{\Gamma(n\to \hat\nu\pi^0)} & \approx 0.022,&
\frac{\Gamma(p\to \hat\nu\pi^+\pi^0)}{\Gamma(n\to \hat\nu\pi^0)} &\approx 0.18.
\end{align}
\end{subequations} 
These ratios indicate that, in the SMEFT or UV scenarios where the dim-6 LEFT operators are dominantly generated, the three-body modes ${\tt N}\to l\pi\pi$ are always subleading to their two-body counterparts ${\tt N}\to l\pi$.

Before closing this section, we emphasize that our global analysis is formulated in the LEFT and does not reply on assumptions about the SMEFT or UV origins of the relevant interactions. In a top-down SMEFT framework, RG evolution can generate flavor mixing among SMEFT WCs and electroweak matching could induce correlations among LEFT WCs. 
As long as the dim-6 LEFT operators considered here capture the dominant contributions from the underlying SMEFT or UV theory, our results remain applicable irrespective of the specific SMEFT RG evolution and flavor mixing above the electroweak scale. If SMEFT RG evolution and SMEFT-to-LEFT matching induce correlations among the dim-6 LEFT WCs, these correlations can restrict the allowed parameter space to specific subregions of the parameter space explored in our analysis. Such restrictions could lead to stronger bounds on three-body nucleon decays, although the extent of any improvement would depend on the specific correlations. 
For instance, if we consider the four dim-6 SMEFT BNV operators and assume that all their independent flavor combinations have independent WCs, our bounds remain valid.
This is because these SMEFT operators match onto the four (two) dim-6 LEFT operators with the $\ell uud$ ($\nu udd$) field configuration, without introducing additional correlations among the WCs within the same field configuration ~\cite{Jenkins:2017jig}.
Meanwhile, the correlations between charged-lepton and neutrino modes would further strengthen the constraints on $\bar\nu_{e,\mu}$-related modes by exploiting the more stringent limits from their charged-lepton counterparts. 
On the other hand, if contributions from dim-7 or higher-dimensional LEFT operators become comparable to or larger than those from the dim-6 operators,  our bounds derived within the dim-6 LEFT framework would no longer be directly applicable without including these additional contributions. 

\section{Summary}
\label{sec:summary}

In this paper, we have conducted a systematic investigation on three-body nucleon decays into a single lepton and two pseudoscalar mesons based on the leading-order dim-6 BNV LEFT interactions. 
Using chiral perturbation theory, we derived the general decay width expressions for 31 three-body modes.  
Since these three-body modes are mediated by the same set of operators as the conventional, experimentally well-constrained two-body nucleon decays involving a single pseudoscalar meson, we use their correlations to derive improved bounds on the former from experimental limits on the latter. 
Two approaches are adopted in the analysis. 
The single-operator-dominance analysis allows us to set stringent operator-dependent bounds on all 31 three-body modes involving either $\pi\pi$, $\pi\eta$, or $\pi K$ in the final state. 
By contrast, the assumption-free global analysis treats all relevant operators on equal footing and constrains 15 three-body modes involving either $\pi\pi$ or $\pi\eta$. 
In both cases, the newly established indirect EFT-consistency bounds are stronger than existing experimental limits by several orders of magnitude.
In addition, our analysis also yields improved bounds on several two-body modes.  
In light of ongoing and upcoming large-fiducial-mass neutrino experiments,
the formalism developed here and the bounds derived for these three-body modes 
provide useful tools and benchmarks for further experimental and theoretical studies.

We emphasize that, within the dim-6 LEFT framework considered here, the three-body decay modes are generally subleading to their corresponding two-body channels. Therefore, in generic scenarios, they primarily serve as complementary diagnostic probes following a potential discovery in two-body nucleon decays. Nevertheless, as demonstrated by the operators involving strange quarks, specific regions of the WC parameter space can exhibit cancellations or suppressions in two-body decays that are not present in the corresponding three-body modes. 
Consequently, dedicated searches for the kaon-related three-body decays can provide complementary probes of BNV interactions.

Although our analysis is formulated in the model-independent LEFT framework, it can also provide useful guidance for investigating UV completions of baryon number violation.
Different mediator structures, such as heavy gauge bosons or scalar fields with diquark and leptoquark couplings, can generate different combinations of BNV operators and flavor structures when the heavy degrees of freedom are integrated out. These differences lead to distinct patterns of LEFT WCs and, consequently, different correlations among nucleon-decay channels. In particular, while two-body nucleon decays provide stringent constraints on the relevant WCs, the complementary three-body modes studied here can probe additional combinations of these coefficients and their interference effects. Comparing the observed rates of two- and three-body modes could therefore help discriminate among classes of UV scenarios and constrain their underlying coupling structures. 
Moreover, the momentum distributions of the final-state particles in three-body decays may help distinguish operator structures that yield similar integrated decay rates.   
A quantitative connection to specific UV scenarios would require matching the underlying models onto SMEFT operators, evolving the WCs to the electroweak scale, and subsequently matching onto the LEFT operators considered in this work.

\acknowledgments
We thank two anonymous referees for several helpful comments and suggestions
that have helped us improve the presentation of our work.
This work was supported in part 
by Grants No.\,NSFC-12305110 
and No.\,NSFC-12035008.  

\appendix
\section{Chiral Lagrangian terms}
\label{app:chira_lLag}
In this appendix, we summarize all relevant hadron-level interactions
responsible for nucleon three-body decays into one lepton and two pseudoscalar mesons.

\subsection{Baryon-number-violating sector}
\label{app:BNV_terms}

\textbf{Two-point baryon-lepton mixing terms}.~ 
By expanding \cref{eq:BNV_ChPT} to the zeroth order in the pseudoscalar meson fields, we obtain the two-point vertices involving a lepton and an octet baryon:
\begin{subequations}
\label{eq:VertexBl}
\begin{align}
{\cal L}_{Bl}\supset
& \Big[ ( c_1 C^{\tL \tR,x}_{\ell uud} + 
 c_2 C^{\tL \tL,x}_{\ell uud } )
\overline{\ell_{\tL x}^{\C}} p_{\tL}
\\
&+( c_1 C^{\tL \tR,x}_{\ell usu} + 
c_2 C^{\tL \tL,x}_{\ell usu} )
\overline{\ell_{\tL x}^{\C}} \Sigma^+_{\tL}
\\
&
+ ( c_1 C^{\tL \tR,x}_{ \bar\ell dds} 
+ c_2 C^{\tL \tL,x}_{ \bar\ell dds} ) 
\overline{\ell_{\tR x}} \Sigma_\tL^-
\\
&+( c_1 C^{\tL \tR,x}_{\nu dud} + 
c_2 C^{\tL \tL,x}_{\nu dud} )
\overline{\nu_{\tL x}^{\C}} n_{\tL} 
\\
& +\frac{1}{\sqrt{6}} \big[ c_1 
(C^{\tL \tR,x}_{\nu uds} + C^{\tL \tR,x}_{\nu dsu} 
- 2 C^{\tL \tR,x}_{\nu sud} )
+ c_2 ( C^{\tL \tL,x}_{\nu dsu} 
- 2 C^{\tL \tL,x}_{ \nu sud}) \big] 
 \overline{\nu_{\tL x}^{\C}} \Lambda^0_{\tL}
\nonumber\\
&+\frac{1}{\sqrt{2}} \big[ c_1 
(C^{\tL \tR,x}_{\nu uds} 
- C^{\tL \tR,x}_{\nu dsu})
 - c_2 C^{\tL \tL,x}_{\nu dsu} \big]
\overline{\nu_{\tL x}^{\C}} \Sigma^0_{\tL} \Big] 
 - (\tL,\nu_\tL^{\C})\leftrightarrow (\tR,\nu_\tL).
\end{align}
\end{subequations}
Note that there are no terms involving a $\Xi^-$ or $\Xi^0$, since these states correspond to operators with the $dss$ and $uss$ quark configurations containing two strange quarks and therefore do not contribute to two- and three-body nucleon decays at leading order.

\textbf{Three-point baryon-lepton-meson terms}.~ 
Similarly, expanding \cref{eq:BNV_ChPT} to the linear order in the pseudoscalar meson field yields the following three-point baryon-lepton-meson vertices:
\begin{subequations}
\label{eq:VertexBlM}
\begin{align}
{\cal L}_{Bl M} 
\supset &\,
\frac{i}{\sqrt{2}F_0}\Big\{
(c_1 C_{\ell uud}^{\tL\tR,x} + c_2  C_{\ell uud}^{\tL\tL,x})
(\overline{\ell_{\tL x}^\C} n_\tL) \pi^+
+ \frac{1}{\sqrt{2}}
(c_1 C_{\ell uud}^{\tL\tR,x} + c_2  C_{\ell uud}^{\tL\tL,x})
(\overline{\ell_{\tL x}^\C} p_\tL)\pi^0 
\nonumber\\
&- \frac{1}{\sqrt{6}}
(c_1 C_{\ell uud}^{\tL\tR,x} - 3 c_2  C_{\ell uud}^{\tL\tL,x})
(\overline{\ell_{\tL x}^\C} p_\tL)\eta
\\%
&+ (c_1 C_{\ell usu}^{\tL\tR,x} - c_2  C_{\ell usu}^{\tL\tL,x})
(\overline{\ell_{\tL x}^\C} p_\tL) \bar K^0   
+ \sqrt{2} c_2 C_{\ell usu}^{\tL\tL,x} 
(\overline{\ell_{\tL x}^{\C}} \Sigma^+_\tL )  \pi^0
\nonumber\\
&+\sqrt{\frac{2}{3}} c_1 C_{\ell usu}^{\tL\tR,x} 
(\overline{\ell_{\tL x}^{\C}} \Lambda^0_\tL) \pi^+ 
- \sqrt{2} c_2 C_{\ell usu}^{\tL\tL,x} 
(\overline{\ell_{\tL x}^{\C}} \Sigma^0_\tL) \pi^+ 
\\
& + (c_1 C_{\bar \ell dds}^{\tL\tR,x} - c_2 C_{\bar \ell dds}^{\tL\tL,x})
(\overline{\ell_{\tR x}} n_\tL) K^-
- \sqrt{2} c_2 C_{\bar \ell dds}^{\tL\tL,x} 
(\overline{\ell_{\tR x}} \Sigma^-_\tL)\pi^0
\nonumber\\
&+\sqrt{\frac{2}{3}} c_1 C_{\bar \ell dds}^{\tL\tR,x} 
(\overline{\ell_{\tR x}} \Lambda^0_\tL)\pi^-
+\sqrt{2} c_2 C_{\bar \ell dds}^{\tL\tL,x} 
(\overline{\ell_{\tR x}} \Sigma^0_\tL) \pi^-
\\%
&+(c_1  C_{\nu dud}^{\tL\tR,x} + c_2 C_{\nu dud}^{\tL\tL,x}) 
(\overline{\nu_{\tL x}^\C} p_\tL) \pi^-
-\frac{1}{\sqrt{2}} 
(c_1  C_{\nu dud}^{\tL\tR,x} + c_2 C_{\nu dud}^{\tL\tL,x}) 
(\overline{\nu_{\tL x}^\C} n_\tL) \pi^0
\nonumber
\\
&
- \frac{1}{\sqrt{6}} 
(c_1  C_{\nu dud}^{\tL\tR,x} - 3 c_2 C_{\nu dud}^{\tL\tL,x}) 
(\overline{\nu_{\tL x}^\C} n_\tL) \eta
\\%
&+\big[c_1  (C_{\nu uds}^{\tL\tR,x}+C_{\nu sud}^{\tL\tR,x}) 
+ c_2 C_{\nu sud}^{\tL\tL,x} \big]
(\overline{\nu_{\tL x}^\C} p_\tL) K^-
\nonumber
\\
&+ \big[c_1  (C_{\nu dsu}^{\tL\tR,x} + C_{\nu sud}^{\tL\tR,x}) 
- c_2 ( C_{\nu dsu}^{\tL\tL,x}-C_{\nu sud}^{\tL\tL,x}) \big]
(\overline{\nu_{\tL x}^\C} n_\tL) \bar K^0
\nonumber
\\
& +\frac{1}{\sqrt{3}}
c_1 ( C_{\nu uds}^{\tL\tR,x} -  C_{\nu dsu}^{\tL\tR,x}) 
(\overline{\nu_{\tL x}^\C} \Lambda^0_\tL) \pi^0
+c_1 (C_{\nu uds}^{\tL\tR,x} + C_{\nu dsu}^{\tL\tR,x}) 
(\overline{\nu_{\tL x}^\C} \Sigma^0_\tL) \pi^0
\nonumber\\
&+\big[c_1 (C_{\nu uds}^{\tL\tR,x} + C_{\nu dsu}^{\tL\tR,x}) 
+ c_2 C_{\nu dsu}^{\tL\tL,x} \big]
(\overline{\nu_{\tL x}^\C} \Sigma^+_\tL) \pi^-
\nonumber\\
&+\big[c_1 (C_{\nu uds}^{\tL\tR,x}+C_{\nu dsu}^{\tL\tR,x}) 
- c_2 C_{\nu dsu}^{\tL\tL,x} \big]
(\overline{\nu_{\tL x}^\C} \Sigma^-_\tL) \pi^+
\Big\}
+ (\tL,\nu_\tL^{\C})\leftrightarrow (\tR,\nu_\tL).
\end{align}
\end{subequations}
In the above, we have omitted terms involving both a $K$ or $\eta$ meson and a hyperon, as they are irrelevant. Such terms would contribute to nucleon decay channels with two kaons or a kaon plus an $\eta$, but these modes are kinematically forbidden. 

\textbf{Four-point baryon-lepton-meson-meson terms}.~
These terms induce contact contributions to the nucleon three-body decays, as depicted in \cref{fig:Feyndiagram}\,(d). Therefore, only those containing a nucleon and at most an $\eta$ or a kaon are needed.
Expanding \cref{eq:BNV_ChPT} to the second order in the meson fields leads to the following relevant terms:
\begin{subequations}
\label{eq:VertexNBlM}
\begin{align}
{\cal L}_{{\tt N}l M M} \supset\,
 & \frac{1}{F_0^2} \Big\{
- \frac{1}{8}
(c_1 C_{\ell uud}^{\tL\tR,x} + c_2 C_{\ell uud}^{\tL\tL,x}) (\overline{\ell_{\tL x}^\C} p_\tL) (2\pi^- \pi^+  + \pi^0 \pi^0)
\nonumber\\
&+\frac{1}{4\sqrt{3}} (c_1 C_{\ell uud}^{\tL\tR,x} -3 c_2 C_{\ell uud}^{\tL\tL,x}) (\overline{\ell_{\tL x}^\C} p_\tL ) \pi^0 \eta
\nonumber\\
&+ \frac{1}{2\sqrt{6}} (c_1 C_{\ell uud}^{\tL\tR,x} - 3 c_2 C_{\ell uud}^{\tL\tL,x}) 
(\overline{\ell_{\tL x}^\C} n_\tL ) \pi^+ \eta
\\%
& - \frac{1}{4} (c_1 C_{\ell usu}^{\tL\tR,x} + c_2 C_{\ell usu}^{\tL\tL,x}) (\overline{\ell_{\tL x}^\C} p_\tL ) \pi^+ K^-
- \frac{1}{4\sqrt{2}} (c_1 C_{\ell usu}^{\tL\tR,x} -3 c_2 C_{\ell usu}^{\tL\tL,x}) (\overline{\ell_{\tL x}^\C} p_\tL ) \pi^0 \bar K^0
\nonumber\\
&- \frac{1}{2}(c_1 C_{\ell usu}^{\tL\tR,x} - c_2 C_{\ell usu}^{\tL\tL,x}) (\overline{\ell_{\tL x}^\C} n_\tL ) \pi^+ \bar K^0
\\%
& - \frac{1}{4}
(c_1 C_{\bar \ell dds}^{\tL\tR,x} + c_2 C_{\bar \ell dds}^{\tL\tL,x}) (\overline{\ell_{\tR x}} n_\tL ) \pi^- \bar K^0
+ \frac{1}{4\sqrt{2}} (c_1 C_{\bar \ell dds}^{\tL\tR,x} - 3 c_2 C_{\bar \ell dds}^{\tL\tL,x}) (\overline{\ell_{\tR x}} n_\tL ) \pi^0 K^-
\nonumber\\
& -\frac{1}{2} (c_1 C_{\bar \ell dds}^{\tL\tR,x} - c_2 C_{\bar \ell dds}^{\tL\tL,x}) (\overline{\ell_{\tR x}} p_\tL ) \pi^- K^-
\\%
& -\frac{1}{8} (c_1 C_{\nu dud}^{\tL\tR,x} 
+c_2 C_{\nu dud}^{\tL\tL,x} ) 
(\overline{\nu_{\tL x}^\C} n_\tL ) (2\pi^- \pi^+  + \pi^0 \pi^0)
\nonumber\\
&- \frac{1}{4\sqrt{3}} (c_1 C_{\nu dud}^{\tL\tR,x} 
-3c_2 C_{\nu dud}^{\tL\tL,x} ) 
(\overline{\nu_{\tL x}^\C} n_\tL ) \pi^0 \eta
\nonumber\\
&+ \frac{1}{2\sqrt{6}} (c_1 C_{\nu dud}^{\tL\tR,x} 
-3c_2 C_{\nu dud}^{\tL\tL,x} ) 
(\overline{\nu_{\tL x}^\C} p_\tL ) \pi^- \eta
\\%
& - \frac{1}{4} \big[ c_1 (C_{\nu uds}^{\tL\tR,x} + 2C_{\nu dsu}^{\tL\tR,x} + C_{\nu sud}^{\tL\tR,x}) 
- c_2 (2 C_{\nu dsu}^{\tL\tL,x} -  C_{\nu sud}^{\tL\tL,x}) \big] (\overline{\nu_{\tL x}^\C} p_\tL ) \pi^- \bar K^0
\nonumber\\
&- \frac{1}{4\sqrt{2}} \big[ c_1 (
3 C_{\nu uds}^{\tL\tR,x} + C_{\nu sud}^{\tL\tR,x}) + c_2 C_{\nu sud}^{\tL\tL,x} \big] (\overline{\nu_{\tL x}^\C} p_\tL ) \pi^0 K^-
\nonumber\\
&- \frac{1}{4} \big[ c_1 (
2 C_{\nu uds}^{\tL\tR,x} +C_{\nu dsu}^{\tL\tR,x} + C_{\nu sud}^{\tL\tR,x}) 
+ c_2 ( C_{\nu dsu}^{\tL\tL,x} + C_{\nu sud}^{\tL\tL,x}) \big] (\overline{\nu_{\tL x}^\C} n_\tL ) \pi^+ K^-
\nonumber\\
&+ \frac{1}{4\sqrt{2}} \big[ c_1 (
3 C_{\nu dsu}^{\tL\tR,x}+C_{\nu sud}^{\tL\tR,x} )  
- c_2 ( C_{\nu dsu}^{\tL\tL,x} - C_{\nu sud}^{\tL\tL,x})  \big] (\overline{\nu_{\tL x}^\C} n_\tL ) \pi^0 \bar K^0
\Big\}
\\
&- (\tL,\nu_\tL^{\C})\leftrightarrow (\tR,\nu_\tL).
\nonumber
\end{align}
\end{subequations}

\subsection{Baryon-number-conserving sector}
\label{app:BNC_terms}

\textbf{Three-point baryon-baryon-meson terms}.~  
By expanding the pseudoscalar meson matrices in \cref{eq:LChPT} to the linear order, we obtain the following relevant BNC three-point vertices
\begin{align}
\label{eq:LBBM}
{\cal L}_{\bar BBM} \supset &  
\frac{1}{F_0} \Big\{ 
\frac{D+F}{2} \big[ (\overline{p} \gamma^\mu \gamma_5 p) \partial_\mu \pi^0 
+\sqrt{2} (\overline{n} \gamma^\mu \gamma_5 p) \partial_\mu \pi^-\big]
+\frac{3F-D}{2\sqrt{3}} (\overline{p} \gamma^\mu \gamma_5 p) \partial_\mu \eta
\nonumber
\\
&+\frac{D-F}{2}
\big[ (\overline{\Sigma^0} \gamma^\mu \gamma_5 p)  \partial_\mu K^-
+ \sqrt{2}(\overline{\Sigma^+}\gamma^\mu \gamma_5 p) 
\partial_\mu \bar K^0 \big]
- \frac{D+3F}{2 \sqrt{3}} (\overline{\Lambda^0} \gamma^\mu \gamma_5 p) \partial_\mu K^-
\nonumber
\\
&+\frac{D+F}{2} 
\big[\sqrt{2}(\overline{p} \gamma^\mu \gamma_5 n) \partial_\mu \pi^+ - (\overline{n} \gamma^\mu \gamma_5 n) \partial_\mu \pi^0 \big]
+\frac{3F-D}{2\sqrt{3}} (\overline{n} \gamma^\mu \gamma_5 n) \partial_\mu \eta
\nonumber
\\
&+\frac{D-F}{2}
\big[ \sqrt{2}(\overline{\Sigma^-}\gamma^\mu \gamma_5 n) \partial_\mu K^-
- (\overline{\Sigma^0} \gamma^\mu \gamma_5 n)  
\partial_\mu \bar K^0 \big]
- \frac{D+3F}{2 \sqrt{3}} (\overline{\Lambda^0} \gamma^\mu \gamma_5 n) \partial_\mu \bar K^0
\nonumber
\\
&+ F \big[(\overline{\Sigma^+}\gamma^\mu \gamma_5 \Sigma^+) \partial_\mu \pi^0 
-  (\overline{\Sigma^0}\gamma^\mu \gamma_5 \Sigma^+) 
\partial_\mu \pi^- \big]
+\frac{D}{\sqrt{3}} (\overline{\Lambda^0}\gamma^\mu \gamma_5 \Sigma^+) 
\partial_\mu \pi^- 
\nonumber
\\
&+ F \big[(\overline{\Sigma^0}\gamma^\mu \gamma_5 \Sigma^-) \partial_\mu \pi^+ 
-  (\overline{\Sigma^-}\gamma^\mu \gamma_5 \Sigma^-) 
\partial_\mu \pi^0 \big]
+\frac{D}{\sqrt{3}} (\overline{\Lambda^0}\gamma^\mu \gamma_5 \Sigma^-) 
\partial_\mu \pi^+ 
\nonumber
\\
&+ F \big[(\overline{\Sigma^-}\gamma^\mu \gamma_5 \Sigma^0) \partial_\mu \pi^-
-  (\overline{\Sigma^+}\gamma^\mu \gamma_5 \Sigma^0) 
\partial_\mu \pi^+ \big]
+\frac{D}{\sqrt{3}} (\overline{\Lambda^0}\gamma^\mu \gamma_5 \Sigma^0) 
\partial_\mu \pi^0 
\nonumber
\\
&+  \frac{D}{\sqrt{3}} 
\big[ (\overline{\Sigma^0}\gamma^\mu \gamma_5 \Lambda^0) \partial_\mu \pi^0
+ (\overline{\Sigma^+}\gamma^\mu \gamma_5 \Lambda^0) 
\partial_\mu \pi^+
+ (\overline{\Sigma^-}\gamma^\mu \gamma_5 \Lambda^0) 
\partial_\mu \pi^- \big]\Big\}.
\end{align}
For operators that do not involve nucleon fields, only those coupled to a pion can contribute [the last four lines in \cref{eq:LBBM}]. 
This is because such vertices can appear only as the middle vertex in \cref{fig:Feyndiagram}(a),  where the other BNC vertex must be of the form 
$(\Sigma,\Lambda){\tt N}K$.
Moreover, terms involving a pair of $\Xi$ baryons do not contribute, as strangeness conservation forbids $\Xi{\tt N}M$ couplings.

\textbf{Four-point baryon-baryon-meson-meson terms}.~ 
When combined with baryon-lepton mixing terms in \cref{eq:VertexBl}, the four-point baryon-baryon-meson-meson vertices can contribute to nucleon three-body decays via the noncontact diagram in \cref{fig:Feyndiagram}(b).
Consequently, the relevant terms must involve a nucleon, a non-$\Xi$ baryon, and at most an eta or a kaon. 
Expanding \cref{eq:LChPT} to the second order in the pseudoscalar meson fields yields the following relevant terms
\begin{align}
\label{eq:LBNMM}
{\cal L}_{\bar B{\tt N} MM} \supset\,& 
\frac{1}{4F^2_0} \Big\{
- (\overline{p} \gamma^\mu p ) 
(\pi^- i\overleftrightarrow{\partial_\mu} \pi^+) 
+ \sqrt{2} (\overline{n} \gamma^\mu p ) 
(\pi^- i\overleftrightarrow{\partial_\mu} \pi^0) 
\nonumber
\\
&+ \frac{1}{2} \big[ \overline{\Sigma^0} \gamma^\mu p + \sqrt{3}(\overline{\Lambda^0} \gamma^\mu p) \big]
\Big(\pi^0 i\overleftrightarrow{\partial_\mu} K^- 
+\sqrt{2} \pi^- i\overleftrightarrow{\partial_\mu} \bar K^0 \Big)
\nonumber\\
& + (\overline{\Sigma^+} \gamma^\mu p ) 
\Big(\pi^+ i\overleftrightarrow{\partial_\mu}K^-  
-\frac{1}{\sqrt{2}} \pi^0 i\overleftrightarrow{\partial_\mu} \bar K^0 \Big)
\nonumber\\
&+(\overline{n} \gamma^\mu n ) 
(\pi^- i\overleftrightarrow{\partial_\mu} \pi^+) 
-\sqrt{2} (\overline{p} \gamma^\mu n ) 
(\pi^+ i\overleftrightarrow{\partial_\mu} \pi^0)
\nonumber\\
&+ \frac{1}{2} \big[ \overline{\Sigma^0} \gamma^\mu n - \sqrt{3}
(\overline{\Lambda^0} \gamma^\mu n ) \big]
\Big( \pi^0 i\overleftrightarrow{\partial_\mu} \bar K^0  
- \sqrt{2} \pi^+ i\overleftrightarrow{\partial_\mu} K^- \Big) 
\nonumber\\
& + (\overline{\Sigma^-} \gamma^\mu n ) 
\Big(\pi^- i\overleftrightarrow{\partial_\mu} \bar K^0  
+\frac{1}{\sqrt{2}} \pi^0 i\overleftrightarrow{\partial_\mu} K^- \Big) 
\Big\}.
\end{align}

\section{Decay width expressions for three-body modes ${\tt N}\to l M_1 M_2$}
\label{app:DeW_three_body}

In this appendix, we present the complete numerical results for the decay widths of various three-body nucleon decay modes involving one lepton and two pseudoscalar mesons, induced  at leading order by dim-6 BNV LEFT operators.
The results are fully parameterized in terms of the LEFT WCs.

\subsection{Decay modes involving a charged lepton}

For the three-body modes with a charged lepton, the decay widths are
\begin{subequations}
\begin{align}
\frac{\Gamma_{p \to e^+ \pi^+\pi^-} }{(0.1\, \rm GeV)^5} 
=&~ 2.52 \big|C^{\tL\tR,e}_{\ell uud}\big|^2 + 
2.57 \big|C^{\tL \tL,e}_{\ell uud}\big|^2 
-5.08\,\Re\big( C^{\tL \tR,e}_{\ell uud} C^{\tL \tL,e*}_{\ell uud} \big)
+ 0.002 C^{\tL \tR,e}_{\ell uud} C^{\tR \tL,e*}_{\ell uud}
\nonumber\\
& -0.0021 \big( C^{\tL \tR,e}_{\ell uud} C^{\tR \tR,e*}_{\ell uud} + 
C^{\tL \tL,e}_{\ell uud} C^{\tR \tL,e*}_{\ell uud} 
- C^{\tL \tL,e}_{\ell uud} C^{\tR \tR,e*}_{\ell uud} \big)
+\tL \leftrightarrow \tR,
\\
\frac{\Gamma_{p \to \mu^+ \pi^+\pi^-} }{(0.1\, \rm GeV)^5} 
=&~ 2.24 \big|C^{\tL\tR,\mu}_{\ell uud}\big|^2 + 
2.29 \big|C^{\tL \tL,\mu}_{\ell uud}\big|^2 
-4.53\,\Re\big( C^{\tL \tR,\mu}_{\ell uud} C^{\tL \tL,\mu*}_{\ell uud} \big)
+ 0.318 C^{\tL \tR,\mu}_{\ell uud} C^{\tR \tL,\mu*}_{\ell uud}
\nonumber\\
& -0.321 \big( C^{\tL \tR,\mu}_{\ell uud} C^{\tR \tR,\mu*}_{\ell uud} + C^{\tL \tL,\mu}_{\ell uud} C^{\tR \tL,\mu*}_{\ell uud} \big)
+ 0.324 C^{\tL \tL,\mu}_{\ell uud} C^{\tR \tR,\mu*}_{\ell uud}
+ \tL \leftrightarrow \tR,
\\
\frac{\Gamma_{p \to e^+ \pi^0\pi^0} }{(0.1\, \rm GeV)^5} 
=&~ 0.42 \big|C^{\tL\tR,e}_{\ell uud}\big|^2 + 
0.43 \big|C^{\tL \tL,e}_{\ell uud}\big|^2 
-0.85 \,\Re\big( C^{\tL \tR,e}_{\ell uud} C^{\tL \tL,e*}_{\ell uud} \big)
\nonumber\\
&+ 1.31\cdot 10^{-4} C^{\tL \tR,e}_{\ell uud} C^{\tR \tL,e*}_{\ell uud}
-1.32\cdot 10^{-4} \big( C^{\tL \tR,e}_{\ell uud} C^{\tR \tR,e*}_{\ell uud} + 
C^{\tL \tL,e}_{\ell uud} C^{\tR \tL,e*}_{\ell uud}  \big)
\nonumber\\
&+1.33\cdot 10^{-4} C^{\tL \tL,e}_{\ell uud} C^{\tR \tR,e*}_{\ell uud}
+\tL \leftrightarrow \tR,
\\
\frac{\Gamma_{p \to \mu^+ \pi^0\pi^0} }{(0.1\, \rm GeV)^5} 
=&~ 0.39 \big|C^{\tL\tR,\mu}_{\ell uud}\big|^2 + 
0.4 \big|C^{\tL \tL,\mu}_{\ell uud}\big|^2 
-0.79 \,\Re\big( C^{\tL \tR,\mu}_{\ell uud} C^{\tL \tL,\mu*}_{\ell uud} \big)
+ 0.024 C^{\tL \tR,\mu}_{\ell uud} C^{\tR \tL,\mu*}_{\ell uud}
\nonumber\\
& -0.024 \big( C^{\tL \tR,\mu}_{\ell uud} C^{\tR \tR,\mu*}_{\ell uud} + C^{\tL \tL,\mu}_{\ell uud} C^{\tR \tL,\mu*}_{\ell uud} 
-C^{\tL \tL,\mu}_{\ell uud} C^{\tR \tR,\mu*}_{\ell uud} \big)
+ \tL \leftrightarrow \tR,
\\
\frac{\Gamma_{p \to e^+ \pi^0\eta} }{(0.1\, \rm GeV)^5} 
=&~ 0.0043 \big|C^{\tL\tR,e}_{\ell uud}\big|^2 + 
0.064 \big|C^{\tL \tL,e}_{\ell uud}\big|^2 
+0.032 \,\Re\big( C^{\tL \tR,e}_{\ell uud} C^{\tL \tL,e*}_{\ell uud} \big)
\nonumber\\
& -2.19 \cdot 10^{-6} C^{\tL \tR,e}_{\ell uud} C^{\tR \tL,e*}_{\ell uud}
-7.86 \cdot 10^{-6} \big( C^{\tL \tR,e}_{\ell uud} C^{\tR \tR,e*}_{\ell uud} + C^{\tL \tL,e}_{\ell uud} C^{\tR \tL,e*}_{\ell uud} \big)
\nonumber\\
& -2.04 \cdot 10^{-5} C^{\tL \tL,e}_{\ell uud} C^{\tR \tR,e*}_{\ell uud}
+\tL \leftrightarrow \tR,
\\
\frac{\Gamma_{p \to \mu^+ \pi^0\eta} }{(0.1\, \rm GeV)^5} 
=&~ 0.0027 \big|C^{\tL\tR,\mu}_{\ell uud}\big|^2 + 
0.039 \big|C^{\tL \tL,\mu}_{\ell uud}\big|^2 
+ 0.02 \,\Re\big( C^{\tL \tR,\mu}_{\ell uud} C^{\tL \tL,\mu*}_{\ell uud} \big)
\nonumber\\
& -3.52 \cdot 10^{-4} C^{\tL \tR,\mu}_{\ell uud} C^{\tR \tL,\mu*}_{\ell uud}
-0.0013 \big( C^{\tL \tR,\mu}_{\ell uud} C^{\tR \tR,\mu*}_{\ell uud} + C^{\tL \tL,\mu}_{\ell uud} C^{\tR \tL,\mu*}_{\ell uud} \big)
\nonumber\\
& -0.0039 C^{\tL \tL,\mu}_{\ell uud} C^{\tR \tR,\mu*}_{\ell uud}
+ \tL \leftrightarrow \tR,
\\
\frac{\Gamma_{n \to e^+ \pi^-\pi^0} }{(0.1\, \rm GeV)^5} 
=&~ 3.55 \big|C^{\tL\tR,e}_{\ell uud}\big|^2 
+ 3.61 \big|C^{\tL \tL,e}_{\ell uud}\big|^2 
-7.16 \,\Re\big( C^{\tL \tR,e}_{\ell uud} C^{\tL \tL,e*}_{\ell uud} \big)
+ 0.0037 C^{\tL \tR,e}_{\ell uud} C^{\tR \tL,e*}_{\ell uud}
\nonumber\\
& -0.0037 \big( C^{\tL \tR,e}_{\ell uud} C^{\tR \tR,e*}_{\ell uud} + 
C^{\tL \tL,e}_{\ell uud} C^{\tR \tL,e*}_{\ell uud} 
- C^{\tL \tL,e}_{\ell uud} C^{\tR \tR,e*}_{\ell uud} \big)
+\tL \leftrightarrow \tR,
\\
\frac{\Gamma_{n \to \mu^+ \pi^-\pi^0} }{(0.1\, \rm GeV)^5} 
=&~ 3.12 \big|C^{\tL\tR,\mu}_{\ell uud}\big|^2 
+ 3.18 \big|C^{\tL \tL,\mu}_{\ell uud}\big|^2 
-6.29\,\Re\big( C^{\tL \tR,\mu}_{\ell uud} C^{\tL \tL,\mu*}_{\ell uud} \big)
+ 0.559 C^{\tL \tR,\mu}_{\ell uud} C^{\tR \tL,\mu*}_{\ell uud}
\nonumber\\
& -0.564 \big( C^{\tL \tR,\mu}_{\ell uud} C^{\tR \tR,\mu*}_{\ell uud} + C^{\tL \tL,\mu}_{\ell uud} C^{\tR \tL,\mu*}_{\ell uud} \big)
+ 0.569 C^{\tL \tL,\mu}_{\ell uud} C^{\tR \tR,\mu*}_{\ell uud}
+ \tL \leftrightarrow \tR,
\\
\frac{\Gamma_{n \to e^+ \pi^-\eta} }{(0.1\, \rm GeV)^5} 
=&~ 0.0083 \big|C^{\tL\tR,e}_{\ell uud}\big|^2 
+ 0.124 \big|C^{\tL \tL,e}_{\ell uud}\big|^2 
+ 0.062 \,\Re\big( C^{\tL \tR,e}_{\ell uud} C^{\tL \tL,e*}_{\ell uud} \big)
\nonumber\\
& -4.57 \cdot 10^{-6} C^{\tL \tR,e}_{\ell uud} C^{\tR \tL,e*}_{\ell uud}
-1.62 \cdot 10^{-5} \big( C^{\tL \tR,e}_{\ell uud} C^{\tR \tR,e*}_{\ell uud} + C^{\tL \tL,e}_{\ell uud} C^{\tR \tL,e*}_{\ell uud} \big)
\nonumber\\
& -4.22 \cdot 10^{-5} C^{\tL \tL,e}_{\ell uud} C^{\tR \tR,e*}_{\ell uud}
+\tL \leftrightarrow \tR,
\\
\frac{\Gamma_{n \to \mu^+ \pi^-\eta} }{(0.1\, \rm GeV)^5} 
=&~ 0.0052 \big|C^{\tL\tR,\mu}_{\ell uud}\big|^2 
+ 0.074 \big|C^{\tL \tL,\mu}_{\ell uud}\big|^2 
+ 0.038\,\Re\big( C^{\tL \tR,\mu}_{\ell uud} C^{\tL \tL,\mu*}_{\ell uud} \big)
\nonumber\\
& -7.15 \cdot 10^{-4} C^{\tL \tR,\mu}_{\ell uud} C^{\tR \tL,\mu*}_{\ell uud}
-0.0026 \big( C^{\tL \tR,\mu}_{\ell uud} C^{\tR \tR,\mu*}_{\ell uud} + C^{\tL \tL,\mu}_{\ell uud} C^{\tR \tL,\mu*}_{\ell uud} \big)
\nonumber\\
& -0.0079 C^{\tL \tL,\mu}_{\ell uud} C^{\tR \tR,\mu*}_{\ell uud}
+ \tL \leftrightarrow \tR,
\\
\frac{\Gamma_{p \to e^+ \pi^0 K^0} }{(0.1\, \rm GeV)^5} 
=&~ 0.049 \big|C^{\tL\tR,e}_{\ell usu}\big|^2 + 
0.1 \big|C^{\tL \tL,e}_{\ell usu}\big|^2 
+0.11 \,\Re\big( C^{\tL \tR,e}_{\ell usu} C^{\tL \tL,e*}_{\ell usu} \big)
\nonumber\\
& +1.31 \cdot 10^{-4} C^{\tL \tR,e}_{\ell usu} C^{\tR \tL,e*}_{\ell usu}
+5.47 \cdot 10^{-5} \big( C^{\tL \tR,e}_{\ell usu} C^{\tR \tR,e*}_{\ell usu} + C^{\tL \tL,e}_{\ell usu} C^{\tR \tL,e*}_{\ell usu} \big)
\nonumber\\
& -1.78 \cdot 10^{-4} C^{\tL \tL,e}_{\ell usu} C^{\tR \tR,e*}_{\ell usu}
+\tL \leftrightarrow \tR,
\\
\frac{\Gamma_{p \to \mu^+ \pi^0 K^0} }{(0.1\, \rm GeV)^5} 
=&~ 0.033 \big|C^{\tL\tR,\mu}_{\ell usu}\big|^2 + 
0.077 \big|C^{\tL \tL,\mu}_{\ell usu}\big|^2 
+ 0.078 \,\Re\big( C^{\tL \tR,\mu}_{\ell usu} C^{\tL \tL,\mu*}_{\ell usu} \big)
\nonumber\\
& +0.014 C^{\tL \tR,\mu}_{\ell usu} C^{\tR \tL,\mu*}_{\ell usu}
+0.0049 \big( C^{\tL \tR,\mu}_{\ell usu} C^{\tR \tR,\mu*}_{\ell usu} + C^{\tL \tL,\mu}_{\ell usu} C^{\tR \tL,\mu*}_{\ell usu} \big)
\nonumber\\
& -0.023 C^{\tL \tL,\mu}_{\ell usu} C^{\tR \tR,\mu*}_{\ell usu}
+ \tL \leftrightarrow \tR,
\\
\frac{\Gamma_{p \to e^+ \pi^- K^+} }{(0.1\, \rm GeV)^5} 
=&~ 0.0083 \big|C^{\tL\tR,e}_{\ell usu}\big|^2 + 
0.019 \big|C^{\tL \tL,e}_{\ell usu}\big|^2 
-0.023 \,\Re\big( C^{\tL \tR,e}_{\ell usu} C^{\tL \tL,e*}_{\ell usu} \big)
\nonumber\\
& -4.25 \cdot 10^{-5} C^{\tL \tR,e}_{\ell usu} C^{\tR \tL,e*}_{\ell usu}
+6.32 \cdot 10^{-5} \big( C^{\tL \tR,e}_{\ell usu} C^{\tR \tR,e*}_{\ell usu} + C^{\tL \tL,e}_{\ell usu} C^{\tR \tL,e*}_{\ell usu} \big)
\nonumber\\
& -8.1 \cdot 10^{-5} C^{\tL \tL,e}_{\ell usu} C^{\tR \tR,e*}_{\ell usu}
+\tL \leftrightarrow \tR,
\\
\frac{\Gamma_{p \to \mu^+ \pi^- K^+} }{(0.1\, \rm GeV)^5} 
=&~ 0.0074 \big|C^{\tL\tR,\mu}_{\ell usu}\big|^2 + 
0.015 \big|C^{\tL \tL,\mu}_{\ell usu}\big|^2 
- 0.02 \,\Re\big( C^{\tL \tR,\mu}_{\ell usu} C^{\tL \tL,\mu*}_{\ell usu} \big)
\nonumber\\
&-0.0054 C^{\tL \tR,\mu}_{\ell usu} C^{\tR \tL,\mu*}_{\ell usu}
+0.0078 \big( C^{\tL \tR,\mu}_{\ell usu} C^{\tR \tR,\mu*}_{\ell usu} + C^{\tL \tL,\mu}_{\ell usu} C^{\tR \tL,\mu*}_{\ell usu} \big)
\nonumber\\
&-0.01 C^{\tL \tL,\mu}_{\ell usu} C^{\tR \tR,\mu*}_{\ell usu}
+ \tL \leftrightarrow \tR,
\\
\frac{\Gamma_{n \to e^+ \pi^- K^0} }{(0.1\, \rm GeV)^5} 
=&~ 0.14 \big|C^{\tL\tR,e}_{\ell usu}\big|^2 + 
0.11 \big|C^{\tL \tL,e}_{\ell usu}\big|^2 
+0.23 \,\Re\big( C^{\tL \tR,e}_{\ell usu} C^{\tL \tL,e*}_{\ell usu} \big)
\nonumber\\
& +2.05 \cdot 10^{-4} C^{\tL \tR,e}_{\ell usu} C^{\tR \tL,e*}_{\ell usu}
+7.22 \cdot 10^{-5} \big( C^{\tL \tR,e}_{\ell usu} C^{\tR \tR,e*}_{\ell usu} + C^{\tL \tL,e}_{\ell usu} C^{\tR \tL,e*}_{\ell usu} \big)
\nonumber\\
&-2.79 \cdot 10^{-5} C^{\tL \tL,e}_{\ell usu} C^{\tR \tR,e*}_{\ell usu}
+\tL \leftrightarrow \tR,
\\
\frac{\Gamma_{n \to \mu^+ \pi^- K^0} }{(0.1\, \rm GeV)^5} 
=&~ 0.098 \big|C^{\tL\tR,\mu}_{\ell usu}\big|^2 
+ 0.078 \big|C^{\tL \tL,\mu}_{\ell usu}\big|^2 
+ 0.17 \,\Re\big( C^{\tL \tR,\mu}_{\ell usu} C^{\tL \tL,\mu*}_{\ell usu} \big)
\nonumber\\
&+0.018 C^{\tL \tR,\mu}_{\ell usu} C^{\tR \tL,\mu*}_{\ell usu}
+0.0047 \big( C^{\tL \tR,\mu}_{\ell usu} C^{\tR \tR,\mu*}_{\ell usu} + C^{\tL \tL,\mu}_{\ell usu} C^{\tR \tL,\mu*}_{\ell usu} \big)
\nonumber\\
& -0.0057 C^{\tL \tL,\mu}_{\ell usu} C^{\tR \tR,\mu*}_{\ell usu}
+ \tL \leftrightarrow \tR,
\\
\frac{\Gamma_{p \to e^- \pi^+ K^+} }{(0.1\, \rm GeV)^5} 
=&~ 0.15 \big|C^{\tL\tR,e}_{\bar \ell dds}\big|^2 + 
0.11 \big|C^{\tL \tL,e}_{\bar \ell dds}\big|^2 
+0.24 \,\Re\big( C^{\tL \tR,e}_{\bar \ell dds} C^{\tL \tL,e*}_{\bar \ell dds} \big)
\nonumber\\
& +2.04 \cdot 10^{-4} C^{\tL \tR,e}_{\bar \ell dds} C^{\tR \tL,e*}_{\bar \ell dds}
+6.99 \cdot 10^{-5} \big( C^{\tL \tR,e}_{\bar \ell dds} C^{\tR \tR,e*}_{\bar \ell dds} + 
C^{\tL \tL,e}_{\bar \ell dds} C^{\tR \tL,e*}_{\bar \ell dds} \big)
\nonumber\\
& -3.1 \cdot 10^{-5} C^{\tL \tL,e}_{\bar \ell dds} C^{\tR \tR,e*}_{\bar \ell dds}
+\tL \leftrightarrow \tR,
\\
\frac{\Gamma_{p \to \mu^- \pi^+ K^+} }{(0.1\, \rm GeV)^5} 
=&~ 0.1 \big|C^{\tL\tR,\mu}_{\bar \ell dds}\big|^2 
+ 0.081 \big|C^{\tL \tL,\mu}_{\bar \ell dds}\big|^2 
+ 0.17 \,\Re\big( C^{\tL \tR,\mu}_{\bar \ell dds} C^{\tL \tL,\mu*}_{\bar \ell dds} \big)
\nonumber\\
& +0.018 C^{\tL \tR,\mu}_{\bar \ell dds} C^{\tR \tL,\mu*}_{\bar \ell dds}
+0.0044 \big( C^{\tL \tR,\mu}_{\bar \ell dds} C^{\tR \tR,\mu*}_{\bar \ell dds} + C^{\tL \tL,\mu}_{\bar \ell dds} C^{\tR \tL,\mu*}_{\bar \ell dds} \big)
\nonumber\\
& -0.0062 C^{\tL \tL,\mu}_{\bar \ell dds} C^{\tR \tR,\mu*}_{\bar \ell dds}
+ \tL \leftrightarrow \tR,
\\
\frac{\Gamma_{n \to e^- \pi^+ K^0} }{(0.1\, \rm GeV)^5} 
=&~0.008 \big|C^{\tL\tR,e}_{\bar \ell dds}\big|^2 + 
0.018 \big|C^{\tL \tL,e}_{\bar \ell dds}\big|^2 
-0.023 \,\Re\big( C^{\tL \tR,e}_{\bar \ell dds} C^{\tL \tL,e*}_{\bar \ell dds} \big)
\nonumber\\
& -4.13 \cdot 10^{-5} C^{\tL \tR,e}_{\bar \ell dds} C^{\tR \tL,e*}_{\bar \ell dds}
+6.15 \cdot 10^{-5} \big( C^{\tL \tR,e}_{\bar \ell dds} C^{\tR \tR,e*}_{\bar \ell dds} + 
C^{\tL \tL,e}_{\bar \ell dds} C^{\tR \tL,e*}_{\bar \ell dds} \big)
\nonumber\\
& -7.91 \cdot 10^{-5} C^{\tL \tL,e}_{\bar \ell dds} C^{\tR \tR,e*}_{\bar \ell dds}
+\tL \leftrightarrow \tR,
\\
\frac{\Gamma_{n \to \mu^- \pi^+ K^0} }{(0.1\, \rm GeV)^5} 
=&~ 0.0071 \big|C^{\tL\tR,\mu}_{\bar \ell dds}\big|^2 
+ 0.015 \big|C^{\tL \tL,\mu}_{\bar \ell dds}\big|^2 
- 0.019 \,\Re\big( C^{\tL \tR,\mu}_{\bar \ell dds} C^{\tL \tL,\mu*}_{\bar \ell dds} \big)
\nonumber\\
& -0.0052 C^{\tL \tR,\mu}_{\bar \ell dds} C^{\tR \tL,\mu*}_{\bar \ell dds}
+0.0075 \big( C^{\tL \tR,\mu}_{\bar \ell dds} C^{\tR \tR,\mu*}_{\bar \ell dds} + C^{\tL \tL,\mu}_{\bar \ell dds} C^{\tR \tL,\mu*}_{\bar \ell dds} \big)
\nonumber\\
& -0.0097 C^{\tL \tL,\mu}_{\bar \ell dds} C^{\tR \tR,\mu*}_{\bar \ell dds}
+ \tL \leftrightarrow \tR,
\\
\frac{\Gamma_{n \to e^- \pi^0 K^+} }{(0.1\, \rm GeV)^5} 
=&~ 0.053 \big|C^{\tL\tR,e}_{\bar \ell dds}\big|^2 + 
0.11 \big|C^{\tL \tL,e}_{\bar \ell dds}\big|^2 
+0.12 \,\Re\big( C^{\tL \tR,e}_{\bar \ell dds} C^{\tL \tL,e*}_{\bar \ell dds} \big)
\nonumber\\
& +1.36 \cdot 10^{-4} C^{\tL \tR,e}_{\bar \ell dds} C^{\tR \tL,e*}_{\bar \ell dds}
+5.72 \cdot 10^{-5} \big( C^{\tL \tR,e}_{\bar \ell dds} C^{\tR \tR,e*}_{\bar \ell dds} + 
C^{\tL \tL,e}_{\bar \ell dds} C^{\tR \tL,e*}_{\bar \ell dds} \big)
\nonumber\\
& -1.84 \cdot 10^{-4} C^{\tL \tL,e}_{\bar \ell dds} C^{\tR \tR,e*}_{\bar \ell dds}
+\tL \leftrightarrow \tR,
\\
\frac{\Gamma_{n \to \mu^- \pi^0 K^+} }{(0.1\, \rm GeV)^5} 
=&~ 0.036 \big|C^{\tL\tR,\mu}_{\bar \ell dds}\big|^2 
+ 0.082 \big|C^{\tL \tL,\mu}_{\bar \ell dds}\big|^2 
+ 0.085 \,\Re\big( C^{\tL \tR,\mu}_{\bar \ell dds} C^{\tL \tL,\mu*}_{\bar \ell dds} \big)
\nonumber\\
& +0.015 C^{\tL \tR,\mu}_{\bar \ell dds} C^{\tR \tL,\mu*}_{\bar \ell dds}
+0.0053 \big( C^{\tL \tR,\mu}_{\bar \ell dds} C^{\tR \tR,\mu*}_{\bar \ell dds} + C^{\tL \tL,\mu}_{\bar \ell dds} C^{\tR \tL,\mu*}_{\bar \ell dds} \big)
\nonumber\\
& -0.024 C^{\tL \tL,\mu}_{\bar \ell dds} C^{\tR \tR,\mu*}_{\bar \ell dds}
+ \tL \leftrightarrow \tR.
\end{align}
\end{subequations}

\subsection{Decay modes involving a neutrino or an antineutrino}

For the three-body modes involving an antineutrino, the decay widths are
\begin{subequations}
\begin{align}
\frac{\Gamma_{p \to \bar\nu_x \pi^+\pi^0} }{(0.1\, \rm GeV)^5} 
=&~ 3.495 \big|C^{\tL\tR,x}_{ \nu dud}\big|^2 
+ 3.563\big|C^{\tL \tL,x}_{\nu dud}\big|^2 
-7.057\,\Re\big( C^{\tL \tR,x}_{\nu dud} C^{\tL \tL,x*}_{\nu dud} \big) ,
\\
\frac{\Gamma_{p \to \bar\nu_x \pi^+\eta} }{(0.1\, \rm GeV)^5}
=&~0.008 \big|C^{\tL \tR,x}_{\nu dud}\big|^2 
+ 0.12\big|C^{\tL \tL,x}_{\nu dud}\big|^2 
+0.06\,\Re\big( C^{\tL \tR,x}_{\nu dud} C^{\tL \tL,x*}_{\nu dud} \big),
\\
\frac{\Gamma_{n \to \bar\nu_x \pi^+\pi^-} }{(0.1\, \rm GeV)^5}
=&~2.561\big|C^{\tL \tR,x}_{\nu dud}\big|^2
+ 2.61\big|C^{\tL \tL,x}_{\nu dud}\big|^2 
- 5.17\,\Re\big( C^{\tL \tR,x}_{\nu dud} C^{\tL \tL,x*}_{\nu dud} \big),
\\
\frac{\Gamma_{n \to \bar\nu_x \pi^0\pi^0} }{(0.1\, \rm GeV)^5}
=&~ 0.425\big|C^{\tL \tR,x}_{\nu dud}\big|^2
+ 0.434\big|C^{\tL \tL,x}_{\nu dud}\big|^2 
- 0.858 \,\Re\big( C^{\tL \tR,x}_{\nu dud} C^{\tL \tL,x*}_{\nu dud} \big),
\\
\frac{\Gamma_{n \to \bar\nu_x \pi^0\eta} }{(0.1\, \rm GeV)^5}
=&~ 0.0043\big|C^{\tL \tR,x}_{\nu dud}\big|^2
+ 0.065\big|C^{\tL \tL,x}_{\nu dud}\big|^2 
+ 0.033 \,\Re\big( C^{\tL \tR,x}_{\nu dud} C^{\tL \tL,x*}_{\nu dud} \big),
\\
\frac{\Gamma_{p \to \bar\nu_x \pi^+K^0} }{(0.1\, \rm GeV)^5}
=&~0.051\big|C^{\tL \tR,x}_{\nu uds}\big|^2 
+ 0.137\big|C^{\tL \tR,x}_{\nu dsu}\big|^2 
+ 0.135\big|C^{\tL \tR,x}_{\nu sud}\big|^2 
+ 0.143\big|C^{\tL \tL,x}_{\nu dsu}\big|^2
\nonumber\\
&+ 0.137\big|C^{\tL \tL,x}_{\nu sud}\big|^2 
+ 0.095 \,\Re\big( C^{\tL \tR,x}_{\nu uds} C^{\tL \tR,x*}_{\nu dsu} \big) 
-0.022 \,\Re\big( C^{\tL \tR,x}_{\nu uds} C^{\tL \tR,x*}_{\nu sud} \big) 
\nonumber\\
& + 0.2 \,\Re\big( C^{\tL \tR,x}_{\nu dsu} C^{\tL \tR,x*}_{\nu sud} \big) 
+ 0.046 \,\Re\big( C^{\tL \tR,x}_{\nu uds} C^{\tL \tL,x*}_{\nu dsu} \big)
+ 0.023 \,\Re\big( C^{\tL \tR,x}_{\nu uds} C^{\tL \tL,x*}_{\nu sud} \big)
\nonumber\\
&+ 0.26 \,\Re\big( C^{\tL \tR,x}_{\nu dsu} C^{\tL \tL,x*}_{\nu dsu} \big) 
- 0.2 \,\Re\big( C^{\tL \tR,x}_{\nu dsu} C^{\tL \tL,x*}_{\nu sud} \big)
+ 0.25 \,\Re\big( C^{\tL \tR,x}_{\nu sud} C^{\tL \tL,x*}_{\nu dsu} \big)
\nonumber\\
&- 0.27 \,\Re\big( C^{\tL \tR,x}_{\nu sud} C^{\tL \tL,x*}_{\nu sud} \big)
- 0.26 \,\Re\big( C^{\tL \tL,x}_{\nu dsu} C^{\tL \tL,x*}_{\nu sud} \big),
\\
\frac{\Gamma_{p \to \bar\nu_x \pi^0 K^+} }{(0.1\, \rm GeV)^5}
=&~ 0.11\big|C^{\tL \tR,x}_{\nu uds}\big|^2 
+ 0.019\big|C^{\tL \tR,x}_{\nu dsu}\big|^2 
+ 0.075\big|C^{\tL \tR,x}_{\nu sud}\big|^2 
+ 0.0037\big|C^{\tL \tL,x}_{\nu dsu}\big|^2
\nonumber\\
&+ 0.077\big|C^{\tL \tL,x}_{\nu sud}\big|^2 
+ 0.021\,\Re\big( C^{\tL \tR,x}_{\nu uds} C^{\tL \tR,x*}_{\nu dsu} \big) 
+0.13 \,\Re\big( C^{\tL \tR,x}_{\nu uds} C^{\tL \tR,x*}_{\nu sud} \big)
\nonumber\\
&- 0.037 \,\Re\big( C^{\tL \tR,x}_{\nu dsu} C^{\tL \tR,x*}_{\nu sud} \big)
+ 0.019 \,\Re\big( C^{\tL \tR,x}_{\nu uds} C^{\tL \tL,x*}_{\nu dsu} \big)
- 0.14 \,\Re\big( C^{\tL \tR,x}_{\nu uds} C^{\tL \tL,x*}_{\nu sud} \big) 
\nonumber\\
&- 0.012 \,\Re\big( C^{\tL \tR,x}_{\nu dsu} C^{\tL \tL,x*}_{\nu dsu} \big) 
+ 0.037 \,\Re\big( C^{\tL \tR,x}_{\nu dsu} C^{\tL \tL,x*}_{\nu sud} \big)
+ 0.031 \,\Re\big( C^{\tL \tR,x}_{\nu sud} C^{\tL \tL,x*}_{\nu dsu} \big)
\nonumber\\
&- 0.15 \,\Re\big( C^{\tL \tR,x}_{\nu sud} C^{\tL \tL,x*}_{\nu sud} \big)
- 0.032 \,\Re\big( C^{\tL \tL,x}_{\nu dsu} C^{\tL \tL,x*}_{\nu sud} \big),
\\
\frac{\Gamma_{n \to \bar\nu_x \pi^- K^+} }{(0.1\, \rm GeV)^5}
=&~ 0.147\big|C^{\tL \tR,x}_{\nu uds}\big|^2 
+ 0.054\big|C^{\tL \tR,x}_{\nu dsu}\big|^2 
+ 0.146\big|C^{\tL \tR,x}_{\nu sud}\big|^2 
+ 0.025\big|C^{\tL \tL,x}_{\nu dsu}\big|^2
\nonumber\\
&+ 0.149\big|C^{\tL \tL,x}_{\nu sud}\big|^2 
+ 0.099 \,\Re\big( C^{\tL \tR,x}_{\nu uds} C^{\tL \tR,x*}_{\nu dsu} \big) 
+0.22 \,\Re\big( C^{\tL \tR,x}_{\nu uds} C^{\tL \tR,x*}_{\nu sud} \big) 
\nonumber\\
&- 0.025 \,\Re\big( C^{\tL \tR,x}_{\nu dsu} C^{\tL \tR,x*}_{\nu sud} \big) 
- 0.065 \,\Re\big( C^{\tL \tR,x}_{\nu uds} C^{\tL \tL,x*}_{\nu dsu} \big)
- 0.22 \,\Re\big( C^{\tL \tR,x}_{\nu uds} C^{\tL \tL,x*}_{\nu sud} \big)
\nonumber\\
&- 0.073 \,\Re\big( C^{\tL \tR,x}_{\nu dsu} C^{\tL \tL,x*}_{\nu dsu} \big) 
+ 0.025 \,\Re\big( C^{\tL \tR,x}_{\nu dsu} C^{\tL \tL,x*}_{\nu sud} \big)
+ 0.019 \,\Re\big( C^{\tL \tR,x}_{\nu sud} C^{\tL \tL,x*}_{\nu dsu} \big)
\nonumber\\
&- 0.29 \,\Re\big( C^{\tL \tR,x}_{\nu sud} C^{\tL \tL,x*}_{\nu sud} \big)
- 0.02 \,\Re\big( C^{\tL \tL,x}_{\nu dsu} C^{\tL \tL,x*}_{\nu sud} \big),
\\
\frac{\Gamma_{n \to \bar\nu_x \pi^0 K^0} }{(0.1\, \rm GeV)^5}
=&~ 0.018\big|C^{\tL \tR,x}_{\nu uds}\big|^2 
+ 0.11\big|C^{\tL \tR,x}_{\nu dsu}\big|^2 
+ 0.073\big|C^{\tL \tR,x}_{\nu sud}\big|^2 
+ 0.047\big|C^{\tL \tL,x}_{\nu dsu}\big|^2
\nonumber\\
&+ 0.074\big|C^{\tL \tL,x}_{\nu sud}\big|^2 
+ 0.021 \,\Re\big( C^{\tL \tR,x}_{\nu uds} C^{\tL \tR,x*}_{\nu dsu} \big) 
- 0.036 \,\Re\big( C^{\tL \tR,x}_{\nu uds} C^{\tL \tR,x*}_{\nu sud} \big) 
\nonumber\\
&+ 0.13 \,\Re\big( C^{\tL \tR,x}_{\nu dsu} C^{\tL \tR,x*}_{\nu sud} \big)
- 0.024 \,\Re\big( C^{\tL \tR,x}_{\nu uds} C^{\tL \tL,x*}_{\nu dsu} \big)
+ 0.036 \,\Re\big( C^{\tL \tR,x}_{\nu uds} C^{\tL \tL,x*}_{\nu sud} \big) 
\nonumber\\
&+ 0.11 \,\Re\big( C^{\tL \tR,x}_{\nu dsu} C^{\tL \tL,x*}_{\nu dsu} \big) 
- 0.13 \,\Re\big( C^{\tL \tR,x}_{\nu dsu} C^{\tL \tL,x*}_{\nu sud} \big)
+ 0.12 \,\Re\big( C^{\tL \tR,x}_{\nu sud} C^{\tL \tL,x*}_{\nu dsu} \big)
\nonumber\\
&- 0.15 \,\Re\big( C^{\tL \tR,x}_{\nu sud} C^{\tL \tL,x*}_{\nu sud} \big)
- 0.12 \,\Re\big( C^{\tL \tL,x}_{\nu dsu} C^{\tL \tL,x*}_{\nu sud} \big).
\end{align}
\end{subequations}
For those modes with a neutrino in the final state, the corresponding expressions can be directly obtained from the above by
simultaneously interchanging $\nu\leftrightarrow\bar\nu$ and $\tL\leftrightarrow \tR$.

\section{Estimate of contributions due to octet vector mesons}
\label{app:vectorM}

\begin{figure}[t]
\centering
\includegraphics[width=0.3\textwidth]{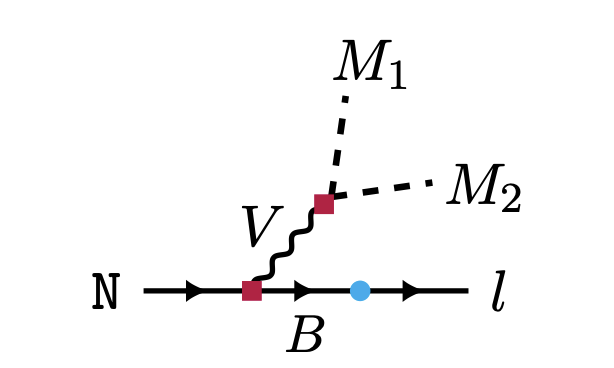}
\caption{Feynman diagram for nucleon three-body decays mediated by vector mesons. The cyan blob represents the BNV interactions in \cref{eq:VertexBl}, while the purple squares denote interactions between vector mesons and octet baryons or pseudoscalar mesons.}
\label{fig:Feyndiagram_vectormeson}
\end{figure}

In this appendix, we provide a detailed estimation for contributions mediated by octet vector mesons. 
In the resonance ChPT, the vector mesons couple to both a pair of pseudoscalar mesons~\cite{Ecker:1989yg} and a pair of baryons~\cite{Unal:2015hea}.
Combining with the BNV baryon-lepton mixing vertices in \cref{eq:VertexBl}, the $\rho$ and $K^*$ mesons can generate additional contributions to the three-body modes involving two different pions or a pion and a kaon.
The corresponding Feynman diagram is shown in \cref{fig:Feyndiagram_vectormeson}. 
In this estimation, we omit contributions from BNV baryon-lepton-vector-meson interactions in the local
chiral Lagrangian involving vector mesons~\cite{Liao:2025sqt}, as the relevant LECs are currently unknown. These contributions are expected to be no larger than those shown in \cref{fig:Feyndiagram_vectormeson}. 

We organize the octet vector mesons in matrix form,
\begin{align}
V_\mu(x) &=
\begin{pmatrix}
\frac{\rho_\mu^{0}}{\sqrt{2}}+\frac{\phi_\mu^{(8)}}{\sqrt{6}} & \rho_\mu^+ & K_\mu^{*+} \\
\rho_\mu^- & -\frac{\rho_\mu^{0}}{\sqrt{2}}+\frac{\phi_\mu^{(8)}}{\sqrt{6}} & K_\mu^{*0} \\ 
K_\mu^{*-} & \bar K_\mu^{*0} & - \sqrt{\frac{2}{3}}\phi_\mu^{(8)}
\end{pmatrix},    
\end{align}
then the leading $p^3$-order Lagrangian between $V_\mu$ and the pseudoscalar mesons takes the form~\cite{Ecker:1989yg}
\begin{align}
\label{eq:LVM}
{\cal L}_{V\text{-}M}^{(3)} = \frac{-i}{2\sqrt{2}} g_V 
\tr (V_{\mu\nu}  [ u^\mu,u^\nu]),
\end{align}
where $V_{\mu\nu}= \nabla_\mu V_\nu - \nabla_\nu V_\mu$ with $\nabla_\mu = \partial_\mu +[\Gamma_\mu, \,\,\,]$.
The LEC $g_V=G_V/M_V$~\cite{Ecker:1989yg}, where $M_V$ denotes the vector meson mass in the chiral limit and $G_V$ is the coupling constant in the tensor-field scheme.
Numerically, we approximate $M_V$ by the $\rho$-meson mass, $M_V \simeq M_\rho$,  
and adopt the value $G_V=66\,\text{MeV}$~\cite{Ecker:1989yg}.
By expanding \cref{eq:LVM} to the second order in the pseudoscalar
meson fields, we obtain the following relevant terms
\begin{align}
\label{eq:L_VMM}
{\cal L}_{VMM} \supset & -i\frac{g_V}{F_0^2} \Big\{
2 (\partial^\mu \pi^+ \partial^\nu \pi^- - \partial^\mu \pi^- \partial^\nu \pi^+)    \partial_\mu \rho^0_\nu  
\nonumber
\\
&+ 2 (\partial^\mu \pi^0 \partial^\nu \pi^+ - \partial^\mu \pi^+ \partial^\nu \pi^0) \partial_\mu \rho^-_\nu  
+ 2 (\partial^\mu \pi^- \partial^\nu \pi^0 - \partial^\mu \pi^0 \partial^\nu \pi^-)  \partial_\mu \rho^+_\nu 
\nonumber
\\
&+ \big[ (\partial^\mu K^- \partial^\nu \pi^0 - \partial^\mu \pi^0 \partial^\nu K^-)  
+ \sqrt{2} (\partial^\mu \bar{K}^0 \partial^\nu \pi^- - \partial^\mu \pi^- \partial^\nu \bar{K}^0)  \big]  \partial_\mu K^{*+}_\nu
\nonumber
\\
&+  \big[ (\partial^\mu \pi^0 \partial^\nu \bar{K}^0 - \partial^\mu \bar{K}^0 \partial^\nu \pi^0)  
+ \sqrt{2} (\partial^\mu K^- \partial^\nu \pi^+ - \partial^\mu \pi^+ \partial^\nu K^-) \big]  \partial_\mu K^{*0}_\nu
\Big\}.
\end{align}
Note that $K^{*-}$ and $\bar K^{*0}$ do not contribute to our calculation,
as strangeness conservation forces them to couple simultaneously to a $\pi/\eta$ and a $K^+/K^0$. In addition, $\phi_\mu^{(8)}$ is irrelevant, since it couples exclusively to a pair of kaons. 
This implies that there are no vector-meson corrections to $n\to \ell^+\pi^- K^0$ or $p\to \ell^-\pi^+ K^+$.

The leading-order baryon-vector-meson Lagrangian is~\cite{Unal:2015hea}, 
\begin{align}
{\cal L}_{\bar B B V} =
{\textsc g}_{\textsc d} \tr (\bar{B} \gamma_\mu \{ V^\mu,B\}) 
+ {\textsc g}_{\textsc f} \tr (\bar{B} \gamma_\mu [ V^\mu,B]),
\end{align}
where the curly and square brackets represent anticommutator and commutator between $V_\mu$ and $B$, respectively. 
For the LECs ${\textsc g}_{\textsc d}$ and ${\textsc g}_{\textsc f}$, we adopt the estimates given in~\cite{Unal:2015hea} with ${\textsc g}_{\textsc d}=0$ and ${\textsc g}_{\textsc f}=g/\sqrt{2}$, 
where the factor $1/\sqrt{2}$ accounts for different parameterizations of the interaction. $g$ is related to the $\rho$-meson mass and pion decay constant via the relation
$g^2=m_\rho^2/(2F_0^2)$~\cite{Kawarabayashi:1966kd,Riazuddin:1966sw}.
From the above Lagrangian, we obtain the following relevant vertices involving a nucleon and a vector meson 
\begin{align}
\label{eq:L_BNV}
{\cal L}_{\bar B{\tt N} V} \supset & {\textsc g}_{\textsc f}
\Big\{ \frac{1}{\sqrt{2}} (\bar{p} \gamma^\mu p) \rho_\mu^0 
+(\bar{n} \gamma^\mu p) \rho_\mu^-
- \frac{1}{\sqrt{2}} (\bar{n} \gamma^\mu n)  \rho_\mu^0
+(\bar{p} \gamma^\mu n) \rho_\mu^+
\nonumber\\
&-\frac{1}{\sqrt{2}} (\overline{\Sigma^0} \gamma^\mu p) K_\mu^{*-}
- \sqrt{\frac{3}{2}} (\overline{\Lambda^0} \gamma^\mu p) K_\mu^{*-}
-(\overline{\Sigma^+} \gamma^\mu p) \bar{K}_\mu^{*0}
\nonumber\\
&+\frac{1}{\sqrt{2}} (\overline{\Sigma^0} \gamma^\mu n) \bar{K}_\mu^{*0}
- \sqrt{\frac{3}{2}} (\overline{\Lambda^0} \gamma^\mu n) \bar{K}_\mu^{*0} 
-(\overline{\Sigma^-} \gamma^\mu n) K_\mu^{*-}
\Big\}.
\end{align}

Based on \cref{eq:VertexBl,eq:L_VMM,eq:L_BNV}, 
the vertices involved in \cref{fig:Feyndiagram_vectormeson} for a generic three-body nucleon decay ${\tt N}\to l M_1 M_2$ can be parameterized as follows,
\begin{align}
\label{eq:L2lM1M2_V}
{\cal L}_{{\tt N} \to l M_1 M_2} =\,&
\mathbb A_{{\tt N} B V} (\overline{B}\gamma^\mu {\tt N}) \bar V_\mu 
+ i \mathbb A_{ V M_1 M_2} \partial_\mu (\partial^\mu \bar M_1 \partial^\nu \bar M_2 - \partial^\mu \bar M_2 \partial^\nu \bar M_1)  V_\nu
\nonumber\\
&+ \overline{l}(\mathbb B_{Bl}^\tL P_\tL + \mathbb B_{Bl}^\tR P_\tR) B.
\end{align}
Here, for convenience the derivatives acting on the vector mesons in \cref{eq:L_VMM} have been transferred to the pseudoscalar part using integration by parts.
The coefficients for each relevant mode can then be extracted from \cref{eq:VertexBl,eq:L_VMM,eq:L_BNV} and are summarized in \cref{tab:coeff_v}.

\begin{table}[t]
\centering
\resizebox{\linewidth}{!}{
\renewcommand{\arraystretch}{1.2}
\begin{tabular}{|l|l|l|c|l|l|l|c|}
\hline
\quad~ Mode
& \multicolumn{1}{c|}{$\mathbb A_{{\tt N}BV}$} 
& \multicolumn{1}{c|}{$\mathbb A_{V M_1 M_2}$}  
& \multicolumn{1}{c|}{$\mathbb B_{B l}$} 
& \multicolumn{1}{c|}{$\rm Mode$}
& \multicolumn{1}{c|}{$\mathbb A_{{\tt N}BV}$} 
& \multicolumn{1}{c|}{$\mathbb A_{V M_1 M_2}$} 
& \multicolumn{1}{c|}{$\mathbb B_{B l}$} 
\\
\hline
$p\to \ell^+ \pi^+ \pi^-$ 
& $\mathbb A_{pp\rho^0} = \frac{{\textsc g}_{\textsc f}}{\sqrt{2}}$
& $\mathbb A_{\rho^0\pi^+\pi^-} = \frac{-2g_V}{F_0^2}$
& $\mathbb B_{p\ell^+}$
& $n\to \ell^+ \pi^- \pi^0$
& $\mathbb A_{np\rho^-} = {\textsc g}_{\textsc f} $
& $\mathbb A_{\rho^- \pi^-\pi^0} =\frac{-2g_V}{F_0^2}$
& $\mathbb B_{p\ell^+}$
\\\hline
$p\to \ell^+ \pi^0 K^0$
& $\mathbb A_{p\Sigma^+ K^{*0}} = -{\textsc g}_{\textsc f}$
& $\mathbb A_{K^{*0}\pi^0K^0} = \frac{g_V}{F_0^2}$
& $\mathbb B_{\Sigma^+\ell^+}$
& $n\to \ell^- \pi^0 K^+$
& $\mathbb A_{n\Sigma^-K^{*+}} = - {\textsc g}_{\textsc f}$
& $\mathbb A_{K^{*+}\pi^0K^+} = \frac{-g_V}{F_0^2}$
& $\mathbb B_{\Sigma^-\ell^-}$
\\\hline
$p\to \ell^+ \pi^- K^+$
& $\mathbb A_{p\Sigma^+ K^{*0}} = -{\textsc g}_{\textsc f}$
& $\mathbb A_{K^{*0}\pi^-K^+} = \frac{-\sqrt{2}g_V}{F_0^2}$
& $\mathbb B_{\Sigma^+\ell^+}$
& $n\to \ell^- \pi^+ K^0$
& $\mathbb A_{n\Sigma^-K^{*+}} = - {\textsc g}_{\textsc f}$
& $\mathbb A_{K^{*+}\pi^+K^0} = \frac{-\sqrt{2}g_V}{F_0^2}$
& $\mathbb B_{\Sigma^-\ell^-}$
\\\hline
$p\to \hat \nu \pi^+ \pi^0$ 
& $\mathbb A_{pn\rho^+} = {\textsc g}_{\textsc f}$
& $\mathbb A_{\rho^+\pi^+\pi^0} = \frac{2g_V}{F_0^2}$
& $\mathbb B_{n\hat \nu}$
& $n\to \hat \nu \pi^+ \pi^-$
& $\mathbb A_{nn\rho^0}= \frac{-{\textsc g}_{\textsc f}}{\sqrt{2}}$
& $\mathbb A_{\rho^0\pi^+\pi^-} = \frac{-2g_V}{F_0^2}$
& $\mathbb B_{n \hat \nu}$
\\\hline
$p\to \hat \nu \pi^0 K^+$
& \makecell[l]{ $\mathbb A_{p\Sigma^0 K^{*+}} = \frac{-{\textsc g}_{\textsc f}}{\sqrt{2}}$ \\
$\mathbb A_{p\Lambda^0 K^{*+}} = -\sqrt{\frac{3}{2}}{\textsc g}_{\textsc f}$ }
& $\mathbb A_{K^{*+}\pi^0 K^+} = \frac{-g_V}{F_0^2}$
& \makecell[l]{ $\mathbb B_{\Sigma^0 \hat \nu}$ \\
$\mathbb B_{\Lambda^0 \hat \nu}$ }
& $n\to \hat \nu \pi^0 K^0$
& \makecell[l]{ $\mathbb A_{n\Sigma^0 K^{*0}} = \frac{{\textsc g}_{\textsc f}}{\sqrt{2}}$ \\
$\mathbb A_{n\Lambda^0 K^{*0}} = - \sqrt{\frac{3}{2}}{\textsc g}_{\textsc f}$ }
& $\mathbb A_{K^{*0}\pi^0 K^0} = \frac{g_V}{F_0^2}$
& \makecell[l]{ $\mathbb B_{\Sigma^0 \hat \nu}$ \\
$\mathbb B_{\Lambda^0 \hat \nu}$ }
\\\hline
$p\to \hat \nu \pi^+ K^0$
& \makecell[l]{ $\mathbb A_{p\Sigma^0 K^{*+}} = \frac{-{\textsc g}_{\textsc f}}{\sqrt{2}}$ \\
$\mathbb A_{p\Lambda^0 K^{*+}} = -\sqrt{\frac{3}{2}}{\textsc g}_{\textsc f}$ }
& $\mathbb A_{K^{*+}\pi^+ K^0} = \frac{-\sqrt{2}g_V}{F_0^2}$
& \makecell[l]{ $\mathbb B_{\Sigma^0 \hat \nu}$ \\
$\mathbb B_{\Lambda^0 \hat \nu}$ }
& $n\to \hat \nu \pi^- K^+$
& \makecell[l]{ $\mathbb A_{n\Sigma^0 K^{*0}} = \frac{{\textsc g}_{\textsc f}}{\sqrt{2}}$ \\
$\mathbb A_{n\Lambda^0 K^{*0}} = - \sqrt{\frac{3}{2}}{\textsc g}_{\textsc f}$ }
& $\mathbb A_{K^{*0}\pi^- K^+} = \frac{-\sqrt{2}g_V}{F_0^2}$
& \makecell[l]{ $\mathbb B_{\Sigma^0 \hat \nu}$ \\
$\mathbb B_{\Lambda^0 \hat \nu}$ }
\\\hline
\end{tabular}}
\caption{The relevant couplings for vector-meson–mediated three-body modes.}
\label{tab:coeff_v}
\end{table}

The amplitude for \cref{fig:Feyndiagram_vectormeson} takes the form:
\begin{align}
{\cal M}^{(V)} =\,&  \overline{u_l}(\mathbb B_{Bl}^\tL P_\tL + \mathbb B_{Bl}^\tR P_\tR)
\frac{\slashed{k}+m_{B}}{m_{B}^2-k^2}
\mathbb A_{{\tt N} B V} \gamma^\mu
 u_{\tt N}
\big[ (m_1^2 + p_1\cdot p_2) p_2^\nu - (m_2^2 + p_1\cdot p_2) p_1^\nu \big]
\notag\\
& \times 
\mathbb A_{{V} M_1 M_2}
\frac{1}{s-m_{V}^2+im_{V} \Gamma_{V} } \left(g_{\mu\nu} - \frac{(p_1+p_2)_\mu (p_1+p_2)_\nu}{m_{V}^2} \right),
\label{eq:amp_v}
\end{align} 
where $\Gamma_V$ represents the decay width of the vector meson. 
Numerically, $\Gamma_\rho = 149.1 \pm 0.8 \,{\rm MeV}$  and $\Gamma_{K^*} = 52\,{\rm MeV}$~\cite{ParticleDataGroup:2024cfk}\footnote{The $T$‑matrix pole position of the $K^*$ meson is reported as $(890 \pm 14) - i (26 \pm 6)$ MeV~\cite{ParticleDataGroup:2024cfk}. The real part corresponds to the mass, while minus twice the imaginary part gives the width.
In our calculation, We adopt the central values $m_{K^*} = 890\,\rm MeV$  and $\Gamma_{K^*} = 52\,\rm MeV$.}. 
Using the properties of Dirac gamma matrices together with the equations of motion, the amplitude in \cref{eq:amp_v} reduces to the same form as \cref{eq:amp3body}, with the following additions to the $\mathbb D_i$-coefficients,
\begin{subequations}
\label{eq:deltaD_vectormeson}
\begin{align}
\delta\mathbb D_{\texttt{N};l M_1M_2}^{{\tt S}\tL} =\,&
\frac{\mathbb A_{{\tt N} B V} \mathbb A_{V M_1 M_2}  (m_1^2 - m_2^2) }{2(m_{B}^2 - m_l^2) (s - m_V^2 +i m_V\Gamma_V)}
\nonumber\\
&\times \Big[ m_l(m_\texttt{N} -m_{B})\mathbb B_{B l}^\tR + (m_B m_\texttt{N} -m_l^2) \mathbb B_{B l}^\tL \Big],
\\
\delta\mathbb D_{\texttt{N};l M_1M_2}^{{\tt S}\tR} =\,&
\frac{\mathbb A_{{\tt N} B V} \mathbb A_{V M_1 M_2} (m_1^2 - m_2^2)}{2(m_{B}^2 - m_l^2) (s - m_V^2 +i m_V \Gamma_V)}
\nonumber\\
&\times \Big[ m_l(m_\texttt{N} -m_{B})\mathbb B_{B l}^\tL + (m_B m_\texttt{N} -m_l^2) \mathbb B_{B l}^\tR \Big] ,
\\
\delta\mathbb  D_{\texttt{N};l M_1M_2}^{{\tt V}\tL} =\,& 
\frac{\mathbb A_{{\tt N} B V} \mathbb A_{ V M_1 M_2}}{2(m_{B}^2 - m_l^2) (s - m_V^2 +i m_V\Gamma_V)}
\Big[ m_l \mathbb B_{B l}^\tL + m_B \mathbb B_{B l}^\tR \Big] s ,
\\
\delta\mathbb  D_{\texttt{N};l M_1M_2}^{{\tt V}\tR } =\,& 
\frac{\mathbb A_{{\tt N} B V} \mathbb A_{V M_1 M_2}}{2(m_{B}^2 - m_l^2) (s - m_V^2 +i m_V \Gamma_V)}
\Big[ m_l \mathbb B_{B l}^\tR + m_B \mathbb B_{B l}^\tL \Big] s.
\end{align}
\end{subequations}

Adding \cref{eq:deltaD_vectormeson} to \cref{eq:Dcoeffs} and substituting the result into \cref{eq:Msquared},
we obtain the full squared matrix element  including the vector-meson contributions. The corresponding decay widths are then computed by performing the phase space integration.  
For the three-body charged lepton modes involving two nonidentical pions or a pion plus a kaon, the updated results are:
\begin{subequations}
\begin{align}
\frac{\Gamma_{p \to e^+ \pi^+\pi^-} }{(0.1\, \rm GeV)^5} 
=&~ 3.34 \big|C^{\tL\tR,e}_{\ell uud}\big|^2 + 
3.4 \big|C^{\tL \tL,e}_{\ell uud}\big|^2 
-6.74\,\Re\big( C^{\tL \tR,e}_{\ell uud} C^{\tL \tL,e*}_{\ell uud} \big)
+ 0.0026 C^{\tL \tR,e}_{\ell uud} C^{\tR \tL,e*}_{\ell uud}
\nonumber\\
& -0.0026 \big( C^{\tL \tR,e}_{\ell uud} C^{\tR \tR,e*}_{\ell uud} + 
C^{\tL \tL,e}_{\ell uud} C^{\tR \tL,e*}_{\ell uud} 
- C^{\tL \tL,e}_{\ell uud} C^{\tR \tR,e*}_{\ell uud} \big)
+\tL \leftrightarrow \tR,
\\
\frac{\Gamma_{p \to \mu^+ \pi^+\pi^-} }{(0.1\, \rm GeV)^5} 
=&~ 2.96 \big|C^{\tL\tR,\mu}_{\ell uud}\big|^2 + 
3.02 \big|C^{\tL \tL,\mu}_{\ell uud}\big|^2 
-5.98\,\Re\big( C^{\tL \tR,\mu}_{\ell uud} C^{\tL \tL,\mu*}_{\ell uud} \big)
+ 0.46 C^{\tL \tR,\mu}_{\ell uud} C^{\tR \tL,\mu*}_{\ell uud}
\nonumber\\
& -0.46 \big( C^{\tL \tR,\mu}_{\ell uud} C^{\tR \tR,\mu*}_{\ell uud} + C^{\tL \tL,\mu}_{\ell uud} C^{\tR \tL,\mu*}_{\ell uud} \big)
+ 0.47 C^{\tL \tL,\mu}_{\ell uud} C^{\tR \tR,\mu*}_{\ell uud}
+ \tL \leftrightarrow \tR,
\\
\frac{\Gamma_{n \to e^+ \pi^-\pi^0} }{(0.1\, \rm GeV)^5} 
=&~ 5.22 \big|C^{\tL\tR,e}_{\ell uud}\big|^2 
+ 5.32 \big|C^{\tL \tL,e}_{\ell uud}\big|^2 
-10.5 \,\Re\big( C^{\tL \tR,e}_{\ell uud} C^{\tL \tL,e*}_{\ell uud} \big)
+ 0.0048 C^{\tL \tR,e}_{\ell uud} C^{\tR \tL,e*}_{\ell uud}
\nonumber\\
& -0.0048 \big( C^{\tL \tR,e}_{\ell uud} C^{\tR \tR,e*}_{\ell uud} + 
C^{\tL \tL,e}_{\ell uud} C^{\tR \tL,e*}_{\ell uud}  \big)
+0.0049 C^{\tL \tL,e}_{\ell uud} C^{\tR \tR,e*}_{\ell uud}
+\tL \leftrightarrow \tR,
\\
\frac{\Gamma_{n \to \mu^+ \pi^-\pi^0} }{(0.1\, \rm GeV)^5} 
=&~ 4.59 \big|C^{\tL\tR,\mu}_{\ell uud}\big|^2 
+ 4.68 \big|C^{\tL \tL,\mu}_{\ell uud}\big|^2 
-9.28\,\Re\big( C^{\tL \tR,\mu}_{\ell uud} C^{\tL \tL,\mu*}_{\ell uud} \big)
+ 0.85 C^{\tL \tR,\mu}_{\ell uud} C^{\tR \tL,\mu*}_{\ell uud}
\nonumber\\
& -0.85 \big( C^{\tL \tR,\mu}_{\ell uud} C^{\tR \tR,\mu*}_{\ell uud} + C^{\tL \tL,\mu}_{\ell uud} C^{\tR \tL,\mu*}_{\ell uud} \big)
+ 0.86 C^{\tL \tL,\mu}_{\ell uud} C^{\tR \tR,\mu*}_{\ell uud}
+ \tL \leftrightarrow \tR,
\\
\frac{\Gamma_{p \to e^+ \pi^0 K^0} }{(0.1\, \rm GeV)^5} 
=&~ 0.11 \big|C^{\tL\tR,e}_{\ell usu}\big|^2 + 
0.11 \big|C^{\tL \tL,e}_{\ell usu}\big|^2 
+0.048 \,\Re\big( C^{\tL \tR,e}_{\ell usu} C^{\tL \tL,e*}_{\ell usu} \big)
-0.035 \,\Im\big( C^{\tL \tR,e}_{\ell usu} C^{\tL \tL,e*}_{\ell usu} \big)
\nonumber\\
& +4.16 \cdot 10^{-4} C^{\tL \tR,e}_{\ell usu} C^{\tR \tL,e*}_{\ell usu}
-2.8  \cdot 10^{-4}  \,\Re\big( C^{\tL \tR,e}_{\ell usu} C^{\tR \tR,e*}_{\ell usu} \big)
\nonumber\\
& -2.5  \cdot 10^{-4}  \,\Im\big( C^{\tL \tR,e}_{\ell usu} C^{\tR \tR,e*}_{\ell usu} \big)
-7.4 \cdot 10^{-5} C^{\tL \tL,e}_{\ell usu} C^{\tR \tR,e*}_{\ell usu}
+\tL \leftrightarrow \tR,
\\
\frac{\Gamma_{p \to \mu^+ \pi^0 K^0} }{(0.1\, \rm GeV)^5} 
=&~ 0.052 \big|C^{\tL\tR,\mu}_{\ell usu}\big|^2 + 
0.069 \big|C^{\tL \tL,\mu}_{\ell usu}\big|^2 
+0.067 \,\Re\big( C^{\tL \tR,\mu}_{\ell usu} C^{\tL \tL,\mu*}_{\ell usu} \big)
\nonumber\\
& -0.0063 \,\Im\big( C^{\tL \tR,\mu}_{\ell usu} C^{\tL \tL,\mu*}_{\ell usu} \big)
+0.024 C^{\tL \tR,\mu}_{\ell usu} C^{\tR \tL,\mu*}_{\ell usu}
+0.0044  \,\Re\big( C^{\tL \tR,\mu}_{\ell usu} C^{\tR \tR,\mu*}_{\ell usu} \big)
\nonumber\\
& -0.0041  \,\Im\big( C^{\tL \tR,\mu}_{\ell usu} C^{\tR \tR,\mu*}_{\ell usu} \big)
-0.028 C^{\tL \tL,\mu}_{\ell usu} C^{\tR \tR,\mu*}_{\ell usu}
+\tL \leftrightarrow \tR,
\\
\frac{\Gamma_{p \to e^+ \pi^- K^+} }{(0.1\, \rm GeV)^5} 
=&~ 0.053 \big|C^{\tL\tR,e}_{\ell usu}\big|^2 + 
0.073 \big|C^{\tL \tL,e}_{\ell usu}\big|^2 
-0.12 \,\Re\big( C^{\tL \tR,e}_{\ell usu} C^{\tL \tL,e*}_{\ell usu} \big)
\nonumber\\
& +0.0058 \,\Im\big( C^{\tL \tR,e}_{\ell usu} C^{\tL \tL,e*}_{\ell usu} \big)
+2.9 \cdot 10^{-4} C^{\tL \tR,e}_{\ell usu} C^{\tR \tL,e*}_{\ell usu}
-5.8  \cdot 10^{-4}  \,\Re\big( C^{\tL \tR,e}_{\ell usu} C^{\tR \tR,e*}_{\ell usu} \big)
\nonumber\\
& +4.27  \cdot 10^{-5}  \,\Im\big( C^{\tL \tR,e}_{\ell usu} C^{\tR \tR,e*}_{\ell usu} \big)
+2.9 \cdot 10^{-4} C^{\tL \tL,e}_{\ell usu} C^{\tR \tR,e*}_{\ell usu}
+\tL \leftrightarrow \tR,
\\
\frac{\Gamma_{p \to \mu^+ \pi^- K^+} }{(0.1\, \rm GeV)^5} 
=&~ 0.01 \big|C^{\tL\tR,\mu}_{\ell usu}\big|^2 + 
0.02 \big|C^{\tL \tL,\mu}_{\ell usu}\big|^2 
-0.029 \,\Re\big( C^{\tL \tR,\mu}_{\ell usu} C^{\tL \tL,\mu*}_{\ell usu} \big)
\nonumber\\
& +9.06 \cdot 10^{-4}\,\Im\big( C^{\tL \tR,\mu}_{\ell usu} C^{\tL \tL,\mu*}_{\ell usu} \big)
-0.004 C^{\tL \tR,\mu}_{\ell usu} C^{\tR \tL,\mu*}_{\ell usu}
+0.01  \,\Re\big( C^{\tL \tR,\mu}_{\ell usu} C^{\tR \tR,\mu*}_{\ell usu} \big)
\nonumber\\
& +6  \cdot 10^{-4}  \,\Im\big( C^{\tL \tR,\mu}_{\ell usu} C^{\tR \tR,\mu*}_{\ell usu} \big)
-0.0064 C^{\tL \tL,\mu}_{\ell usu} C^{\tR \tR,\mu*}_{\ell usu}
+\tL \leftrightarrow \tR,
\\
\frac{\Gamma_{n \to e^- \pi^+ K^0} }{(0.1\, \rm GeV)^5} 
=&~0.053 \big|C^{\tL\tR,e}_{\bar \ell dds}\big|^2 + 
0.072 \big|C^{\tL \tL,e}_{\bar \ell dds}\big|^2 
-0.12 \,\Re\big( C^{\tL \tR,e}_{\bar \ell dds} C^{\tL \tL,e*}_{\bar \ell dds} \big)
+0.0058 \,\Im\big( C^{\tL \tR,e}_{\bar \ell dds} C^{\tL \tL,e*}_{\bar \ell dds} \big)
\nonumber\\
& +2.9 \cdot 10^{-4} C^{\tL \tR,e}_{\bar \ell dds} C^{\tR \tL,e*}_{\bar \ell dds}
- 5.77 \cdot 10^{-4} \,\Re\big( C^{\tL \tR,e}_{\bar \ell dds} C^{\tR \tR,e*}_{\bar \ell dds} \big)
\nonumber\\
& + 4.2 \cdot 10^{-5} \,\Im\big( C^{\tL \tR,e}_{\bar \ell dds} C^{\tR \tR,e*}_{\bar \ell dds} \big)
+2.88 \cdot 10^{-4} C^{\tL \tL,e}_{\bar \ell dds} C^{\tR \tR,e*}_{\bar \ell dds}
+\tL \leftrightarrow \tR,
\\
\frac{\Gamma_{n \to \mu^- \pi^+ K^0} }{(0.1\, \rm GeV)^5} 
=&~0.0097 \big|C^{\tL\tR,\mu}_{\bar \ell dds}\big|^2 + 
0.021 \big|C^{\tL \tL,\mu}_{\bar \ell dds}\big|^2 
-0.028 \,\Re\big( C^{\tL \tR,\mu}_{\bar \ell dds} C^{\tL \tL,\mu*}_{\bar \ell dds} \big)
\nonumber\\
& +8.9 \cdot 10^{-4}\,\Im\big( C^{\tL \tR,\mu}_{\bar \ell dds} C^{\tL \tL,\mu*}_{\bar \ell dds} \big)
-0.0038 C^{\tL \tR,\mu}_{\bar \ell dds} C^{\tR \tL,\mu*}_{\bar \ell dds}
+ 0.01 \,\Re\big( C^{\tL \tR,\mu}_{\bar \ell dds} C^{\tR \tR,\mu*}_{\bar \ell dds} \big)
\nonumber\\
& + 5.9 \cdot 10^{-4} \,\Im\big( C^{\tL \tR,\mu}_{\bar \ell dds} C^{\tR \tR,\mu*}_{\bar \ell dds} \big)
-0.0062 C^{\tL \tL,\mu}_{\bar \ell dds} C^{\tR \tR,\mu*}_{\bar \ell dds}
+\tL \leftrightarrow \tR,
\\
\frac{\Gamma_{n \to e^- \pi^0 K^+} }{(0.1\, \rm GeV)^5} 
=&~0.11 \big|C^{\tL\tR,e}_{\bar \ell dds}\big|^2 + 
0.11 \big|C^{\tL \tL,e}_{\bar \ell dds}\big|^2 
+0.052 \,\Re\big( C^{\tL \tR,e}_{\bar \ell dds} C^{\tL \tL,e*}_{\bar \ell dds} \big)
-0.037 \,\Im\big( C^{\tL \tR,e}_{\bar \ell dds} C^{\tL \tL,e*}_{\bar \ell dds} \big)
\nonumber\\
& +4.3 \cdot 10^{-4} C^{\tL \tR,e}_{\bar \ell dds} C^{\tR \tL,e*}_{\bar \ell dds}
- 2.9 \cdot 10^{-4} \,\Re\big( C^{\tL \tR,e}_{\bar \ell dds} C^{\tR \tR,e*}_{\bar \ell dds} \big)
\nonumber\\
& -2.58 \cdot 10^{-4} \,\Im\big( C^{\tL \tR,e}_{\bar \ell dds} C^{\tR \tR,e*}_{\bar \ell dds} \big)
-7.5 \cdot 10^{-5} C^{\tL \tL,e}_{\bar \ell dds} C^{\tR \tR,e*}_{\bar \ell dds}
+\tL \leftrightarrow \tR,
\\
\frac{\Gamma_{n \to \mu^- \pi^0 K^+} }{(0.1\, \rm GeV)^5} 
=&~0.056 \big|C^{\tL\tR,\mu}_{\bar \ell dds}\big|^2 + 
0.074 \big|C^{\tL \tL,\mu}_{\bar \ell dds}\big|^2 
+0.073 \,\Re\big( C^{\tL \tR,\mu}_{\bar \ell dds} C^{\tL \tL,\mu*}_{\bar \ell dds} \big)
\nonumber\\
& -0.0068 \,\Im\big( C^{\tL \tR,\mu}_{\bar \ell dds} C^{\tL \tL,\mu*}_{\bar \ell dds} \big)
+0.026 C^{\tL \tR,\mu}_{\bar \ell dds} C^{\tR \tL,\mu*}_{\bar \ell dds}
+ 0.0047 \,\Re\big( C^{\tL \tR,\mu}_{\bar \ell dds} C^{\tR \tR,\mu*}_{\bar \ell dds} \big)
\nonumber\\
& -0.0044 \,\Im\big( C^{\tL \tR,\mu}_{\bar \ell dds} C^{\tR \tR,\mu*}_{\bar \ell dds} \big)
-0.03 C^{\tL \tL,\mu}_{\bar \ell dds} C^{\tR \tR,\mu*}_{\bar \ell dds}
+\tL \leftrightarrow \tR.
\end{align}
\end{subequations}

For the decay modes involving an antineutrino, the improved decay widths are:
\begin{subequations}
\begin{align}
\frac{\Gamma_{p \to \bar\nu_x \pi^+\pi^0} }{(0.1\, \rm GeV)^5} 
=&~ 5.153 \big|C^{\tL\tR,x}_{ \nu dud}\big|^2 
+ 5.252\big|C^{\tL \tL,x}_{\nu dud}\big|^2 
-10.4\,\Re\big( C^{\tL \tR,x}_{\nu dud} C^{\tL \tL,x*}_{\nu dud} \big) ,
\\
\frac{\Gamma_{n \to \bar\nu_x \pi^+\pi^-} }{(0.1\, \rm GeV)^5}
=&~3.387\big|C^{\tL \tR,x}_{\nu dud}\big|^2
+ 3.452\big|C^{\tL \tL,x}_{\nu dud}\big|^2 
- 6.838\,\Re\big( C^{\tL \tR,x}_{\nu dud} C^{\tL \tL,x*}_{\nu dud} \big),
\\
\frac{\Gamma_{p \to \bar\nu_x \pi^+K^0} }{(0.1\, \rm GeV)^5}
=&~0.134\big|C^{\tL \tR,x}_{\nu uds}\big|^2 
+ 0.136\big|C^{\tL \tR,x}_{\nu dsu}\big|^2 
+ 0.277\big|C^{\tL \tR,x}_{\nu sud}\big|^2 
+ 0.145\big|C^{\tL \tL,x}_{\nu dsu}\big|^2
\nonumber\\
&+ 0.282\big|C^{\tL \tL,x}_{\nu sud}\big|^2 
+ 0.063 \,\Re\big( C^{\tL \tR,x}_{\nu uds} C^{\tL \tR,x*}_{\nu dsu} \big)
+0.024 \,\Im\big( C^{\tL \tR,x}_{\nu uds} C^{\tL \tR,x*}_{\nu dsu} \big)
\nonumber\\
&-0.25 \,\Re\big( C^{\tL \tR,x}_{\nu uds} C^{\tL \tR,x*}_{\nu sud} \big) 
+0.033 \,\Im\big( C^{\tL \tR,x}_{\nu uds} C^{\tL \tR,x*}_{\nu sud} \big)
+ 0.23 \,\Re\big( C^{\tL \tR,x}_{\nu dsu} C^{\tL \tR,x*}_{\nu sud} \big) 
\nonumber\\
&+ 0.026 \,\Im\big( C^{\tL \tR,x}_{\nu dsu} C^{\tL \tR,x*}_{\nu sud} \big) 
- 0.02 \,\Re\big( C^{\tL \tR,x}_{\nu uds} C^{\tL \tL,x*}_{\nu dsu} \big)
+ 0.04 \,\Im\big( C^{\tL \tR,x}_{\nu uds} C^{\tL \tL,x*}_{\nu dsu} \big)
\nonumber\\
&+ 0.25 \,\Re\big( C^{\tL \tR,x}_{\nu uds} C^{\tL \tL,x*}_{\nu sud} \big)
-0.033 \,\Im\big( C^{\tL \tR,x}_{\nu uds} C^{\tL \tL,x*}_{\nu sud} \big)
+ 0.26 \,\Re\big( C^{\tL \tR,x}_{\nu dsu} C^{\tL \tL,x*}_{\nu dsu} \big) 
\nonumber\\
&+ 0.002 \,\Im\big( C^{\tL \tR,x}_{\nu dsu} C^{\tL \tL,x*}_{\nu dsu} \big) 
- 0.23 \,\Re\big( C^{\tL \tR,x}_{\nu dsu} C^{\tL \tL,x*}_{\nu sud} \big)
- 0.026 \,\Im\big( C^{\tL \tR,x}_{\nu dsu} C^{\tL \tL,x*}_{\nu sud} \big)
\nonumber\\
&+ 0.33 \,\Re\big( C^{\tL \tR,x}_{\nu sud} C^{\tL \tL,x*}_{\nu dsu} \big)
- 0.04 \,\Im\big( C^{\tL \tR,x}_{\nu sud} C^{\tL \tL,x*}_{\nu dsu} \big)
- 0.56 \,\Re\big( C^{\tL \tR,x}_{\nu sud} C^{\tL \tL,x*}_{\nu sud} \big)
\nonumber\\
&- 0.33 \,\Re\big( C^{\tL \tL,x}_{\nu dsu} C^{\tL \tL,x*}_{\nu sud} \big)
- 0.04 \,\Im\big( C^{\tL \tL,x}_{\nu dsu} C^{\tL \tL,x*}_{\nu sud} \big),
\\
\frac{\Gamma_{p \to \bar\nu_x \pi^0 K^+} }{(0.1\, \rm GeV)^5}
=&~0.119\big|C^{\tL \tR,x}_{\nu uds}\big|^2 
+ 0.02\big|C^{\tL \tR,x}_{\nu dsu}\big|^2 
+ 0.152\big|C^{\tL \tR,x}_{\nu sud}\big|^2 
+ 0.004\big|C^{\tL \tL,x}_{\nu dsu}\big|^2
\nonumber\\
&+ 0.155\big|C^{\tL \tL,x}_{\nu sud}\big|^2 
+ 0.042 \,\Re\big( C^{\tL \tR,x}_{\nu uds} C^{\tL \tR,x*}_{\nu dsu} \big)
-0.014 \,\Im\big( C^{\tL \tR,x}_{\nu uds} C^{\tL \tR,x*}_{\nu dsu} \big)
\nonumber\\
&+0.057 \,\Re\big( C^{\tL \tR,x}_{\nu uds} C^{\tL \tR,x*}_{\nu sud} \big) 
+0.044 \,\Im\big( C^{\tL \tR,x}_{\nu uds} C^{\tL \tR,x*}_{\nu sud} \big)
-0.061 \,\Re\big( C^{\tL \tR,x}_{\nu dsu} C^{\tL \tR,x*}_{\nu sud} \big) 
\nonumber\\
&- 0.013 \,\Im\big( C^{\tL \tR,x}_{\nu dsu} C^{\tL \tR,x*}_{\nu sud} \big) 
+0.0071 \,\Re\big( C^{\tL \tR,x}_{\nu uds} C^{\tL \tL,x*}_{\nu dsu} \big)
+ 0.0071 \,\Im\big( C^{\tL \tR,x}_{\nu uds} C^{\tL \tL,x*}_{\nu dsu} \big)
\nonumber\\
&-0.057 \,\Re\big( C^{\tL \tR,x}_{\nu uds} C^{\tL \tL,x*}_{\nu sud} \big)
-0.044 \,\Im\big( C^{\tL \tR,x}_{\nu uds} C^{\tL \tL,x*}_{\nu sud} \big)
-0.014 \,\Re\big( C^{\tL \tR,x}_{\nu dsu} C^{\tL \tL,x*}_{\nu dsu} \big) 
\nonumber\\
&- 2.2\cdot 10^{-4} \,\Im\big( C^{\tL \tR,x}_{\nu dsu} C^{\tL \tL,x*}_{\nu dsu} \big) 
+0.06 \,\Re\big( C^{\tL \tR,x}_{\nu dsu} C^{\tL \tL,x*}_{\nu sud} \big)
+ 0.013 \,\Im\big( C^{\tL \tR,x}_{\nu dsu} C^{\tL \tL,x*}_{\nu sud} \big)
\nonumber\\
&+ 0.046 \,\Re\big( C^{\tL \tR,x}_{\nu sud} C^{\tL \tL,x*}_{\nu dsu} \big)
- 0.0058 \,\Im\big( C^{\tL \tR,x}_{\nu sud} C^{\tL \tL,x*}_{\nu dsu} \big)
- 0.31 \,\Re\big( C^{\tL \tR,x}_{\nu sud} C^{\tL \tL,x*}_{\nu sud} \big)
\nonumber\\
&- 0.046 \,\Re\big( C^{\tL \tL,x}_{\nu dsu} C^{\tL \tL,x*}_{\nu sud} \big)
- 0.0059 \,\Im\big( C^{\tL \tL,x}_{\nu dsu} C^{\tL \tL,x*}_{\nu sud} \big),
\\
\frac{\Gamma_{n \to \bar\nu_x \pi^- K^+} }{(0.1\, \rm GeV)^5}
=&~0.145\big|C^{\tL \tR,x}_{\nu uds}\big|^2 
+ 0.143\big|C^{\tL \tR,x}_{\nu dsu}\big|^2 
+ 0.298\big|C^{\tL \tR,x}_{\nu sud}\big|^2 
+ 0.105\big|C^{\tL \tL,x}_{\nu dsu}\big|^2
\nonumber\\
&+ 0.304\big|C^{\tL \tL,x}_{\nu sud}\big|^2 
+ 0.065 \,\Re\big( C^{\tL \tR,x}_{\nu uds} C^{\tL \tR,x*}_{\nu dsu} \big)
-0.026 \,\Im\big( C^{\tL \tR,x}_{\nu uds} C^{\tL \tR,x*}_{\nu dsu} \big)
\nonumber\\
&+0.25 \,\Re\big( C^{\tL \tR,x}_{\nu uds} C^{\tL \tR,x*}_{\nu sud} \big) 
+0.028 \,\Im\big( C^{\tL \tR,x}_{\nu uds} C^{\tL \tR,x*}_{\nu sud} \big)
-0.26 \,\Re\big( C^{\tL \tR,x}_{\nu dsu} C^{\tL \tR,x*}_{\nu sud} \big) 
\nonumber\\
&+ 0.035 \,\Im\big( C^{\tL \tR,x}_{\nu dsu} C^{\tL \tR,x*}_{\nu sud} \big) 
-0.031 \,\Re\big( C^{\tL \tR,x}_{\nu uds} C^{\tL \tL,x*}_{\nu dsu} \big)
+ 0.026 \,\Im\big( C^{\tL \tR,x}_{\nu uds} C^{\tL \tL,x*}_{\nu dsu} \big)
\nonumber\\
&-0.25 \,\Re\big( C^{\tL \tR,x}_{\nu uds} C^{\tL \tL,x*}_{\nu sud} \big)
-0.028 \,\Im\big( C^{\tL \tR,x}_{\nu uds} C^{\tL \tL,x*}_{\nu sud} \big)
-0.24 \,\Re\big( C^{\tL \tR,x}_{\nu dsu} C^{\tL \tL,x*}_{\nu dsu} \big) 
\nonumber\\
&- 0.0073 \,\Im\big( C^{\tL \tR,x}_{\nu dsu} C^{\tL \tL,x*}_{\nu dsu} \big) 
+0.27 \,\Re\big( C^{\tL \tR,x}_{\nu dsu} C^{\tL \tL,x*}_{\nu sud} \big)
- 0.036 \,\Im\big( C^{\tL \tR,x}_{\nu dsu} C^{\tL \tL,x*}_{\nu sud} \big)
\nonumber\\
&+ 0.25 \,\Re\big( C^{\tL \tR,x}_{\nu sud} C^{\tL \tL,x*}_{\nu dsu} \big)
+ 0.043 \,\Im\big( C^{\tL \tR,x}_{\nu sud} C^{\tL \tL,x*}_{\nu dsu} \big)
- 0.6 \,\Re\big( C^{\tL \tR,x}_{\nu sud} C^{\tL \tL,x*}_{\nu sud} \big)
\nonumber\\
&- 0.25 \,\Re\big( C^{\tL \tL,x}_{\nu dsu} C^{\tL \tL,x*}_{\nu sud} \big)
+ 0.044 \,\Im\big( C^{\tL \tL,x}_{\nu dsu} C^{\tL \tL,x*}_{\nu sud} \big),
\\
\frac{\Gamma_{n \to \bar\nu_x \pi^0 K^0} }{(0.1\, \rm GeV)^5}
=&~0.019\big|C^{\tL \tR,x}_{\nu uds}\big|^2 
+ 0.116\big|C^{\tL \tR,x}_{\nu dsu}\big|^2 
+ 0.149\big|C^{\tL \tR,x}_{\nu sud}\big|^2 
+ 0.111\big|C^{\tL \tL,x}_{\nu dsu}\big|^2
\nonumber\\
&+ 0.151\big|C^{\tL \tL,x}_{\nu sud}\big|^2 
+ 0.041 \,\Re\big( C^{\tL \tR,x}_{\nu uds} C^{\tL \tR,x*}_{\nu dsu} \big)
+0.014 \,\Im\big( C^{\tL \tR,x}_{\nu uds} C^{\tL \tR,x*}_{\nu dsu} \big)
\nonumber\\
&-0.059 \,\Re\big( C^{\tL \tR,x}_{\nu uds} C^{\tL \tR,x*}_{\nu sud} \big) 
-0.013 \,\Im\big( C^{\tL \tR,x}_{\nu uds} C^{\tL \tR,x*}_{\nu sud} \big)
+0.052 \,\Re\big( C^{\tL \tR,x}_{\nu dsu} C^{\tL \tR,x*}_{\nu sud} \big) 
\nonumber\\
&+ 0.044 \,\Im\big( C^{\tL \tR,x}_{\nu dsu} C^{\tL \tR,x*}_{\nu sud} \big) 
-0.047 \,\Re\big( C^{\tL \tR,x}_{\nu uds} C^{\tL \tL,x*}_{\nu dsu} \big)
- 0.013 \,\Im\big( C^{\tL \tR,x}_{\nu uds} C^{\tL \tL,x*}_{\nu dsu} \big)
\nonumber\\
&+0.06 \,\Re\big( C^{\tL \tR,x}_{\nu uds} C^{\tL \tL,x*}_{\nu sud} \big)
+0.013 \,\Im\big( C^{\tL \tR,x}_{\nu uds} C^{\tL \tL,x*}_{\nu sud} \big)
+0.046 \,\Re\big( C^{\tL \tR,x}_{\nu dsu} C^{\tL \tL,x*}_{\nu dsu} \big) 
\nonumber\\
&+ 0.037 \,\Im\big( C^{\tL \tR,x}_{\nu dsu} C^{\tL \tL,x*}_{\nu dsu} \big) 
-0.053 \,\Re\big( C^{\tL \tR,x}_{\nu dsu} C^{\tL \tL,x*}_{\nu sud} \big)
- 0.044 \,\Im\big( C^{\tL \tR,x}_{\nu dsu} C^{\tL \tL,x*}_{\nu sud} \big)
\nonumber\\
&+ 0.25 \,\Re\big( C^{\tL \tR,x}_{\nu sud} C^{\tL \tL,x*}_{\nu dsu} \big)
+ 0.0059 \,\Im\big( C^{\tL \tR,x}_{\nu sud} C^{\tL \tL,x*}_{\nu dsu} \big)
- 0.3 \,\Re\big( C^{\tL \tR,x}_{\nu sud} C^{\tL \tL,x*}_{\nu sud} \big)
\nonumber\\
&- 0.26 \,\Re\big( C^{\tL \tL,x}_{\nu dsu} C^{\tL \tL,x*}_{\nu sud} \big)
+ 0.006 \,\Im\big( C^{\tL \tL,x}_{\nu dsu} C^{\tL \tL,x*}_{\nu sud} \big).
\end{align}
\end{subequations}
For those modes with a neutrino in the final state, the corresponding expressions can be obtained from the above by
simultaneously interchanging $\nu\leftrightarrow\bar\nu$ and $\tL\leftrightarrow \tR$.

The results above indicate that, for decay modes involving a pion and a kaon, an additional imaginary part of the WC product appears in their decay widths, originating from the imaginary part of the vector-meson propagator.
Moreover, a comparison with the corresponding results in appendix \ref{app:DeW_three_body} shows that the corrections due to vector‑meson contributions are at a level of about 20--30\,\% for modes involving either two nonidentical pions or a kaon and a $\mu^\pm$. In contrast, for modes involving a kaon and an $e^\pm$ or a (anti)neutrino, the corrections are generally larger, reaching roughly 1 to 5 times the contribution shown in \cref{fig:Feyndiagram}. 
This difference is the joint result of the vector kaon having a much smaller width than the rho meson, and the muon mass being significantly larger than that of the electron or neutrino.

\section{Decay width expressions for two-body modes ${\tt N}\to l M$}
\label{app:DeW_two_body}

For the two-body nucleon decay ${\tt N}\to lM$, the general expression for the decay width takes the form~\cite{Liao:2025sqt}, 
\begin{align}
\label{eq:GammaN2lM}
\Gamma_{{\tt N} \to l M} 
=& \frac{m_{\tt N} \lambda^{1/2}(1,x_l, x_M)} {32\pi } 
\Big[ (1 + x_l -x_M)
\big( | \mathbb  D_{{\tt N};lM}^\tL|^2 
+ | \mathbb  D_{{\tt N};lM}^\tR|^2\big) 
\nonumber\\
&+ 4 \sqrt{x_l} \Re(\mathbb  D_{{\tt N};lM}^\tL \mathbb  D_{{\tt N};lM}^{\tR,*}) \Big],
\end{align}
where $\lambda(x,y,z)\equiv x^2+y^2+z^2-2(xy+yz+zx)$ 
is the triangle function, $x_M= m_M^2/m_{\tt N}^2$, 
and $\mathbb D_{{\tt N};lM}^{\tL,\tR}$ are related to the parameters defined in \cref{eq:LN2lMM} via,
\begin{subequations}
\begin{align}
\mathbb  D_{{\tt N};lM}^\tL & 
= \mathbb  B_{{\tt N}lM}^\tL  
+ \frac{\mathbb A_{{\tt N}BM}}{m_B^2 - m_l^2} 
\left[ (m_{\tt N} m_B + m_l^2) \mathbb B_{Bl}^\tL + m_l(m_{\tt N} +m_B) \mathbb B_{Bl}^\tR  \right] ,
\\
\mathbb  D_{{\tt N};lM}^\tR & 
= \mathbb  B_{{\tt N}lM}^\tR  
- \frac{ \mathbb A_{{\tt N}BM} }{m_B^2 - m_l^2} 
\left[ (m_{\tt N} m_B + m_l^2) \mathbb B_{Bl}^\tR + m_l(m_{\tt N} +m_B) \mathbb B_{Bl}^\tL  \right],
\end{align}
\end{subequations}
where summation over the intermediate baryon $B$ is implied. 
For each specific decay mode ${\tt N}\to l M$, the relevant quantities $\mathbb A_{{\tt N}BM}$, $\mathbb B_{Bl}^{\tL(\tR)}$, and $\mathbb B_{{\tt N}lM}^{\tL(\tR)}$ can be found in Table S3 of~\cite{Liao:2025sqt}, where they correspond to $C_{{\tt N}\to BM}/F_0$, $C_{Bl}^{\tL(\tR)}$, and $C_{{\tt N}lM}^{\tL(\tR)}/F_0$, respectively.

\subsection{Decay modes involving a charged lepton}
For the two-body modes with a charged lepton, the decay widths expressed in terms of LEFT WCs are
\begin{subequations}
\begin{align}
\frac{\Gamma_{p \to e^+ \pi^0} }{(0.1\, \rm GeV)^5} 
=&~ 19.7 \big|C^{\tL\tR,e}_{\ell uud}\big|^2 + 
20.1 \big|C^{\tL \tL,e}_{\ell uud}\big|^2 
-39.8 \,\Re\big( C^{\tL \tR,e}_{\ell uud} C^{\tL \tL,e*}_{\ell uud} \big)
-0.0013 C^{\tL \tR,e}_{\ell uud} C^{\tR \tL,e*}_{\ell uud}
\nonumber\\
& +0.0013 \big( C^{\tL \tR,e}_{\ell uud} C^{\tR \tR,e*}_{\ell uud} + 
C^{\tL \tL,e}_{\ell uud} C^{\tR \tL,e*}_{\ell uud} 
- C^{\tL \tL,e}_{\ell uud} C^{\tR \tR,e*}_{\ell uud} \big)
+\tL \leftrightarrow \tR,
\\
\frac{\Gamma_{p \to \mu^+ \pi^0} }{(0.1\, \rm GeV)^5} 
=&~19.4\big|C^{\tL\tR,\mu}_{\ell uud}\big|^2 
+ 19.8\big|C^{\tL \tL,\mu}_{\ell uud}\big|^2 
-39.2\,\Re\big( C^{\tL \tR,\mu}_{\ell uud} C^{\tL \tL,\mu*}_{\ell uud} \big)
-0.26 C^{\tL \tR,\mu}_{\ell uud} C^{\tR \tL,\mu*}_{\ell uud}
\nonumber\\
& +0.26 \big( C^{\tL \tR,\mu}_{\ell uud} C^{\tR \tR,\mu*}_{\ell uud} + C^{\tL \tL,\mu}_{\ell uud} C^{\tR \tL,\mu*}_{\ell uud} \big)
- 0.27 C^{\tL \tL,\mu}_{\ell uud} C^{\tR \tR,\mu*}_{\ell uud}
+ \tL \leftrightarrow \tR,
\\
\frac{\Gamma_{p \to e^+ \eta} }{(0.1\, \rm GeV)^5} 
=&~0.095 \big|C^{\tL\tR,e}_{\ell uud}\big|^2 
+ 8.3 \big|C^{\tL \tL,e}_{\ell uud}\big|^2 
+1.8 \,\Re\big( C^{\tL \tR,e}_{\ell uud} C^{\tL \tL,e*}_{\ell uud} \big)
+4.8 \cdot 10^{-4} C^{\tL \tR,e}_{\ell uud} C^{\tR \tL,e*}_{\ell uud}
\nonumber\\
& +0.0028 \big( C^{\tL \tR,e}_{\ell uud} C^{\tR \tR,e*}_{\ell uud} + 
C^{\tL \tL,e}_{\ell uud} C^{\tR \tL,e*}_{\ell uud} \big)
+ 0.011 C^{\tL \tL,e}_{\ell uud} C^{\tR \tR,e*}_{\ell uud}
+\tL \leftrightarrow \tR,
\\
\frac{\Gamma_{p \to \mu^+ \eta} }{(0.1\, \rm GeV)^5} 
=&~ 0.12 \big|C^{\tL\tR,\mu}_{\ell uud}\big|^2 
+ 8.03 \big|C^{\tL \tL,\mu}_{\ell uud}\big|^2 
+ 1.8 \,\Re\big( C^{\tL \tR,\mu}_{\ell uud} C^{\tL \tL,\mu*}_{\ell uud} \big)
+0.097 C^{\tL \tR,\mu}_{\ell uud} C^{\tR \tL,\mu*}_{\ell uud}
\nonumber\\
& +0.56 \big( C^{\tL \tR,\mu}_{\ell uud} C^{\tR \tR,\mu*}_{\ell uud} + C^{\tL \tL,\mu}_{\ell uud} C^{\tR \tL,\mu*}_{\ell uud} \big)
+2.1 C^{\tL \tL,\mu}_{\ell uud} C^{\tR \tR,\mu*}_{\ell uud}
+ \tL \leftrightarrow \tR,
\\
\frac{\Gamma_{n \to e^+ \pi^-} }{(0.1\, \rm GeV)^5} 
=&~ 39.4 \big|C^{\tL\tR,e}_{\ell uud}\big|^2 
+ 40.2 \big|C^{\tL \tL,e}_{\ell uud}\big|^2 
-79.6 \,\Re\big( C^{\tL \tR,e}_{\ell uud} C^{\tL \tL,e*}_{\ell uud} \big)
-0.0026 C^{\tL \tR,e}_{\ell uud} C^{\tR \tL,e*}_{\ell uud}
\nonumber\\
& +0.0026 \big( C^{\tL \tR,e}_{\ell uud} C^{\tR \tR,e*}_{\ell uud} + 
C^{\tL \tL,e}_{\ell uud} C^{\tR \tL,e*}_{\ell uud} 
- C^{\tL \tL,e}_{\ell uud} C^{\tR \tR,e*}_{\ell uud} \big)
+\tL \leftrightarrow \tR,
\\
\frac{\Gamma_{n \to \mu^+ \pi^-} }{(0.1\, \rm GeV)^5} 
=&~ 38.85 \big|C^{\tL\tR,\mu}_{\ell uud}\big|^2 
+ 39.6 \big|C^{\tL \tL,\mu}_{\ell uud}\big|^2 
-78.4 \,\Re\big( C^{\tL \tR,\mu}_{\ell uud} C^{\tL \tL,\mu*}_{\ell uud} \big)
-0.52 C^{\tL \tR,\mu}_{\ell uud} C^{\tR \tL,\mu*}_{\ell uud}
\nonumber\\
& +0.53 \big( C^{\tL \tR,\mu}_{\ell uud} C^{\tR \tR,\mu*}_{\ell uud} + C^{\tL \tL,\mu}_{\ell uud} C^{\tR \tL,\mu*}_{\ell uud} 
- C^{\tL \tL,\mu}_{\ell uud} C^{\tR \tR,\mu*}_{\ell uud} \big)
+ \tL \leftrightarrow \tR,
\\
\frac{\Gamma_{p \to e^+ K^0} }{(0.1\, \rm GeV)^5} 
=&~ 6.7 \big|C^{\tL\tR,e}_{\ell usu}\big|^2 
+ 2.75 \big|C^{\tL \tL,e}_{\ell usu}\big|^2 
+8.6 \,\Re\big( C^{\tL \tR,e}_{\ell usu} C^{\tL \tL,e*}_{\ell usu} \big)
+0.0078 C^{\tL \tR,e}_{\ell usu} C^{\tR \tL,e*}_{\ell usu}
\nonumber\\
& +0.0069 \big( C^{\tL \tR,e}_{\ell usu} C^{\tR \tR,e*}_{\ell usu} + 
C^{\tL \tL,e}_{\ell usu} C^{\tR \tL,e*}_{\ell usu} \big)
+0.0057 C^{\tL \tL,e}_{\ell usu} C^{\tR \tR,e*}_{\ell usu}
+\tL \leftrightarrow \tR,
\\
\frac{\Gamma_{p \to \mu^+ K^0} }{(0.1\, \rm GeV)^5} 
=&~ 6.5 \big|C^{\tL\tR,\mu}_{\ell usu}\big|^2 
+ 2.79 \big|C^{\tL \tL,\mu}_{\ell usu}\big|^2 
+ 8.49 \,\Re\big( C^{\tL \tR,\mu}_{\ell usu} C^{\tL \tL,\mu*}_{\ell usu} \big)
+1.56 C^{\tL \tR,\mu}_{\ell usu} C^{\tR \tL,\mu*}_{\ell usu}
\nonumber\\
& +1.39 \big( C^{\tL \tR,\mu}_{\ell usu} C^{\tR \tR,\mu*}_{\ell usu} + C^{\tL \tL,\mu}_{\ell usu} C^{\tR \tL,\mu*}_{\ell usu} \big)
+1.14 C^{\tL \tL,\mu}_{\ell usu} C^{\tR \tR,\mu*}_{\ell usu}
+ \tL \leftrightarrow \tR,
\\
\frac{\Gamma_{n \to e^- K^+} }{(0.1\, \rm GeV)^5} 
=&~ 6.8 \big|C^{\tL\tR,e}_{\bar \ell dds}\big|^2 
+ 2.8 \big|C^{\tL \tL,e}_{\bar \ell dds}\big|^2 
+8.7 \,\Re\big( C^{\tL \tR,e}_{\bar \ell dds} C^{\tL \tL,e*}_{\bar \ell dds} \big)
+0.0078 C^{\tL \tR,e}_{\bar \ell dds} C^{\tR \tL,e*}_{\bar \ell dds}
\nonumber\\
& +0.007 \big( C^{\tL \tR,e}_{\bar \ell dds} C^{\tR \tR,e*}_{\bar \ell dds} + 
C^{\tL \tL,e}_{\bar \ell dds} C^{\tR \tL,e*}_{\bar \ell dds} \big)
+0.0058 C^{\tL \tL,e}_{\bar \ell dds} C^{\tR \tR,e*}_{\bar \ell dds}
+\tL \leftrightarrow \tR,
\\
\frac{\Gamma_{n \to \mu^- K^+} }{(0.1\, \rm GeV)^5} 
=&~ 6.6\big|C^{\tL\tR,\mu}_{\bar \ell dds}\big|^2 
+ 2.8 \big|C^{\tL \tL,\mu}_{\bar \ell dds}\big|^2 
+ 8.6 \,\Re\big( C^{\tL \tR,\mu}_{\bar \ell dds} C^{\tL \tL,\mu*}_{\bar \ell dds} \big)
+1.57 C^{\tL \tR,\mu}_{\bar \ell dds} C^{\tR \tL,\mu*}_{\bar \ell dds}
\nonumber\\
& +1.4 \big( C^{\tL \tR,\mu}_{\bar \ell dds} C^{\tR \tR,\mu*}_{\bar \ell dds} + C^{\tL \tL,\mu}_{\bar \ell dds} C^{\tR \tL,\mu*}_{\bar \ell dds} \big)
+1.16 C^{\tL \tL,\mu}_{\bar \ell dds} C^{\tR \tR,\mu*}_{\bar \ell dds}
+ \tL \leftrightarrow \tR.
\end{align}
\end{subequations}

\subsection{Decay modes involving a neutrino or an antineutrino}

For the two-body modes involving an antineutrino, we have
\begin{subequations}
\begin{align}
\frac{\Gamma_{p \to \bar\nu_x \pi^+} }{(0.1\, \rm GeV)^5} 
=&~ 39.24 \big|C^{\tL\tR,x}_{ \nu dud}\big|^2 
+ 40 \big|C^{\tL \tL,x}_{\nu dud}\big|^2 
-79.22\,\Re\big( C^{\tL \tR,x}_{\nu dud} C^{\tL \tL,x*}_{\nu dud} \big),
\\
\frac{\Gamma_{n \to \bar\nu_x \pi^0} }{(0.1\, \rm GeV)^5} 
=&~ 19.74 \big|C^{\tL\tR,x}_{ \nu dud}\big|^2 
+ 20.1 \big|C^{\tL \tL,x}_{\nu dud}\big|^2 
-39.8\,\Re\big( C^{\tL \tR,x}_{\nu dud} C^{\tL \tL,x*}_{\nu dud} \big),
\\
\frac{\Gamma_{n \to \bar\nu_x \eta} }{(0.1\, \rm GeV)^5} 
=&~0.095 \big|C^{\tL\tR,x}_{ \nu dud}\big|^2 
+ 8.38 \big|C^{\tL \tL,x}_{\nu dud}\big|^2 
+1.79\,\Re\big( C^{\tL \tR,x}_{\nu dud} C^{\tL \tL,x*}_{\nu dud} \big),
\\
\frac{\Gamma_{p \to \bar\nu_x K^+} }{(0.1\, \rm GeV)^5}
=&~ 3.06 \big|C^{\tL \tR,x}_{\nu uds}\big|^2 
+ 0.73 \big|C^{\tL \tR,x}_{\nu dsu}\big|^2 
+ 11.3 \big|C^{\tL \tR,x}_{\nu sud}\big|^2 
+ 0.74 \big|C^{\tL \tL,x}_{\nu dsu}\big|^2
\nonumber\\
&+ 11.5 \big|C^{\tL \tL,x}_{\nu sud}\big|^2 
- 2.99 \,\Re\big( C^{\tL \tR,x}_{\nu uds} C^{\tL \tR,x*}_{\nu dsu} \big) 
+ 11.77 \,\Re\big( C^{\tL \tR,x}_{\nu uds} C^{\tL \tR,x*}_{\nu sud} \big) 
\nonumber\\
&- 5.76 \,\Re\big( C^{\tL \tR,x}_{\nu dsu} C^{\tL \tR,x*}_{\nu sud} \big)
+ 3.02 \,\Re\big( C^{\tL \tR,x}_{\nu uds} C^{\tL \tL,x*}_{\nu dsu} \big)
- 11.9 \,\Re\big( C^{\tL \tR,x}_{\nu uds} C^{\tL \tL,x*}_{\nu sud} \big) 
\nonumber\\
&- 1.48 \,\Re\big( C^{\tL \tR,x}_{\nu dsu} C^{\tL \tL,x*}_{\nu dsu} \big) 
+5.8 \,\Re\big( C^{\tL \tR,x}_{\nu dsu} C^{\tL \tL,x*}_{\nu sud} \big)
+ 5.8 \,\Re\big( C^{\tL \tR,x}_{\nu sud} C^{\tL \tL,x*}_{\nu dsu} \big)
\nonumber\\
&- 22.87 \,\Re\big( C^{\tL \tR,x}_{\nu sud} C^{\tL \tL,x*}_{\nu sud} \big)
- 5.87 \,\Re\big( C^{\tL \tL,x}_{\nu dsu} C^{\tL \tL,x*}_{\nu sud} \big),
\\
\frac{\Gamma_{n \to \bar\nu_x K^0} }{(0.1\, \rm GeV)^5}
=&~ 0.73 \big|C^{\tL \tR,x}_{\nu uds}\big|^2 
+ 3.03 \big|C^{\tL \tR,x}_{\nu dsu}\big|^2 
+ 11.24 \big|C^{\tL \tR,x}_{\nu sud}\big|^2 
+ 6.37 \big|C^{\tL \tL,x}_{\nu dsu}\big|^2
\nonumber\\
&+ 11.46 \big|C^{\tL \tL,x}_{\nu sud}\big|^2 
- 2.97 \,\Re\big( C^{\tL \tR,x}_{\nu uds} C^{\tL \tR,x*}_{\nu dsu} \big) 
- 5.72 \,\Re\big( C^{\tL \tR,x}_{\nu uds} C^{\tL \tR,x*}_{\nu sud} \big) 
\nonumber\\
&+ 11.67 \,\Re\big( C^{\tL \tR,x}_{\nu dsu} C^{\tL \tR,x*}_{\nu sud} \big)
- 4.3 \,\Re\big( C^{\tL \tR,x}_{\nu uds} C^{\tL \tL,x*}_{\nu dsu} \big)
+ 5.77 \,\Re\big( C^{\tL \tR,x}_{\nu uds} C^{\tL \tL,x*}_{\nu sud} \big) 
\nonumber\\
&+ 8.78 \,\Re\big( C^{\tL \tR,x}_{\nu dsu} C^{\tL \tL,x*}_{\nu dsu} \big) 
- 11.78 \,\Re\big( C^{\tL \tR,x}_{\nu dsu} C^{\tL \tL,x*}_{\nu sud} \big)
+ 16.9 \,\Re\big( C^{\tL \tR,x}_{\nu sud} C^{\tL \tL,x*}_{\nu dsu} \big)
\nonumber\\
&- 22.7 \,\Re\big( C^{\tL \tR,x}_{\nu sud} C^{\tL \tL,x*}_{\nu sud} \big)
- 17.1 \,\Re\big( C^{\tL \tL,x}_{\nu dsu} C^{\tL \tL,x*}_{\nu sud} \big).
\end{align}
\end{subequations}
Similar expressions apply to the modes involving a neutrino upon simultaneously interchanging $\nu\leftrightarrow\bar\nu$ and $\tL\leftrightarrow \tR$. 

Note that, due to approximate isospin symmetry, matrix elements between $p$-to-$\pi$ and 
$n$-to-$\pi$ transitions are related to each other:
$\langle \pi^- | \calO^{\chi\chi^\prime,x}_{\ell uud} | n \rangle = \sqrt{2} \langle \pi^0 | \calO^{\chi\chi^\prime,x}_{\ell uud} | p \rangle$
and $\langle \pi^+ | \calO^{\chi\chi^\prime,x}_{\nu(\bar\nu) dud} | p \rangle = -\sqrt{2} \langle \pi^0 | \calO^{\chi\chi^\prime,x}_{\nu(\bar\nu) dud} | n \rangle$, where $\chi,\chi^\prime = \tL,\tR$.
These relations directly imply the decay width relations, $\Gamma(n\to \ell^+\pi^-)\approx2\Gamma(p\to \ell^+\pi^0)$
and $\Gamma(p\to \hat\nu\pi^+)\approx 2\Gamma(n\to \hat\nu\pi^0)$, 
as can be seen explicitly from the above expressions.

\bibliography{reference}{}
\bibliographystyle{JHEP}
\end{document}